\documentclass[structabstract]{aa}

\usepackage{natbib}
\bibpunct{(}{)}{;}{a}{}{,} 
\usepackage{graphicx}

\usepackage{txfonts}

\newcommand{\Alfven}{Alfv\'en}
\newcommand{\Reyn}{\mathsf{Re}}
\newcommand{\Mach}{\mathsf{M}}
\newcommand{\Alfv}{\mathsf{A}}
\newcommand{\eqref}[1]{Eq.\,(\ref{#1})}

\newcommand{\figref}[1]{Fig.\,\ref{#1}}
\newcommand{\tabref}[1]{Tab.\,\ref{#1}}
\newcommand{\secref}[1]{Sect.\,\ref{#1}}

\newcommand{\absatz}[1]{\paragraph{\textbf{#1:}}}

\begin{document}

\title{
  Local simulations of the magnetized Kelvin-Helmholtz instability in
  neutron-star mergers
}
\author{
  M.~Obergaulinger
  \and
  M.~A.~Aloy
  \and
  E.~M{\"u}ller 
}
\institute{
  Max-Planck-Institut f{\"u}r Astrophysik, Garching bei M{\"u}nchen 
  \and
  Departamento de Astronom{\'i}a y Astrof{\'i}sica, Universidad de Valencia
}
\date{Received day month year / Accepted day month year}

{\abstract
  {
    Global magnetohydrodynamic simulations show the growth of
    Kelvin-Helmholtz instabilities at the contact surface of two
    merging neutron stars.  That region has been identified as the
    site of efficient amplification of magnetic fields. However, these
    global simulations, due to numerical limitations, were unable to
    determine the saturation level of the field strength, and thus the
    possible back-reaction of the magnetic field onto the flow.  }
  {
    We investigate the amplification of initially weak magnetic fields
    in Kelvin-Helmholtz unstable shear flows, and the back-reaction of
    the field onto the flow.  }
  {
    We use a high-resolution finite-volume ideal MHD code to perform
    2D and 3D local simulations of hydromagnetic shear flows, both for
    idealized systems and simplified models of merger flows.  }
  {
    In 2D, the magnetic field is amplified on time scales of less than
    $0.01\,$ms until it reaches locally equipartition with the kinetic
    energy.  Subsequently, it saturates due to resistive instabilities
    that disrupt the Kelvin-Helmholtz unstable vortex and decelerate
    the shear flow on a secular time scale.  We determine scaling laws
    of the field amplification with the initial field strength and the
    grid resolution.  In 3D, the hydromagnetic mechanism seen in 2D
    may be dominated by purely hydrodynamic instabilities leading to
    less filed amplification.  We find maximum magnetic fields $\sim
    10^{16}\,$G locally, and r.m.s.\ maxima within the box $\sim
    10^{15}\,$G.  However, due to the fast decay of the shear flow
    such strong fields exist only for a short period ($< 0.1\,$ms).
    In the saturated state of most models, the magnetic field is
    mainly oriented parallel to the shear flow for rather strong
    initial fields, while weaker initial fields tend to lead to a more
    balanced distribution of the field energy among the components.
    In all models the flow shows small-scale features.  The magnetic
    field is at most in energetic equipartition with the decaying
    shear flow. }
  {
    The magnetic field may be amplified efficiently to very high field
    strengths, the maximum field energy reaching values of the order
    of the kinetic energy associated with the velocity components
    transverse to the interface between the two neutron stars.
    However, the dynamic impact of the field onto the flow is limited
    to the shear layer, and it may not be adequate to produce
    outflows, because the time during which the magnetic field stays
    close to its maximum value is short compared to the time scale for
    launching an outflow (i.e., a few milliseconds).}

  \keywords{MHD - Instabilities - Stars: magnetic fields - Stars: neutron}
}

\maketitle

\section{Introduction}
\label{Sec:Intro}
%
The merger of two neutron stars is considered the most promising
scenario for the generation of short gamma-ray bursts (GRBs).  After a
phase of inspiral due to the loss of angular momentum and orbital
energy by gravitational radiation, the merging neutron stars are
distorted by their mutual tidal forces.  Finally, they touch each
other at a contact surface.  Due to a combination of the orbital
motion and the rotation of the neutron stars, the gas streams along
that surface, the flow directions on either side of the surface being
anti-parallel with respect to each other.

As a consequence of this jump in the tangential velocity, the contact
surface is Kelvin-Helmholtz (KH) unstable.  Growing within a few
milliseconds, the KH instability leads to the formation of typical KH
vortices between the neutron stars.  These vortices can modify the
merger dynamics via the dissipation of kinetic into thermal energy.
The generation of KH vortices is observed in actual merger numerical
simulations
\citep[e.g.,][]{Oechslin_etal__2007__AA__NS-NS-merger--EOS-1}.

The exponential amplification of seed perturbations can lead to very
strong magnetic fields as shown by
\cite{Price_Rosswog__2006__Sci__NS-NS-merger-B-amplif}, and
\cite{Rosswog__2007__RevMex__NS-merger}.  These fields, in turn, can
modify the dynamics of the instability described above, either already
during its linear growth phase or, for weak fields, in the saturated
state.  Exerting stresses and performing work on the fluid, the
magnetic field does lose part of its energy.  Thus, the maximum
attainable field strength is limited by the non-linear dynamics.

In their merger simulations,
\cite{Price_Rosswog__2006__Sci__NS-NS-merger-B-amplif}, and
\cite{Rosswog__2007__RevMex__NS-merger} observed fields exceeding by
far $10^{15}\,$G.  Their numerical resolution, however, did not allow
them to follow the detailed evolution of the KH instability in the
non-linear phase.  Thus, they could not draw any definite conclusions
on the maximum strength of the field nor its back-reaction onto the
fluid.  They observed that the maximum field strength is a function of
the numerical resolution: the better the resolution, the stronger
becomes the field.

Performing numerical convergence tests, these authors did not find an
upper bound for the field strength attainable in the magnetized KH
instability.  Thus,
\citet{Price_Rosswog__2006__Sci__NS-NS-merger-B-amplif} discussed,
based on energetic arguments, but not supported by simulation results,
two different saturation levels: the field growth saturates when the
magnetic energy density equals either the kinetic (\emph{kinetic
  equipartition}) or the internal energy of the gas (\emph{thermal
  equipartition}), corresponding to fields of the order of
$10^{16}\,$G and $\sim 10^{18}\,$G, respectively.  From their
simulations they were not able to identify the saturation mechanism
applying to the KH instability in neutron-star mergers.  Thus, we
address this question here again using highly resolved simulations and
independent numerical methods.

Most simulations of neutron-star mergers, including the ones by
\cite{Price_Rosswog__2006__Sci__NS-NS-merger-B-amplif}, and
\cite{Rosswog__2007__RevMex__NS-merger}, are performed using
smoothed-particle hydrodynamics (SPH)
\citep{Monaghan__1992__ARAA__SPH}.  This free Lagrangian method is
highly adaptive in space, and allows on to follow large density
contrasts without ``wasting'' computational resources in areas of very
low density.  This property of SPH makes it highly advantageous for
the problem of mergers.  On the other hand, its relatively high
numerical viscosity renders SPH inferior compared to Eulerian
grid-based schemes for the treatment of (magneto-)hydrodynamic
instabilities and turbulence
\citep{Agertz_etal__2007__MNRAS__SPH-grid-comparison}.  Moreover, the
spatial resolution of most merger simulations is rather low, i.e., the
reliability of their results concerning the details of the KH
instability is limited.

A grid-based code such as ours is well suited for a study of flow
instabilities and turbulence.  Using it to simulate the entire merger
event, however, is cumbersome due to the large computational costs
required to cover the entire system with an appropriate computational
grid.  In spite of this fact,
\cite{Giacomazzo_Rezzolla_Baiotti__2009__PRL__NS_mergers_MHD} (see
also \cite{Liu_etal__2008__prd__GRMHD_NS_mergers,
  Anderson_etal__2008__PRL__NS_mergers_MHD_GW}) have performed full
general-relativistic MHD simulations using vertex-centered mesh
refinement to assess the influence of magnetic fields on the merger
dynamics and the resulting gravitational waveform.  But, as we shall
show below, even their (presently world-best) grid resolution ($h \sim
350\,$m) is still too crude to properly capture the disruptive
dynamics after the KH amplification of the field. For comparison, we
note here that our merger models employ a grid resolution of $h \sim
0.1\,$m in 2D (\secref{sSek:mrgr-2d}) and $h \sim 0.8\,$m in 3D
(\secref{sSek:mrgr-3d}), respectively.

We performed a set of numerical simulations of the KH instability to
understand the dynamics of magnetized shear flows and to draw
conclusions on the evolution of merging neutron stars.  The main
issues we address in our study are motivated by two different, albeit
related, intentions:
\begin{itemize}
\item We strive for a better understanding of the magneto-hydrodynamic
  (MHD) KH instability.  This includes the influence of numerical
  parameters such as the grid resolution on the dynamics, and generic
  properties of the saturation of the instability.  We address these
  questions by a series of \emph{dimensionless} models that use
  scale-free parameters as most previous studies focusing on the
  generic properties of the KH instability instead of a particular
  astrophysical application.
\item We further want to verify the results of
  \cite{Price_Rosswog__2006__Sci__NS-NS-merger-B-amplif} and reassess
  their estimates of the saturation field strength. Hence, we consider
  the growth time of the instability that has to compete with the
  dynamical time scale of the merger event (a few milliseconds), the
  saturation mechanism, the saturation field strength, and generic
  dynamical features of supersonic shear flows.  Our results should
  also allow us to reassess the findings of global simulations
  extending the ones performed by
  \cite{Price_Rosswog__2006__Sci__NS-NS-merger-B-amplif}, e.g., the
  simulations by \cite{Anderson_etal__2008__PRL__NS_mergers_MHD_GW}
  and \cite{Liu_etal__2008__prd__GRMHD_NS_mergers}.
\end{itemize}

To this end we utilize a newly developed multidimensional MHD code
\citep{Obergaulinger_etal__2009__AA__Semi-global_MRI_CCSN} that
employs various explicit finite-volume algorithms, and that is
particularly well suited for simulating instabilities and turbulent
systems.  As the code is based on Eulerian high-resolution methods
instead of SPH as in
\cite{Price_Rosswog__2006__Sci__NS-NS-merger-B-amplif}, our results
are complementary to theirs, serving as an independent check.

Since we are unable to simulate the entire merger event using fine
resolution, we focus on the evolution of a small, representative
volume around the contact surface.  This \emph{local} simulation
allows us to concentrate on the dynamics of the magnetohydrodynamic KH
instability.  However, as our simulations lack the feedback from the
dynamics occurring on scales larger than the simulated volume, its
influence has to be mimicked by suitably chosen boundary conditions.
We neglect neutrino radiation, and the gas obeys either an ideal-gas
or a hybrid (barotropic and ideal-gas) equation of state (EOS), the
latter serving as a rough model for nuclear matter.

This paper is organized as follows.  We describe the physics of the
magnetohydrodynamic KH instability in \secref{Sek:KH-Phys}, and our
numerical code in \secref{Sek:PhysNum}.  We discuss the simulations
addressing generic properties of the KH instability in two and three
spatial dimensions in \secref{Sek:KH2} and \secref{Sek:3d-mod},
respectively. The results applying to neutron-star mergers are given
in \secref{Sek:mrgr}.  Finally, we present a summary and conclusions
of our work in \secref{Sek:Summary}.

\section{The magnetohydrodynamic KH instability}
\label{Sek:KH-Phys}
%
The KH instability leads to exponential growth of perturbations in a
non-magnetized shear layer of a fluid of background density $\rho$
\citep[e.g.,][]{Chandrasekhar__1961__Buch__HD-HM-stab}.  If a
plane-parallel shear layer extends over a thickness $d$, all modes
with wavelengths $\lambda > d$ are unstable, shorter modes growing
faster.  After a phase of exponential growth, a stable KH vortex
forms.

If the shear layer is threaded by a magnetic field of field strength,
$b$, parallel to the shear flow (the $x$-direction in our
models), magnetic tension stabilizes all modes, if the \emph{Alfv\'en
  number} of the shear flow
\begin{equation}
  \label{Gl:KH-Phys--strong-field-stab}
  \Alfv  \equiv  U_0 / c_\mathrm{A}  < 2,
\label{alfnum}
\end{equation}
where $U_0$ and $c_\mathrm{A} \equiv \sqrt { b^2 / \rho}$ are the
velocity difference across the shear layer, and the Alfv\'en velocity,
respectively.  If the field is weaker, the instability can develop
similarly to the non-magnetic case, but its growth and its non-linear
saturated state are affected significantly
\citep[e.g.,][]{Frank_etal__1996__ApJ__MHD-KH-2d-1,
  Jones_etal__1997__ApJ__MHD-KH-2d-2,
  Jeong_etal__2000__Apj__MHD-KH-2d-3,
  Ryu_etal__2000__ApJ__MHD-KHI-3d}.

A magnetic field \emph{perpendicular} to the shear flow and the
shearing interface (a $b_y$ field in our models) is sheared into a
parallel $b_x$ field. Thus, the resulting flow dynamics is similar.  A
field orthogonal to the shear flow but parallel to the interface (a
$b_z$ field in our models) acts mainly by adding magnetic pressure to
the thermal one, thus modifying the dynamics of the KH instability
only if its strength approaches or exceeds the equipartition field
strength.  Hence, we focus here on fields in the direction of the
flow, only.

Depending on the field strength, the above authors identified three
different regimes concerning the dynamics of the instability.

Rather strong fields with an Alfv\'en number slightly below 2 lead
to \emph{non-linear stabilization}.  Too weak for stabilization
initially, the field is amplified by the instability, and after less
than one turnover of the KH vortex, it is strong enough to suppress
further winding.  The field, concentrated in thin sheets, annihilates
in localized reconnection and, mediating the conversion of kinetic via
magnetic into internal energy, destroys the vortex.  The late phases
of the evolution consist of a very broad transition layer between
those parts of the fluid moving in opposite directions.  The flow is
almost entirely parallel to the initial shear layer, and no vortex is
retained.  The magnetic field has decreased strongly due to
reconnection, and is still concentrated in sheet-like patterns.

Weaker fields give rise to \emph{disruptive dynamics}.  The
amplification process takes longer to produce strong fields, i.e., the
vortex survives several turnover times.  The field is wound up in
increasingly thin sheets, that eventually reconnect due to (numerical)
resistivity.  Afterwards the dynamics is similar to the previous case:
the vortex is disrupted, leading to a broad laminar transition region
threaded by filamentary magnetic fields.

For even weaker fields one encounters the flow regime of
\emph{dissipative dynamics}.  Even after a long phase of
amplification, the field is still too weak to affect the flow.
Reconnection occurs, but due to the weakness of the involved fields,
it leads only to a gradual conversion of kinetic into internal energy.
The global topology of the flow does not change as in the previous
cases, and the vortex exists throughout the evolution. Its velocity
decreases slowly as kinetic energy is extracted from the vortex.

We note that the transition between these three dynamic regimes is not
sharp.  In particular, it is not possible to define a threshold
Alfv\'en number separating disruptive and dissipative dynamics.

Further complications arise in three spatial dimensions.  Here, the KH
vortex can be disrupted even without the presence of a magnetic field
by purely hydrodynamic instabilities
\citep{Ryu_etal__2000__ApJ__MHD-KHI-3d}, and the effects of a magnetic
field overlay with those of the non-magnetic instabilities.

\section{Numerical methods}
\label{Sek:PhysNum}
%
We use a newly developed high-resolution code to solve the equations
of ideal (Newtonian) MHD (Einstein's summation convention applies),
\begin{equation}
  \label{Gl:Phys--MHD-rho}
  \partial_{t} \rho 
  + \nabla_j \left[ \rho v^j \right]
  = 0
  ,
\end{equation}
\begin{equation}
  \label{Gl:Phys--MHD-mom}
  \partial_{t} p^i 
  + \nabla_j 
  \left[ 
    p^i v^j + P_{\star} \delta^{ij} - b^i b^j
  \right]
  = f^i,
\end{equation}
\begin{equation}
  \label{Gl:Phys--MHD-erg}
  \partial_{t} e_{\star}
  + \nabla_j 
  \left[ 
    \left( e_{\star} + P_{\star} \right) v^j
    - b^i v_i b^j
  \right]
  = f_j v^j
  ,  
\end{equation}
\begin{equation}
  \label{Gl:Phys--MHD-ind}
  \partial_{t} \vec b 
  = 
  - c \ \vec \nabla \times \vec E,
\end{equation}
\begin{equation}
  \label{Gl:Phys--MHD-div}
  \nabla_j b^j = 0,
\end{equation}
where the mass density, momentum density, velocity, and total-energy
density of the gas are denoted by $\rho$, $\vec p$, $\vec v$, and
$e_{\star}$, respectively; $\vec b$ is the magnetic field.  The
total-energy density and the total pressure, $P_{\star}$, are composed
of fluid and magnetic contributions: $e_{\star} = \varepsilon + \rho
\vec v^2/2 + \vec b^2/2$, and $P_{\star} = P + \vec b^2/2$, where
$\varepsilon$ and $P = P( \rho, \varepsilon, \dots)$ are the internal
energy density and the gas pressure, respectively.  The electric
field, $\vec E$, is given by $\vec E = - ({\vec v}/c) \times \vec b$
with $c$ being the speed of light in vacuum.  The external force,
arises from gravity, i.e, $\vec f = \vec f_\mathrm{G} = - \rho \vec
\nabla \Phi$, where $\Phi$ is the gravitational potential.

The above equations are implemented into our code in their
finite-volume form.  We use Eulerian high-resolution shock-capturing
methods for their solution \citep[see,
  e.g.,][]{LeVeque_Book_1992__Conservation_Laws}.  To reconstruct the
zone interface values of variables defined as volume averages over
grid zones, we use high-order algorithms of one of the following
types:
\begin{itemize}
  \item \emph{Piecewise-linear} reconstruction using
    \emph{total-variation diminishing (TVD)} methods
    \citep{Harten__1983__JCP__HR_schemes}.  While formally
    $2^\mathrm{nd}$ order accurate in smooth parts of the flow and
    away from local extrema, these methods achieve a stable
    representation of discontinuities by reverting to
    $1^\mathrm{st}$-order accurate piecewise-constant reconstruction.
    The accuracy of the scheme depends on its slope limiter for which
    different choices are possible, e.g., the Minmod, the van Leer, or
    the MC (monotonized central) limiters.
  \item The class of \emph{weighted essentially non-oscillatory
    (WENO)} algorithms \citep{Liu_etal__1994__WENO} offer a way of
    constructing schemes of arbitrarily high order of accuracy.  In
    these methods, an interpolant for a variable at a given point in
    space (e.g., a zone interface) is constructed from a number of
    candidate polynomials by maximizing a measure of the smoothness of
    these polynomials.  In our scheme, based on the one described by
    \cite{Levy_etal__2002__SIAM_JSciC__WENO4}, we use three candidate
    parabolas, leading to a nominal order of accuracy of $4$.
  \item \cite{Suresh_Huynh__1997__JCP__MP-schemes} use a
    generalization of the TVD criterion to construct high-order
    \emph{monotonicity-preserving (MP)} schemes.  The new MP stability
    and accuracy constraints do not lead to the clipping of extrema in
    smooth regions of the flow that is innate to the TVD criteria.
    Thus, they allow for a higher accuracy in smooth flows while
    retaining stability close to discontinuities.
    \cite{Suresh_Huynh__1997__JCP__MP-schemes} give MP schemes of
    formally $5^\mathrm{th}$, $7^\mathrm{th}$, and $9^\mathrm{th}$
    order that we implemented in our code.
\end{itemize}

We compute the fluxes of the MHD equations from the reconstructed
interface states using approximate Riemann solvers.
\cite{Titarev_Toro__2005__IJNMF__MUSTA} and
\cite{Toro_Titarev__2006__JCP__MUSTA} developed \emph{multi-stage
  (MUSTA)} Riemann solvers that are built on a combination of
predictor and corrector steps using simple approximate Riemann
solvers.  These solvers do not require a computationally expensive
decomposition of the MHD state into characteristic variables, yet they
achieve an accuracy comparable to exact solvers.

In MHD simulations, it is important to use a numerical scheme that
keeps the magnetic field divergence-free.  To this end we employ in
our code the \emph{constraint-transport (CT)} scheme of
\citep{Evans_Hawley__1998__ApJ__CTM} that uses a spatial
discretization of the magnetic field consistent with the curl operator
in the induction equation, leading to a staggering of the collocation
points of $\vec b$ with respect to those of the hydrodynamic variables
$\rho$, $\vec p$, and $e_\star$.  
According to the definition of $\vec b$ the electric field, $\vec E$,
is defined as the average over the zone edges.  The staggering of
$\vec b$ requires interpolations between the staggered grids (to
obtain, e.g., the Maxwell stress $b^i b^j$; see
Eqn.\,(\ref{Gl:Phys--MHD-mom})), and special care has to be taken in
the computation of the electric field from the (zone-centered)
velocity and the (zone-interface) magnetic field.
Various implementations of the CT scheme have been devised that differ
mainly in the way the magnetic stress and electric field are
calculated.  Of these, our implementation resembles most closely the
recently developed upwind-CT schemes
\citep{Londrillo_Del_Zanna__2004__JCP__UpwindCT,
  Gardiner_Stone__2005__JCP__Athena-code,
  Gardiner_Stone__2008__JCP__Athena-code}.  We obtain $\vec E$ from
the zone interface values of the velocity and the magnetic field that
are both computed by the (MUSTA) Riemann solver.  This guarantees that
the electric field is consistent with the solution of the Riemann
problem.

Our code is written in FORTRAN\,90 and parallelized for shared or
distributed memory computers using the OpenMP or MPI programming
paradigm, respectively.  The code successfully passed various standard
tests including MHD shock tube problems \citep[e.g., the ones
  published by][]{Ryu_Jones__1995__ApJ__MHD_1d}, the propagation of
MHD waves, and some multi-dimensional flow problems such as the
Orszag-Tang vortex \citep{Orszag_Tang__1979__JFM__MHD_turb}.  These
tests demonstrate the stability and accuracy of the code in handling
flows involving discontinuities and turbulent structures.  According
to the results of the wave-propagation tests, the order of accuracy of
the code is 2, 3.3, and 4.1 for piecewise-linear, MP, and WENO
reconstruction, respectively
\citep{Obergaulinger__2008__PhD__RMHD}. The code has also been used to
study the magneto-rotational instability (MRI) in core collapse
supernovae \citep{Obergaulinger_etal__2009__AA__Semi-global_MRI_CCSN}.

The simulations reported in this paper were performed with MP
reconstruction based on $5^\mathrm{th}$-order polynomials (the
\textit{MP5} method), and the MUSTA solver derived from the HLL
Riemann solver.  This reconstruction method represents a good
trade-off between accuracy and computational costs.  Methods based on
higher-order polynomials increase the accuracy of the code, but at the
expense of a larger stencil, reducing the efficiency of the parallel
code, since the number of ghost zones that have to be communicated
among different processors is larger.  The same adverse effect on the
computational efficiency can be observed when comparing our WENO
reconstruction to MP5.

\section{The KH instability in 2D planar magnetized shear flows}
\label{Sek:KH2}
%
We performed a set of two dimensional simulations to study the
properties of the KH instability in 2D planar magnetized shear flows.
These simulations allow us to validate our numerical tool and to
assess the significance of results obtained in simulations aiming at
an understanding of the KH instability in neutron-star mergers.

As we shall show below we reproduce, but also extend the results
obtained by \cite{Frank_etal__1996__ApJ__MHD-KH-2d-1}, \cite{
  Jones_etal__1997__ApJ__MHD-KH-2d-2},
\cite{Baty_etal__2003__PhysPlas__2dMHD-KH-compress}, and
\cite{Keppens_etal__1999__PP__MHD-KH} which are summarized in
Sect.\,\ref{Sek:KH-Phys}.

We consider both subsonic and supersonic 2D planar shear flows in the
$x-y$ plane in $x$-direction with an initial velocity profile given by
(note that all numerical values are given in dimensionless code units
in the following!)
\begin{equation}
  \label{Gl:KH-2d--shearprof}
  \left( v_x, v_y \right)^T
  =
  \left(v_0 \tanh \frac{y}{a}, 0 \right)^T \, ,
\end{equation}
where $U_0 = 2 v_0$ is the shear velocity, and $a$ is a length scale
characterizing the width of the shear flow.  The background density
and pressure are uniform, and the thermodynamic properties of the
fluid are described by an ideal-gas EOS with an adiabatic index
$\Gamma$,
\begin{equation}
  \label{Gl:KH-2d--idgasEOS}
  P = \left( \Gamma - 1 \right) \varepsilon \, ,
\end{equation}
where $\varepsilon = e_\star - \frac{1}{2} \rho \vec v ^ 2 -
\frac{1}{2} \vec b ^2$ is the internal energy density of the fluid.
Initially, a uniform magnetic field $\vec b (t=0) = \left( b_x^0,
.b_y^0 \right)^T$ threads the shear layer.

To trigger the KH instability we perturb the shear flow by a
transverse velocity
\begin{equation}
  \label{Gl:KH-2d--perturb}
  v_y ( t = 0 ) = v_y^0\, f ( y )\, \sin ( k_x x ) \, ,
\end{equation}
where $f$ (with $f (y) \in [0,1]$) is a function localized at the
shearing interface, i.e., it vanishes beyond a distance $a'$ from the
interface.  We set $a' = 4a$ here.  The maximum perturbation velocity,
$v_y^0$, is typically a factor $10^{6 \dots 8}$ smaller than the shear
velocity.  To test the influence of the form of the perturbations, we
also simulated some models (both magnetic and non-magnetic ones) with
random perturbations which do not select \textit{a priori} a single
sinusoidal unstable mode (see below).

Finally, we introduce the \emph{volume-averaged kinetic energy
  densities}
\begin{equation}
  e_\mathrm{kin}^i \equiv \frac{1}{\mathcal{V}} 
                         \int \mathrm{d} \mathcal{V}\, \frac{1}{2} \rho v_i^2 \,,
\label{kiny}
\end{equation}
and \emph{volume-averaged magnetic energy densities}
\begin{equation}
  e_\mathrm{mag}^i \equiv \frac{1}{\mathcal{V}} 
                         \int \mathrm{d} \mathcal{V}\, \frac{1}{2} b_i^2 \,,
\label{magy}
\end{equation}
with $i \in {x,y,z}$. These quantities will be useful for the
following discussion.

\subsection{Linear growth}
%
Our code reproduces the growth rate of the KH instability very
accurately.  To demonstrate this we recalculated some of the models
studied by \cite{Keppens_etal__1999__PP__MHD-KH} (models grw-$n$ in
Tab.\,\ref{Tab:n2d-grow-models}).  The growth rates for these models
are either given in \cite{Keppens_etal__1999__PP__MHD-KH}, or can be
obtained from the figures of
\cite{Miura_Pritchett__1982__JGR__MHD-KH-stability}.

The models have a uniform background density $\rho_0 = 1$, and a
uniform background pressure $P_0$.  We impose open boundary conditions
in the transverse ($y$) direction, periodic ones in $x$-direction, and
vary the value of the shear velocity, the width of the shear layer,
and the grid resolution.  

\begin{figure}[htbp]
  \centering  \includegraphics[width=7cm]{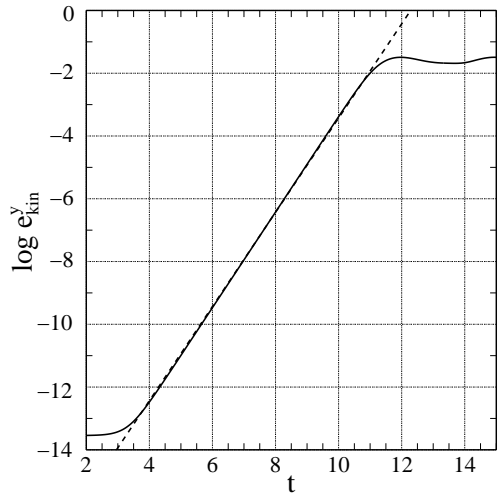}
  \caption{Linear growth phase of the KH instability in model grw-3.
    The solid black line shows the volume-averaged kinetic energy
    density $e_{\mathrm{kin}}^y$, as a function of time $t$. The
    dashed line gives the theoretical growth rate. }
  \label{Fig:grw-1--growth}
\end{figure}

We derive growth rates, $\Gamma_\mathrm{num}$, from the exponential
growth of $e_\mathrm{kin}^y$, and compare these to the values,
$\Gamma_\mathrm{MP}$, given by
\cite{Miura_Pritchett__1982__JGR__MHD-KH-stability} and
\cite{Keppens_etal__1999__PP__MHD-KH}, respectively.  We note in this
respect that $e_\mathrm{kin}^y (t) \propto v_y^2 \propto \left( \exp
{\Gamma t} \right)^2$ (see Eq.\,\ref{kiny}) grows at twice the rate of
the KH instability.  The agreement between the theoretical predictions
and our numerical results is, in general, very good (see
Tab.\,\ref{Tab:n2d-grow-models} and Fig.\,\ref{Fig:grw-1--growth}).

After the initial phase of exponential growth, a roughly circular
vortex develops in the perturbed non-magnetized shear layer which
should be eventually dissipated by (numerical) viscosity.  However,
this process is very slow for our models (we see no sign of
dissipation until the end of our simulations), as the numerical
viscosity of our code is very low.

\begin{figure}
  \centering
  \includegraphics[width=7cm]{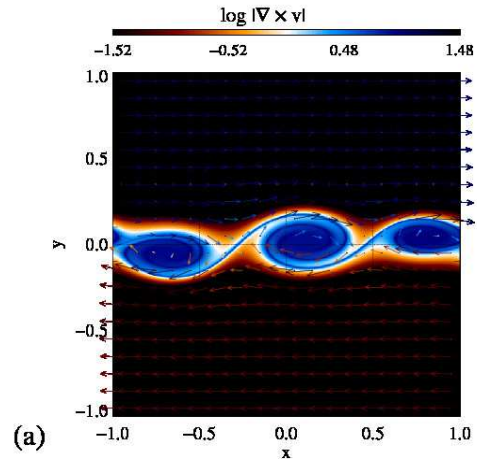}
  \includegraphics[width=7cm]{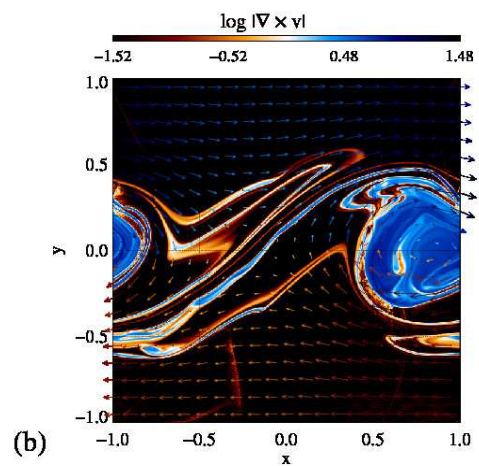}
  \caption{Logarithm of the modulus of the flow vorticity and the
    velocity field (vectors) of a non-magnetic model with $\Mach = 1$
    and a random perturbation at $t = 16$ (panel (a)), and $t = 25.5$
    (panel (b)), respectively.  }
  \label{Fig:KH2-HD--randompert}
\end{figure}

The formation of a \emph{single} KH vortex rather than of a multitude
of small vortices is not an artifact of the form of the initial
perturbation (\eqref{Gl:KH-2d--perturb}).  To demonstrate this, we
simulated a non-magnetic model with random rather than sinusoidal
perturbations of the transverse velocity with an amplitude of
$10^{-6}$ of the shear velocity (see \figref{Fig:KH2-HD--randompert}
for two snapshots of the model simulated with $1024^2$ zones at $t =
16$ and $t = 25.5$ (panels (a) and (b), respectively).  Initially,
three small KH vortices develop (panel (a)), but after two subsequent
mergers of these vortices, only one large vortex remains (panel (b)),
resembling closely the flow field of a model with sinusoidal
perturbation.  Due to this evolution towards a single large-scale
vortex, we focus on models with sinusoidal perturbations in the
following \footnote{Without elaborating in more detail, we note that a
  similar result holds for magnetized models.}.

\subsection{Non-magnetic models}
\label{sSek:2dhydro}
%
We simulated a set of non-magnetic models (summarized in
\tabref{Tab:n2d-HD-models}) to study the influence of the box size and
boundary conditions on the evolution of transonic and supersonic
($\Mach = U_0 / c_\mathrm{s} \ge 1.8$) shear flows.  As noted by
\cite{Miura_Pritchett__1982__JGR__MHD-KH-stability}, there is no
growing mode for a $\Mach \ge 2$ shear flow, but in models with closed
boundaries we find nevertheless a growing instability whose growth
mechanism is, however, different (see below).

\begin{figure}[htbp]
  \centering
  \includegraphics[width=7cm]{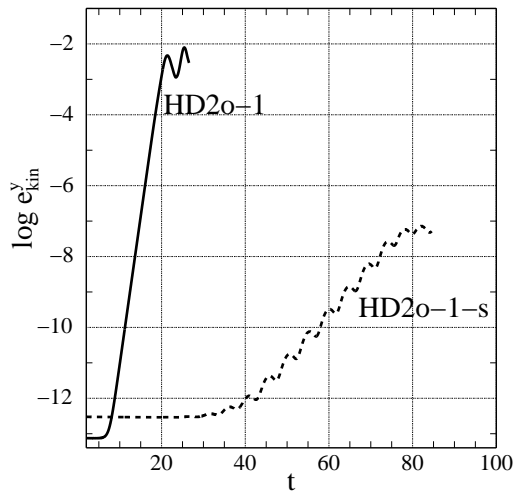}
  \caption{Volume averaged kinetic energy density $e^y_\mathrm{kin}$
    of models HD2o-1 (solid line; $l_y = 2$) and HD2o-1-s (dashed
    line; $l_y = 0.5$) as a function of time illustrating the
    influence of the size of the vertical extent of the computational
    domain $l_y$.  }
  \label{Fig:HD-1--growth}
\end{figure}

We first consider models with $\Mach = 1.8$. For these models the
instability grows faster when the vertical domain size is enlarged,
and open boundaries yield larger growth rates than reflecting ones.
The reason for this behavior is that the instability affects a larger
region of the flow than in the case of slow shear flows.  To
demonstrate this we compare models HD2o-1 and HD2o-1-s that differ
only in the size of the computational domain in $y$-direction: $y \in
[-1; 1]$ and $y \in [-0.25; 0.25]$ for models HD2o-1 and HD2o-1-s,
respectively.  According to Fig.\,\ref{Fig:HD-1--growth} the
volume-averaged kinetic energy density, $e_\mathrm{kin}^y$, grows
faster and leads to much larger values in model HD2o-1 than in model
HD2o-1-s.  Furthermore, in model HD2o-1-s the growth of
$e_\mathrm{kin}^y$ shows superimposed oscillations.  In both models
waves are created at the shear layer which travel outwards in
$y$-direction carrying (transverse) kinetic energy.  If the waves are
allowed to travel over a sufficiently long distance $\delta y$ (which
is the case for model HD2o-1), they steepen into shock waves when the
fluid velocity exceeds the sound speed.  The shocks propagate mainly
in $x$-direction, advected by the shear flow.  Kinetic energy is
dissipated into internal one in these shocks, and the flow develops a
vortex-like structure.  If the boundaries of the computational domain
are too close to the shear layer, the waves leave the domain before
they can affect the flow, i.e., the growth rate is reduced.  Each time
a wave leaves the computational domain, it carries away kinetic energy
giving rise to the oscillations of $e_\mathrm{kin}^y$ visible in
Fig.\,\ref{Fig:HD-1--growth}.

For an intermediate domain size of $y \in [ - 0.5; 0.5 ]$ (model
HD2o-1-s), we find despite the absence of oscillations a smaller
growth rate than for models HD2o-1 ($y \in [ -1; 1]$) and HD2o-1-1 ($y
\in [ -2; 2]$), respectively. The boundaries are sufficiently close to
the shear layer to affect the growth of the instability.  Saturation
occurs by the same mechanism as in case of a larger domain, namely by
the development of shock waves.

The distance the waves travel in transverse direction increases with
increasing Mach number of the shear flow.  For $\Mach = 1$ the waves
are contained essentially in the region $y \in [ -0.25; 0.25 ]$ (a
version of model grw-3 simulated on a smaller grid of $200 \times 100$
zones covering a domain of $[-0.5; 0.5] \times [-0.25; 0.25]$ does not
show oscillation of $e_{\mathrm{kin}}^y$).  For the same reason the
evolution does not depend on whether one imposes reflecting or open
boundary conditions (compare models HD2r-0 in
Tab.\,\ref{Tab:n2d-HD-models} and grw-3 in
Tab.\,\ref{Tab:n2d-grow-models}).  Thus, to encounter a rapidly
growing instability in a fast shear flow, one has to simulate a
sufficiently large domain, or alternatively to use reflecting
boundaries in $y$-direction.  For $\Mach = 1.8$, open and closed
models (i.e., models where open or reflecting boundaries are imposed)
agree in their growth rates if simulated on a sufficiently large
domain.  However, when the extent of the computational domain is small
in the transverse direction ($l_y = 0.5$), we observe a
destabilization of closed models: the growth rate of the closed model
HD2r-1-s exceeds that of the corresponding open model HD2o-1-s by a
factor of $\approx 3.5$.  Furthermore, closed models exhibit a phase
of exponential and oscillatory growth of $e_{\mathrm{kin}}^y (t)$ even
when $\Mach > 2$, whereas open models are stable.

In the KH saturated state the flow consists of a dominant vortex for
shear flows of moderate Mach numbers. At large Mach numbers and when
the growth of the instability is mediated mainly by shock waves, the
flow is characterized by a rather thin and clearly delimited
transition layer oriented along the initial discontinuity (at $y=0$).
This layer is surrounded by two regions of anti-parallel flows.

The shocks created at the supersonic shear layer are initially
oblique, but eventually become planar shocks parallel to the
$y$-direction.  This process happens earlier close to $y = 0$.  The
vertical extent of the planar shock structures varies from a fraction
of the vertical domain size to almost the whole computational box.
When the propagation of the shocks is restricted in $y$-direction, the
fluid tries to avoid these by sliding along the vertical direction.
Thus, the planar shocks very efficiently convert $x$- into $y$-kinetic
energy.

\subsection{Intermediate and weak fields}
\label{Sek:weakfields}
%
Sufficiently strong magnetic fields (Alfv\'en number $\Alfv \le 2$;
see Eq.\,\ref{alfnum}) stabilize the flow according to linear
stability analysis. We indeed observe this stabilization in
simulations of both subsonic and supersonic strongly magnetized shear
flows.  In the following, we thus focus on the more interesting case
of intermediate and weak initial fields, which according to
\cite{Frank_etal__1996__ApJ__MHD-KH-2d-1} can give rise to
\emph{disruptive} and \emph{dissipative} dynamics, respectively.  The
models we describe in this section were computed using a grid with
$l_x \times l_y = 2 \times 2$ and reflecting boundary conditions in
$y$-direction.  We simulated shear flows with $U_0 = 1$, and varied
the Mach number of the flow by setting the pressure either to $P_0 =
0.6$ or $P_0 = 0.0375$ corresponding to Mach numbers of $\Mach = 1$
and $\Mach = 4$, respectively.  The adiabatic index of the gas was
$\Gamma = 4/3$.

\subsubsection{Intermediate fields}
%
For $\Alfv = 2.5$, we find, in agreement with
\cite{Frank_etal__1996__ApJ__MHD-KH-2d-1}, non-linear stabilization.
The magnetic field is amplified during the linear phase, and the
magnetic tension becomes eventually sufficiently strong to prevent
further bending of the field lines.  Thus, the formation of a KH
vortex is suppressed. Instead, the velocity and the magnetic fields
remain essentially aligned with each other and the shear layer
developing only small $y$-components.  After the end of linear growth
a broad shear layer develops inside which the magnetic field has a
sheet-like structure.

If the magnetic field strength is reduced further ($\Alfv = 5$), we
observe a linear growth of the KH instability, and the formation of a
KH vortex.  The overturning vortex continues to amplify the field
until it becomes eventually so strong that it resists further bending,
i.e. the instability saturates in the non-linear phase.  The magnetic
energy, which grows exponentially during the linear phase, reaches a
maximum, and then gradually declines back to almost its initial value.

It is important to note that although we are evolving the equations of
ideal (i.e., non-resistive) MHD numerical resistivity is present and
acts similar as a physical resistivity. Hence, reconnection of field
lines and dissipation of magnetic energy into internal energy occurs.
Though being a purely numerical effect, this dissipation mimics a
physical process: in ideal MHD (or for exceedingly large magnetic
Reynolds number $\Reyn_\mathrm{m}$), energy is transferred to ever
smaller length scales by a turbulent cascade. When the cascade reaches
the scale set by the grid resolution, the physics is no longer
appropriately represented by the discretized magnetic field.  Instead,
the unresolved (sub-grid) magnetic energy is assigned to the internal
energy.  Hence, numerical resistivity (like numerical viscosity) acts
as an unspecific sub-grid model for unresolved dynamics.

As a result of numerical resistivity, our models show the dynamics
discussed by \cite{Jones_etal__1997__ApJ__MHD-KH-2d-2}: the emergence
of coherent flow and field structures, and their subsequent disruption
in intense reconnection events whereby kinetic energy is efficiently
converted into internal energy.  As a consequence, the kinetic energy
decreases more strongly than in the non-magnetic case, and the flow
barely resembles a KH vortex at the end of the simulation.  Instead,
we find a broad transition layer that is embedded into two
anti-parallel flows and that contains thin magnetic flux sheets.  The
flow is rather laminar than turbulent, with elongated streaks of gas
and field stretching across the computational domain.

\subsubsection{Weak fields}

\absatz{Overview}
%
Models with a weak initial magnetic field show \emph{disruptive} or
\emph{dissipative} dynamics
\citep{Jones_etal__1997__ApJ__MHD-KH-2d-2}.  In both regimes, a KH
vortex develops.  The magnetic field forms thin flux sheets while it
is wound up by the vortex.  If two flux sheets of opposite polarity
come to lie close to each other, they suffer the resistive
\emph{tearing-mode instability} which leads to the reconnection of
field lines of different orientation and the conversion of magnetic
into thermal energy.  Since the magnetic energy was previously
amplified at the cost of the kinetic energy, the tearing modes act
essentially as a catalyst facilitating the dissipation of kinetic into
internal energy.  This behavior characterizes the dissipation regime,
while in the disruption regime another effect comes into play: the
magnetic field eventually becomes sufficiently strong to disrupt the
vortex leaving behind a broad transition layer where turbulent flow
and magnetic fields decay slowly.  The dynamics of the flow and the
magnetic field are highly coupled since the field is dominated by flux
sheets where the velocity and the magnetic field are strongly aligned,
reminiscent of the \emph{Alfv\'en effect} in MHD turbulence
\citep{Iroshnikov__1964__SovietAstronomy__Turbulence_of_a_Conducting_Fluid_in_a_Strong_Magnetic_Field,
  Kraichnan__1965__PhysicsofFluids__Inertial-Range_Spectrum_of_Hydromagnetic_Turbulence}.
Accordingly, we also find near equipartition between the transverse
magnetic and kinetic energy densities (see the disruption models
below).

\begin{figure}
  \centering
   \includegraphics[width=7cm]{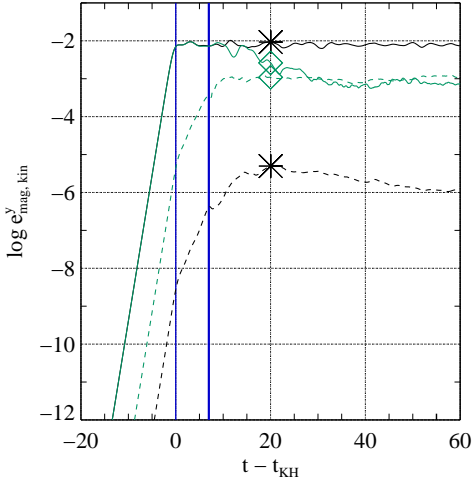}
   \caption{Volume averaged transverse kinetic (solid) and magnetic
     (dashed) energy densities $e^y_\mathrm{kin}$ and
     $e^y_\mathrm{mag}$ versus time for models with an initial Mach
     number $\Mach = 1$, and Alfv\'en numbers $\Alfv = 125$ (green,
     diamond) and $\Alfv = 5000$ (black, asterisk), respectively.
     Both models were computed using a grid of $2048^2$ zones.  The
     blue vertical lines indicate the end of the KH phase,
     $t_{\mathrm{KH}}$, and of the kinematic phase, respectively. }
  \label{Fig:kh2-ekinmag-t}
\end{figure}

The evolution of the simulated weak-field models (summarized in
\tabref{Tab:n2d-weak-models}) consists of three distinct phases:
\begin{itemize}
\item \emph{Linear KH growth phase:} initial perturbations of both
  velocity and magnetic field grow exponentially until a KH vortex
  forms.
\item \emph{Kinematic field amplification phase:} magnetic field is
  wound up by the secularly evolving KH vortex.
\item \emph{Dissipation/disruption phase:} KH vortex looses its energy
  due to magnetic stresses and resistive effects.
\end{itemize}
We discuss these three phases and the transitions between them in more
detail in the following.  The phases can be distinguished best on the
basis of the evolution of the transverse kinetic and magnetic energy
densities $e^y_\mathrm{kin}$ and $e^y_\mathrm{mag}$, respectively (see
\eqref{kiny} and \eqref{magy}; \figref{Fig:kh2-ekinmag-t}).  For this
purpose, we consider a pair of prototype models, with initial Mach
number $\Mach = 1$, and Alfv\'en numbers $\Alfv = 125$ and $\Alfv =
5000$, respectively, computed on a grid of $2048^2$ zones.

\begin{figure}
  \centering
  \includegraphics[width=7cm]{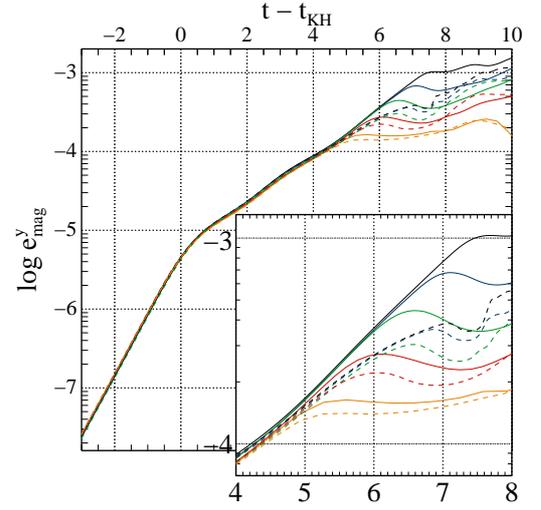}
  \caption{Growth of the volume-averaged turbulent (transverse)
    magnetic energy density $e^y_\mathrm{mag}$ (see \eqref{magy}) with
    time for two different models simulated at five different grid
    resolutions.  At time zero the ratio of $e^y_\mathrm{kin}$ and the
    volume-averaged total magnetic energy density is maximal,
    i.e. this moment corresponds to $t_{\mathrm{KH}}$.  The solid and
    dashed lines refer to a model with initial Alfv\'en numbers $\Alfv
    = 5000$ and 125, respectively. Note that the $f$ values of the
    former model are scaled by the factor $(5000/125 )^2$.  Orange,
    red, green, blue, and black lines refer to simulations with
    $256^2$, $512^2$, $1024^2$, $2048^2$, and $4096^2$ zones,
    respectively.  The insert shows a magnified view of the late
    evolution. }
  \label{Fig:kh2-emag-t}
\end{figure}

\absatz{KH growth phase}
%
Early on during the evolution the seed perturbations imposed on the
initial shearing profile are amplified exponentially, but the magnetic
field remains too weak to affect the evolution.  When the exponential
growth of the KH instability terminates, the total magnetic energy has
grown by about a factor $1.4$ in all models, the contribution of the
transverse field component $b^y$ amounting to about $10\%$.  Due to
the persisting weakness of the magnetic field the growth rate of the
instability and the flow structure after the end of the KH growth
phase are the same as those without any field.

\begin{figure}
 \centering
 \includegraphics[width=6.4cm]{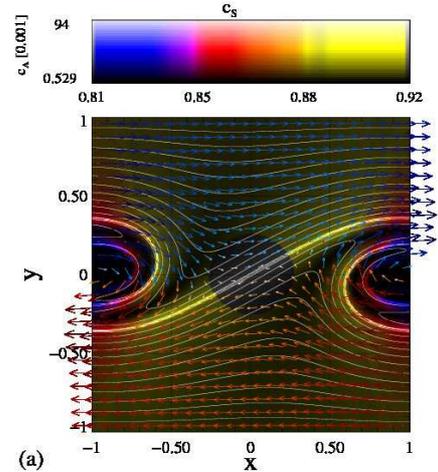}
 \caption{Snapshot of a model with initial Mach number $\Mach = 1$
   and Alfv\'en number $\Alfv = 125$ computed on a grid of $2048^2$
   zones shortly after the end of the KH growth phase.  The hue gives
   the sound speed, $c_{\mathrm{s}}$, and the brightness of the colors
   the Alfv\'en velocity, $c_{\mathrm{A}}$, respectively.  Magnetic
   field lines and flow velocity vectors are shown, too.  The latter
   are color-coded according to the size of the $x$-component of
   $\vec v$, reddish and bluish colors corresponding to matter
   flowing to the left and right, respectively. }
  \label{Fig:kh2-M1-MAma-kinemat--v}
\end{figure}

When the KH instability saturates with the formation of a KH vortex
(see \figref{Fig:kh2-M1-MAma-kinemat--v} for a model with $\Alfv =
125$), the growth of the transverse kinetic energy ceases, too
(\figref{Fig:kh2-ekinmag-t}).  Density, pressure, sound speed, and
magnetic field strength possess a minimum at the center of the vortex,
and the magnetic field is wound up into a long thin sheet surrounding
the vortex.  These findings hold for the models with $\Alfv = 125$ and
$\Alfv = 5000$, respectively. \figref{Fig:kh2-emag-t} shows that the
growth rate of the instability (the slope of the curves) is
independent of the grid resolution and the initial field strength for
$t - t_{\mathrm{KH}} < 0$.

\absatz{Kinematic amplification phase} 
%
After saturation of the essentially hydrodynamic KH instability,
$e^{y}_{\mathrm{kin}} (t)$ exhibits small oscillations about a
constant value.  The initial shearing interface, wound up several
times by the overturning vortex, has become a thin fluid layer
separating flow regions of opposite velocities
\figref{Fig:kh2-M1-MAma-kinemat--v}). The magnetic sheet is being
stretched by the overturning vortex giving rise to an exponential
amplification of the field (instead of a linear one by winding), as
the growth rate due to stretching depends on the field strength
itself. In spite of the growing magnetic field the flow structure as
well as the kinetic and internal energies of the fluid show only minor
changes throughout the entire kinematic amplification phase.

To understand the amplification of the magnetic field in detail we
consider the sources and sinks of magnetic energy.  From the scalar
product of $\partial_{t} \vec b$ (given by the induction equation) and
the magnetic field, $\vec b \cdot \partial_{t} \vec b$, one can derive
the equation for the evolution of the total energy density of the
magnetic field, $e_{\mathrm{mag}}$, which has the form of an advection
equation with source terms,
\begin{equation}
  \label{Gl:Emag}
  \partial_{t} e_{\mathrm{mag}} 
  +
  \vec \nabla \left( e_{\mathrm{mag}} \vec v \right )
  =
  s_{\mathrm{mag}}.
\end{equation}
The source term, 
\begin{equation}
  \label{Gl:Emag-source}
  s_{\mathrm{mag}} = 
  - e_{\mathrm{mag}} \vec \nabla \cdot \vec v
  +
  b^x b^y \left( \partial_{y} v^x - \partial_{x} v^y \right),
\end{equation}
consists of a compression term proportional to the divergence of the
velocity field, and a shear term proportional to the curl of the
velocity field. The sum of both terms (i.e., the source term) is
negative, when the magnetic field does work on the fluid.

\begin{figure}
  \centering
  \includegraphics[width=7cm]{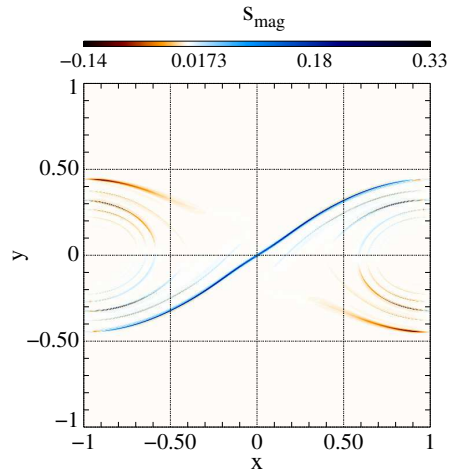}
  \caption{Snapshot of the source term of the total magnetic energy
    density (Eq.\,\ref{Gl:Emag-source}) for the model shown in
    \figref{Fig:kh2-M1-MAma-kinemat--v} taken during the kinematic
    amplification phase.  Reddish (blueish) colors show regions where
    the total magnetic energy density increases (decreases). }
  \label{Fig:kh2-M1-MA125-kinemat--bampl}
\end{figure}

The evolution of $e_{\mathrm{mag}}$ is exemplified in
\figref{Fig:kh2-M1-MA125-kinemat--bampl} for a model with $\Alfv =
125$. As the fluid is nearly incompressible in our models, the first
term on the r.h.s.\, of \eqref{Gl:Emag-source} is small, and field
amplification (blueish areas) occurs predominantly by stretching.  As
there is no back-reaction onto the flow, the volume-averaged
transverse magnetic energy density grows exponentially with
time. Stretching mainly happens in the thin flux sheet passing through
the origin of the grid, and to a lesser extent in the flux sheets
located closer to the center of the vortex.  There even a small
reduction of $e_{\mathrm{mag}}$ can be observed (see
\figref{Fig:kh2-M1-MA125-kinemat--bampl}).  The volume integral of the
source term over the entire computational domain is positive, i.e.,
the magnetic energy of the models is increasing.

Because field amplification is mediated by a well resolved, rather
smooth flow, the growth rate of the turbulent magnetic energy density
$e^y_\mathrm{mag}$ is independent of the grid resolution during the
kinematic amplification phase ($0 \leq t - t_{\mathrm{KH}} \la 5$; see
\figref{Fig:kh2-emag-t}). Models with $\Mach = 0.5$, but otherwise
identical initial conditions and grid resolution, show a slower growth
of the field (see \tabref{Tab:n2d-weak-models}).  As
$e_{\mathrm{kin}}^y$ (monitoring the turnover velocity of the vortex)
shows small variations with time during the kinematic amplification
phase (see \figref{Fig:kh2-ekinmag-t}), the growth rate varies
slightly, too (note the variation of the slope in
\figref{Fig:kh2-emag-t} for $2 \la t - t_{\mathrm{KH}} \la 4$).

The evolution of the turbulent magnetic energy density after the end
of the kinematic amplification phase depends strongly on the grid
resolution and the initial field strength (\figref{Fig:kh2-emag-t}).
Comparing the results for the models with $\Alfv = 125$ and $\Alfv =
5000$ we conclude that the growth of the turbulent magnetic energy
density is less for models with a stronger initial field than for
those with a weaker initial field at the same grid resolution.

For the model with $\Alfv = 125$ the magnetic field eventually reaches
locally (within a factor of a few) equipartition strength, i.e.,
magnetic stresses start to change the flow.  In the model with the
lower initial Alfv\'en velocity (i.e., larger Alfv\'en number), the
magnetic field remains, in spite of a larger amplification, too weak
to cause such an effect.

\begin{figure}
  \centering
  \label{Fig:kh2-M1-MA5000-kinemat--v}
\end{figure}

To quantify the amount of amplification of the magnetic field occurring
during the kinematic amplification phase we introduce the field
\emph{amplification factor}
\begin{equation}
  f_\mathrm{kin} = \mathsf{M}_{xy} (t = t_\mathrm{kin}) / 
                   \mathsf{M}_{xy} (t = t_\mathrm{KH})
            \equiv \mathsf{M}_{xy}^\mathrm{kin} / 
                   \mathsf{M}_{xy}^\mathrm{KH}
\label{fkin}
\end{equation}
defined as the ratio of the off-diagonal volume-integrated Maxwell
stress component $\mathsf{M}_{xy}$ at the end of the kinematic
amplification phase and at the end of the KH growth phase.  

\begin{figure*}
  \centering
  \includegraphics[width=14cm]{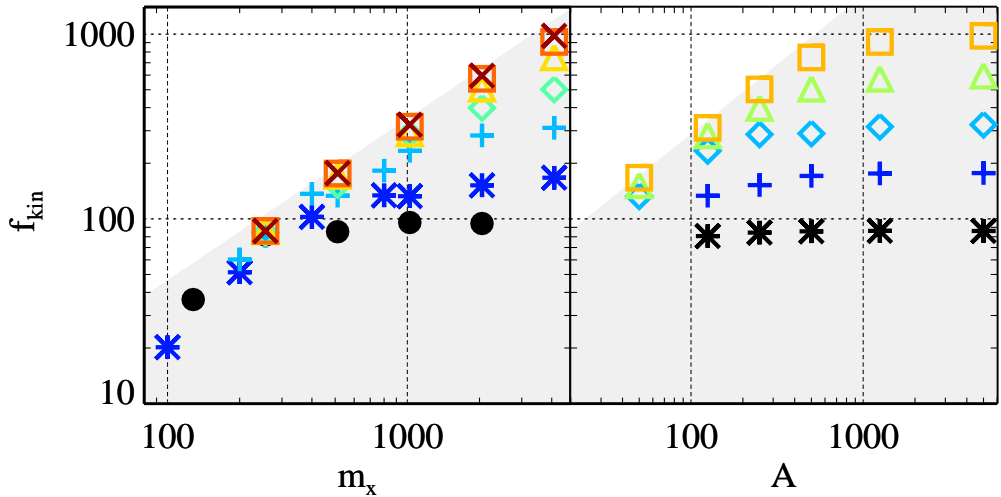}
  \caption{Amplification of the magnetic field during the kinematic
    amplification phase: the amplification factor, $f_\mathrm{kin}$
    (see Eq.\,\ref{fkin}), is shown for the models with an initial
    Mach number $\Mach = 1$ as a function of the number of grid zones,
    $m_x$ (left panel), and of the initial Alfv\'en number, $\Alfv$
    (right panel).  In the left panel, models with $\Alfv = 25$, 50,
    125, 250, 500, 1250, and 5000 correspond to black circles, dark
    blue asterisks, light blue plus signs, green diamonds, yellow
    triangles, orange squares, and red $\times$ signs, respectively.
    In the right panel, models with a grid size of $m_x = 256$, 512,
    1024, 2048, and 4096 zones are displayed by black asterisks, dark
    blue $+$ signs, light blue diamonds, green triangles, and orange
    squares, respectively.  The upper border of the gray shaded
    regions is approximately given by the power laws $m_x^{7/8}$ and
    $\Alfv^{3/4}$, respectively. }
  \label{Fig:kh2-M1-kin-termination}
\end{figure*}

When plotting $f_\mathrm{kin}$ as a function of grid resolution and
initial Alfv\'en number we find that our models populate the lower
right region (shaded in gray) in both diagrams
(\figref{Fig:kh2-M1-kin-termination}).  Both for a given grid
resolution and initial Alfv\'en number, $f_\mathrm{kin}$ converges
towards a maximum value with increasing initial Alfv\'en number
(\figref{Fig:kh2-M1-kin-termination}, left panel), and increasing grid
resolution (\figref{Fig:kh2-M1-kin-termination}, right panel).  This
convergence is also obvious from the graph of $f_\mathrm{kin} (m_X)$
for $\Alfv = const.$ (\figref{Fig:kh2-M1-kin-termination}, left
panel); note that for large values of $\Alfv$ even our finest the grid
spacing was not yet sufficient to show the flattening of
$f_\mathrm{kin} (m_X)$.

The panels further show that the weaker (larger) the initial field
(the value of $\Alfv$), the higher is the amplification factor
$f_\mathrm{kin}$ achievable during the kinematic amplification phase.
The upper border of the gray shaded regions is approximately given by
the power laws $m_x^{7/8}$ and $\Alfv ^ {3/4}$, respectively.

To explain these results and to quantify the effects of the grid
resolution, we define a characteristic length scale of variations of
the magnetic field
\begin{equation}
  l_{b} = \frac{|\vec b|}{|\vec \nabla \times \vec b|} \, , 
\label{lscalmag}
\end{equation}
where the denominator is proportional to the current density.
Initially infinite (the initial magnetic field is curl free), $l_b$
decreases during the KH growth and the kinematic amplification phases.

Due to flux conservation, the amplification of the field occurring
mainly in flux sheets goes along with a decrease of the width of the
sheets orthogonal to the magnetic field, which is roughly given by
$l_{b}$.  In simulations, the decrease of $l_{b}$ can properly be
followed only as long as $l_{b} \ga \Delta_g$, where $\Delta_g$ is the
finite grid spacing. When this limit is reached the exponential
amplification of the field strength and field energy ceases. Further
growth only regards the magnetic energy, which can increase at most
linearly with time due to the increasing length of the sheet (at a
constant width!). This point in the evolution marks the end of the
phase of kinematic amplification.

Consequently, there exists an upper limit for the amplification of the
magnetic field strength attainable by flux-sheet stretching that
depends on the grid resolution.  However, this limit set by the ratio
of the grid spacing and the initial thickness of the flux sheet can
only be reached, if the field strength remains dynamically negligible
(i.e. below equipartition strength) during the kinematic amplification
phase.  This applies to models with weak initial magnetic fields
($\Alfv \ga 1000$), which are located near the upper border of the
gray shaded region in the left panel of
\figref{Fig:kh2-M1-kin-termination}.

\begin{figure*}
  \centering
  \includegraphics[width=7.0cm]{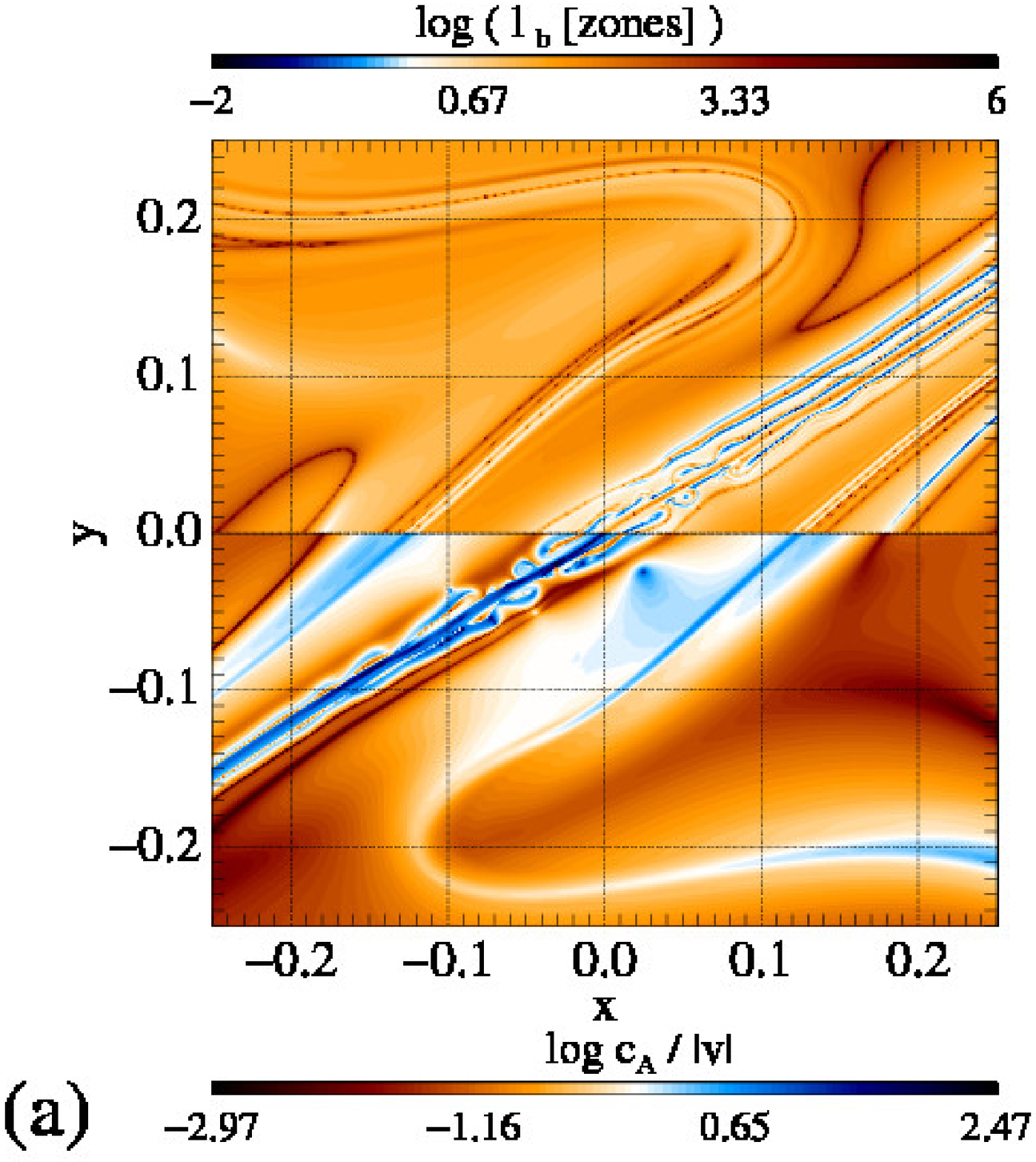}
  \includegraphics[width=11.0cm]{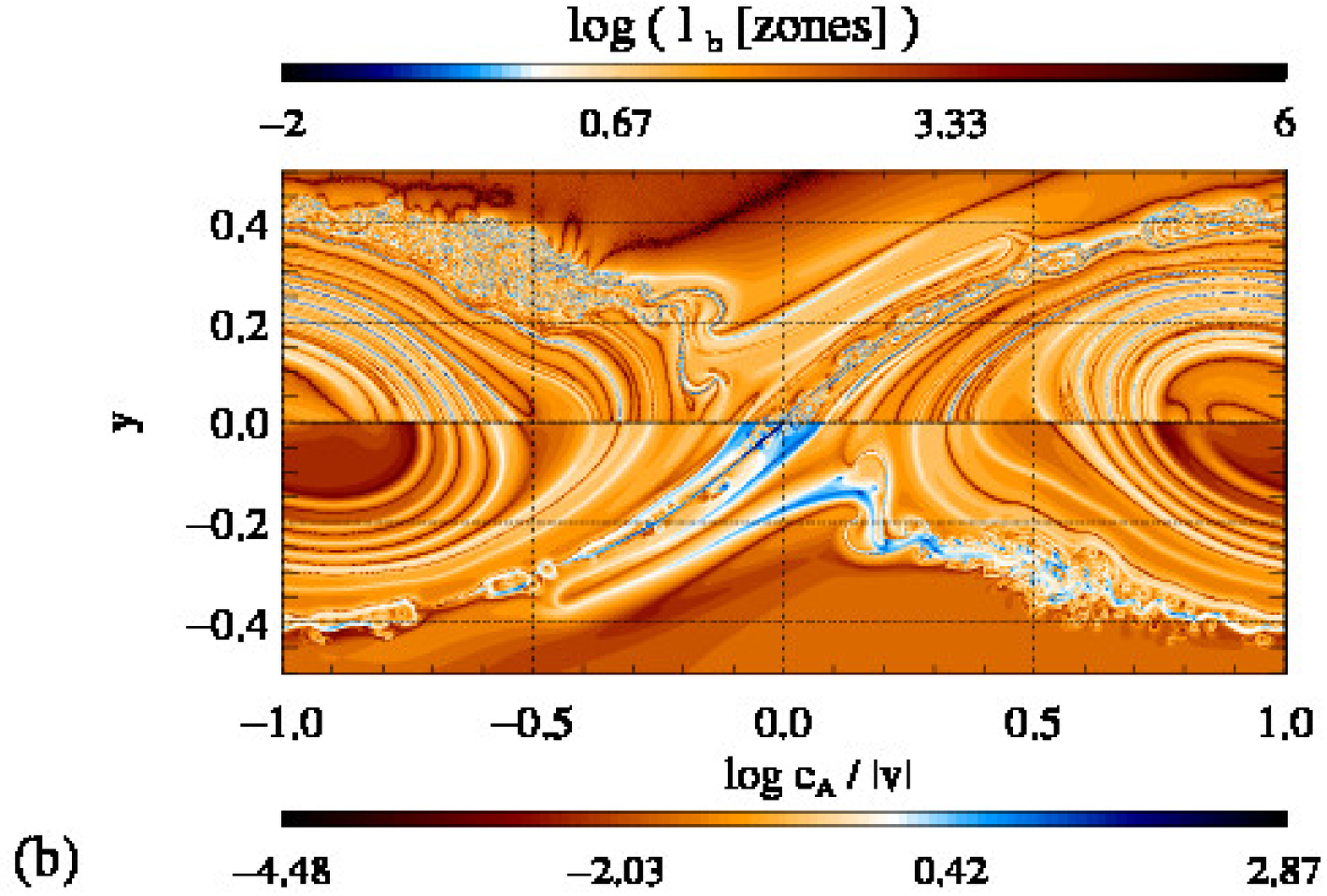}
  \caption{Snapshots of the structure of the model with an initial
    Mach number $\Mach = 1$ and an Alfv\'en number $\Alfv = 125$ taken
    close to the termination of the kinematic amplification phase
    (panel (a)), and shortly afterwards (panel (b)). The top half
    of each panel shows the logarithm of the characteristic length
    scale of the magnetic field, $|\vec b| / | \vec \nabla \times \vec
    b |$ in units of the zone size; reddish colors indicate regions
    where magnetic structures are larger than one computational zone,
    and blueish colors where they are smaller.  The bottom half of
    each panel shows the logarithm of the ratio of the Alfv\'en
    velocity and the modulus of the fluid velocity, blueish and
    reddish colors denoting strongly and weakly magnetized regions,
    respectively.  }
  \label{Fig:kh2-M1-MA125-kinemat-term}
\end{figure*}

If the magnetic field reaches -- within a factor of order unity --
local equipartition strength during the kinematic amplification phase,
the flow dynamics and as a consequence the termination of that phase
show distinct features. This is the case for models with strong
initial magnetic fields ($\Alfv \la 500$) and sufficiently fine
resolution, which are located near the upper border of the gray shaded
region in the right panel of \figref{Fig:kh2-M1-kin-termination}.  For
these models $f_\mathrm{kin} \propto A^{3/4}$, i.e., the amplification
is larger for weaker initial fields.  One factor contributing to this
trend is the back-reaction of the field onto the flow. When locally
the Alfv\'en number approaches the order of unity (see, e.g., the
lower panel of \figref{Fig:kh2-M1-MA125-kinemat-term}), magnetic
stresses start to decelerate the fluid in the flux sheets, and as the
flux sheets partially thread the KH vortex its rotational velocity
decreases, too. Consequently, the amplification factor will be smaller
in this case than for an initially less strongly magnetized model.
Finally, note that for models with weak initial magnetic fields
($\Alfv \ga 1000$) we do not observe effects due to back-reaction, as
this requires larger field amplification factors than reached in our
simulations due to insufficient grid resolution (see discussion above).

A second important issue for understanding our results is the effect
of numerical resistivity.  Although we integrate the equations of
\emph{ideal} MHD, the numerical scheme employed in our code mimics to
some degree the effects of \emph{physical} resistivity due to its
inherent \emph{numerical} resistivity. Thus, the numerical scheme
smooths sharp features in the magnetic field and causes violent
resistive instabilities of, e.g., tearing-mode type. The latter effect
is most pronounced at length scales close to the grid spacing
$\Delta_g$.

When the typical length scales of the magnetic field -- given
approximately by $l_b$ -- are comparable to the grid spacing
$\Delta_g$, we expect numerical resistivity to be important.  For the
model with $\Mach = 1$, $\Alfv = 125$, and $m_x = 2048$ zones $l_b
\approx \Delta_g$ inside the flux sheet near the end of the kinematic
amplification phase (\figref{Fig:kh2-M1-MA125-kinemat-term}, upper
panel).  The magnetic field is dominated by a complex pattern of
sheets partially arranged in pairs or even triplets with anti-parallel
fields.  An example is the triple sheet structure passing roughly
diagonally through the origin from down left to top right
(\figref{Fig:kh2-M1-MA125-kinemat-term}, upper panel). This triplet
consisting of a central sheet with $b^x > 0$ and two parallel "wing"
sheets with $b^x < 0$ is the result of the advection of magnetic flux
towards the central sheet by the flow.

As the advection continues the strength of the magnetic field in the
side sheets increases, while their width decreases leading to intense
currents.  Eventually $l_b \le \Delta_g$, and resistive instabilities
(tearing modes) start to grow, which curl up the two wing sheets and
eventually disrupt them leaving behind only the central sheet
(\figref{Fig:kh2-M1-MA125-kinemat-term}, upper panel). This process
affects the entire triple sheet structure
(\figref{Fig:kh2-M1-MA125-kinemat-term}, lower panel).

Shortly afterwards, the central sheet of the former triplet, still
intact, is disrupted.  From the interior of the vortex further sheets
of magnetic flux are expelled creating new strong currents that again
suffer strong resistive instabilities.  This cycle of processes
repeats every time strong currents build up by approaching flux
sheets.  As a consequence, the large coherent flux sheet structures
are disrupted, and reconnection of magnetic field lines leads to
numerous small-scale field structures including closed field loops,
similarly to those reported in previous simulations
\citep[e.g.,][]{Keppens_etal__1999__PP__MHD-KH}.  

The amplification of the magnetic field terminates due to the
development of these resistive instabilities, because (i) they convert
magnetic energy into thermal energy, and because (ii) the resulting
small-scale field and flow is less efficient in amplifying the
magnetic field than a more coherent flow.

The mechanism just described is responsible for the termination of the
kinematic amplification phase in well resolved models.  All models
with $\Alfv = 50$, $m_x > 256$ and $\Alfv = 125$, $m_x > 1024$ undergo
this evolution.  For even finer grids the results are essentially
converged in terms of the amplification factor $f_\mathrm{kin}$
(\figref{Fig:kh2-M1-kin-termination}).  Finding convergence for a flow
whose behavior depends strongly on numerical resistivity is a
remarkable result that deserves some explanation. Naturally, one would
expect that with finer grid resolution (i.e., decreasing numerical
resistivity) tearing modes are better suppressed, thus enabling the
field to grow stronger.

However, this reasoning does only apply, if the main effect of
numerical resistivity is the disruption of isolated flux sheets.  In
such a situation, the magnetic field in the flux sheet will be
amplified until tearing modes grow faster than the field strength
increases.  As soon as the stretching of the flux sheet leads to a
combination of a sufficiently strong field and a sufficiently thin
sheet (both conditions as well as an increasing resistivity imply
higher growth rates of resistive instabilities; see e.g.,
\citet{Biskamp__2000__Buch__Reconnection}) tearing modes would start
to disrupt the sheet.  The amount of stretching necessary to reach
this state depends on the resistivity, i.e., in our case on the grid
resolution: finer grids require stronger fields and thinner sheets for
disruption. Hence, the maximum field strength achievable at disruption
should grow with increasing grid resolution, but the situation in our
models described above is crucially different. Instead of operating on
an isolated current sheet in a static background, the resistive
instabilities terminating the growth of the magnetic energy act in our
models on a multitude of flux sheets approaching each other closely
due to a dynamic background flow.  Their growth rates can become
faster than the kinematic amplification of the field once the
\emph{distance}, $D_s$, between two sheets rather than the width of
the sheets, $l_b$, becomes sufficiently small, i.e, $D_s \la
\Delta_g$, but $l_b > \Delta_g$.  Contrary to the sheet width $l_b$,
the distance $D_s$ is not related to the magnetic energy stored in the
sheets, but it is determined mainly by the flow field.  Hence, there
exists a relation between the velocity field and the instance of
growth termination.  The velocity field, in turn, depends mainly on
the hydrodynamics of the KH vortex, and only weakly on the grid
resolution, i.e., the moment when the flux sheets break up is
independent of resolution. The latter also holds for the energy
contained in the sheets.  Converged results for the amplification
factor can therefore be obtained despite the presence of a grid
spacing dependent numerical resistivity.

As we saw above, the tearing modes of our models are triggered first
after the formation of multiple sheet structures.  At this point the
central flux sheet of the triplet passing through the origin is still
well resolved by several zones ($l_b \approx \mathrm{a\, few\,} \times
\Delta_g$), but the distance $D_s$ between the side sheets and the
central sheet approaches $\Delta_g$ as the former are advected towards
the latter one.

Some of our model sequences show no convergence behavior
(\figref{Fig:kh2-M1-kin-termination}, left panel), as the grid
resolution necessary for that increases with the initial Alfv\'en
number. For very weak initial fields ($\Alfv \ge 250$) even our finest
grid with $4096^2$ zones does not yield a resolution-independent
amplification factor.  However, as the advection of the flux sheets
does not depend on resolution and only weakly on the strength of the
initial field (except for the sheets feedback is very limited in the
kinematic amplification phase), the formation of unstable multiple
sheets is possible even on coarse grids.  Nevertheless, we do not
observe strong resistive instabilities during this phase for these
models.

We showed above that the growth rate of resistive instabilities during
the kinematic amplification phase depends, apart from the resistivity,
on the width of the flux sheet and the field strength, and that this
phase ends when the tearing modes grow faster than the field is
kinematically amplified by the velocity field.  To match this
condition, sufficiently strong fields are required during close
encounters of flux sheets.  This fact explains why we do not find
resistive instabilities in models with too weak initial fields or too
coarse resolution. In these cases the limitation of the maximum field
strength of a flux sheet imposed by its minimum (resolvable) width
leads to a reduced growth rate of resistive instabilities even when
$D_s \approx \Delta_g$, i.e., the distance between two flux sheets is
reduced to the grid spacing.  Thus, these instabilities cannot
terminate the kinematic field amplification process the same way as
they do it in the case of stronger initial fields or finer grids.

The field strength required for resistive instabilities to terminate
the kinematic amplification phase depends on the flow field: faster
shear flows require stronger fields.  Empirically, we find that the
maximum field strength at termination corresponds roughly to an
Alfv\'en number of order unity, i.e., to field strengths similar to
those required for dynamic feedback.

To summarize, we find that there exist two different mechanisms to
terminate the kinematic amplification phase.
\begin{itemize}
\item \emph{Passive termination:} the magnetic field strength reaches
  a maximum when the decreasing thickness of the flux sheets
  approaches the grid spacing, i.e., when $l_b \approx \Delta_g$.
\item \emph{Resisto-dynamic termination:} the magnetic field reaches
  equipartition strength with the flow field when a combination of
  dynamic and resistive processes terminate further field growth.
  Lorentz forces reduce the rotational velocity of the KH vortex,
  while resistive instabilities develop as flux sheets merge.
\end{itemize}
Whereas passive termination is a numerical artifact due to finite grid
resolution, resisto-dynamic termination can be expected to occur in
nature.  The latter process leads to Alfv\'en velocities that are
locally comparable with the shear velocity, and it is rather
independent of the initial field strength.  The volume average of the
magnetic energy, on the other hand, increases with increasing initial
field strength, because the volume filling factor of the magnetic
field increases with the initial field strength.

\absatz{Total amplification}
%
The total amplification of the magnetic field is given by its growth
during both the KH and the kinematic amplification phases.

According to our results the field amplification factor
$f_\mathrm{kin}$ (Eq.\,\ref{fkin}) scales with the initial Alfv\'en
number, $\Alfv$, approximately as $\Alfv^{3/4}$ (see
\figref{Fig:kh2-M1-kin-termination}).  Consequently, the \emph{maximum
  Maxwell stress obtainable at the end of the kinematic amplification
  phase} scales with the initial magnetic field $b_0 \propto
\Alfv^{-1}$ approximately as
\begin{equation}
  \mathsf{M}^{\mathrm{max}}_{xy} \propto b_0^{5/4} \,,
\label{Gl:KH2-kinemat-E-scaling}
\end{equation}
since $\mathsf{M}^\mathrm{max} = f_\mathrm{kin}\,
\mathsf{M}_{xy}^\mathrm{KH}$ (see Eq.\,\ref{fkin}), and
$\mathsf{M}^\mathrm{KH}_{xy} \propto b_0^2$ (i.e., the growth of the
Maxwell stress during the KH growth phase is practically independent
of the field).  Note that this maximum value is only reached for a
sufficiently fine grid resolution. If the model is under-resolved,
$\mathsf{M}^{\mathrm{max}}_{xy}$ is reduced by a factor approximately
$\propto m_x^{7/8}$ , i.e., the maximum obtainable magnetic field
strength depends on the strength of the initial magnetic field.
Furthermore, as weak initial fields imply weak termination fields,
which modify the dynamics of the flow only weakly, there exists a
hydrodynamic limit of the magnetic KH instability.

\begin{figure}
  \centering
  \includegraphics[width=7cm]{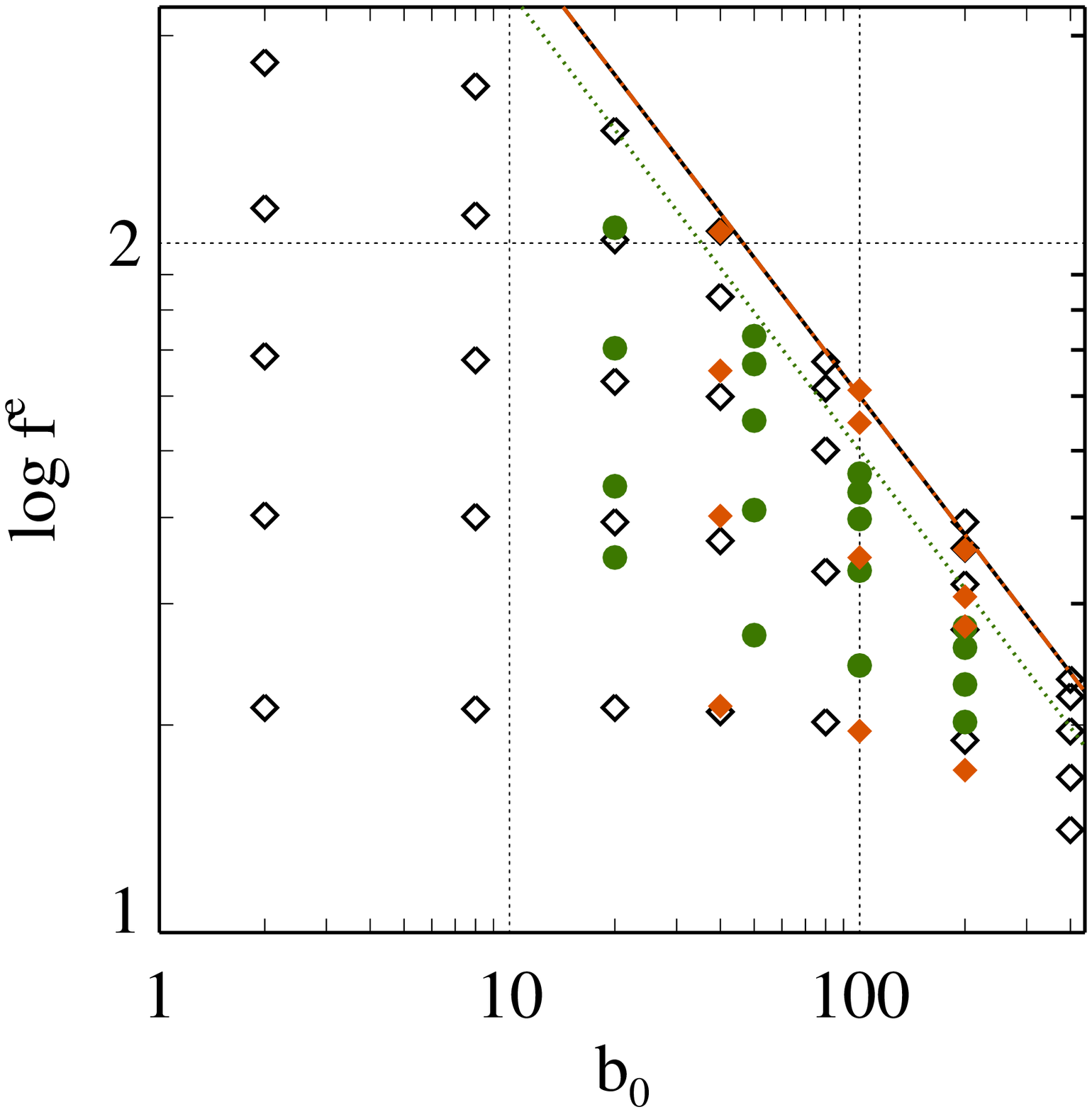}
  \includegraphics[width=7cm]{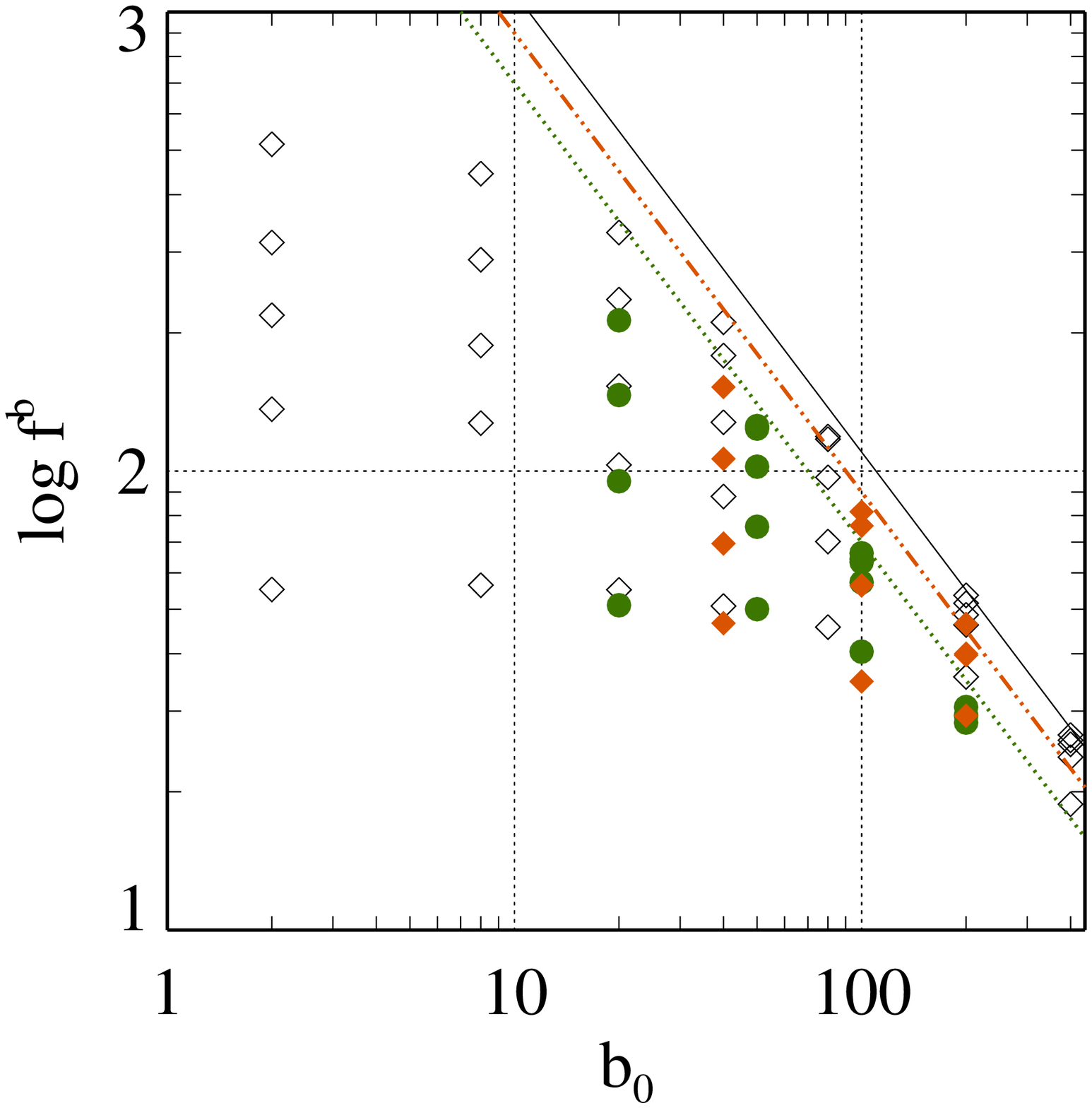}
  \caption{The total amplification factors for the magnetic energy
    $f^{e}$ (top panel) and the magnetic field strength, $f^{b}$
    (bottom panel) as a function of the initial magnetic field, $b_0$,
    for models with different initial shear flows: empty black
    diamonds, filled green circles, and filled red diamonds correspond
    to models with $\Mach = 1$ and $a = 0.05$, $\Mach = 0.5$ and $a =
    0.05$, and $\Mach = 1$ and $a = 0.15$, respectively.  The spread
    in vertical direction reflects different grid resolutions.  To
    indicate the scaling of the amplification factors with the initial
    field strength, the figure also gives power laws $\propto
    b_{0}^{-2/3}$ (top panel), and $\propto b_0^{-1}$ (bottom
    panel). }
  \label{Fig:kh2--fe-fb-scaling}
\end{figure}

The total amplification factors for the magnetic energy, $f^{e}$,
and the magnetic field strength, $f^{b}$, are listed for various
models in \tabref{Tab:n2d-weak-models} and displayed in
\figref{Fig:kh2--fe-fb-scaling}.  The trends described above also hold
here. The amplification factors increase with finer grid resolution
and eventually converge, the resolution required for convergence being
higher for weaker fields.  The converged amplification factors are
larger for weaker magnetic fields, scaling as $f^{e} \propto
b_0^{-2/3}$ and $f^{b} \propto b_0^{-1}$, respectively.  Note that the
latter scaling implies a maximum field strength that is independent of
the initial field strength, consistent with the fact that there exists
a hydrodynamic limit of the magnetic KH instability for weak fields
(see above).

For models differing by their initial hydrodynamic state (i.e, initial
Mach number $\Mach$, and initial shear layer width $a$; see
section\,\ref{Sek:KH2}) both amplification factors scale very similarly
with the initial field strength (\figref{Fig:kh2--fe-fb-scaling}).  In
models with a smaller initial Mach number but the same initial shear
layer width ($\Mach = 0.5$, $a=0.05$; filled green circles), and with
the same initial Mach number but an initially wider shear layer
($\Mach = 1$, $a = 0.15$; red diamonds) the KH instability grows
slower than in standard model ($\Mach = 1$, $a=0.05$) discussed
above. It also saturates at smaller transverse kinetic energies
($\approx 3.3 \times 10^{-3}$ and $\approx 4.2 \times 10^{-3}$,
respectively, instead of $\approx 9.5 \times 10^{-3}$), which implies
a slower kinematic amplification of the field. Hence, $f^{b}$ is
smaller, but its scaling $\propto b_0^{-1}$ is similar to that of the
reference models.  Independent of the properties of the initial shear
flow, we find $f^{e} \propto b_0^{-2/3}$, the proportionality constant
depending, however, in a complex way on the initial state.  For fixed
shear layer width, slower shear flows lead to less efficient field
amplification.  The amplification factor of the magnetic energy
$f^{e}$, on the other hand, is practically independent of the shear
layer width, while $f^{b}$ decreases for narrower initial shear
layers.  However, since the volume where amplification takes place is
larger than that given by the initial shear layer width, overall the
total magnetic energy grows as in the case of a narrower transition
layer.

To summarize, the maximum magnetic field achieved is mainly a function
of the overturning velocity of the KH vortex, corresponding to the
transverse kinetic energy, while the magnetic energy at the
termination of the growth depends on the initial Mach number, on the
width of the shear profile and on the initial magnetic field.

\absatz{Saturation, dissipation and disruption}
%
After termination of the amplification of the magnetic field, the
shear flow enters the saturation phase.  We will discuss in the
following mainly models encountering a resisto-dynamic termination
rather than a passive one, but also briefly mention the behavior of
models suffering a passive termination of the field growth.

\begin{figure}
  \centering
  \includegraphics[width=7cm]{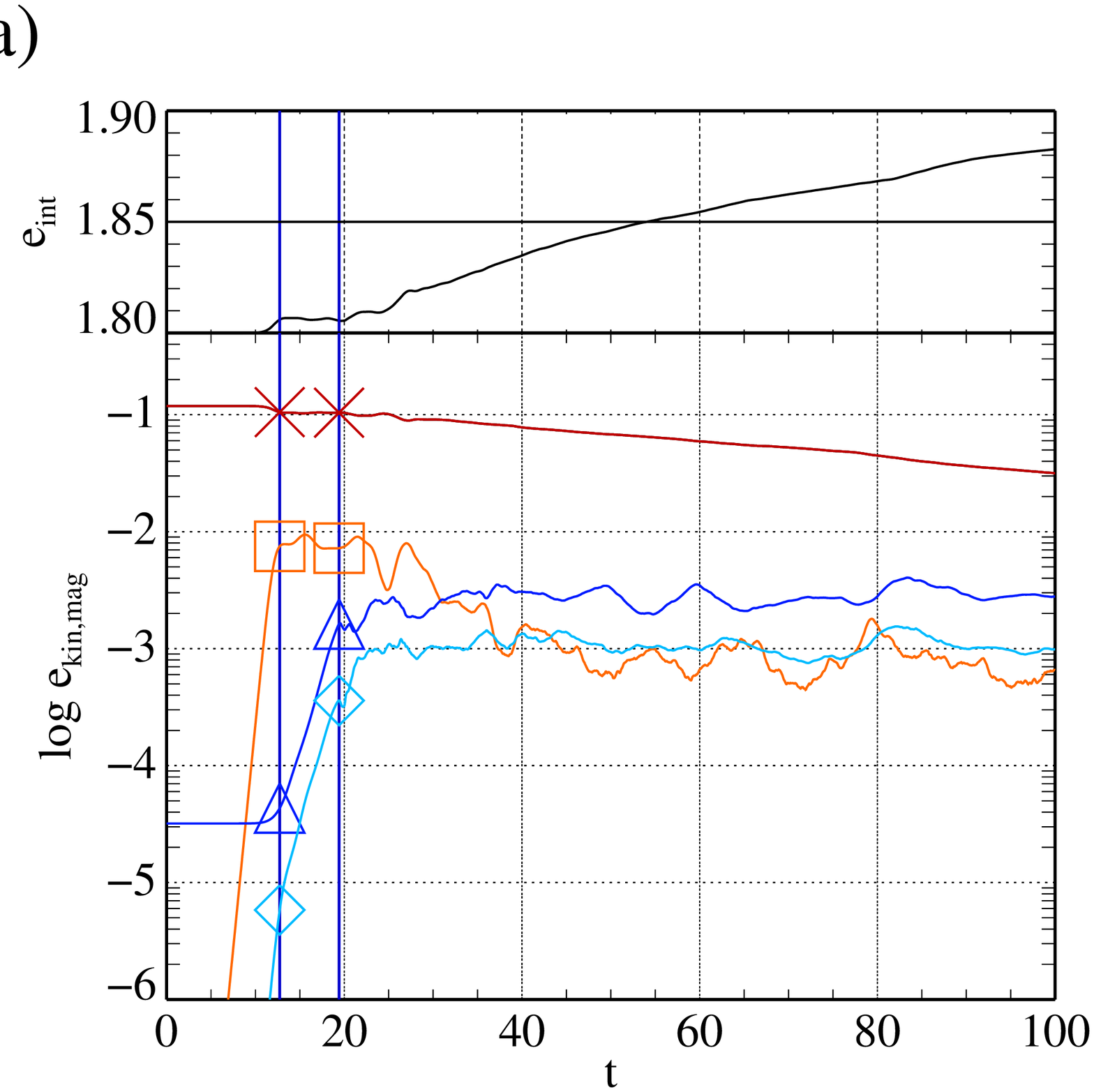}
  \includegraphics[width=7cm]{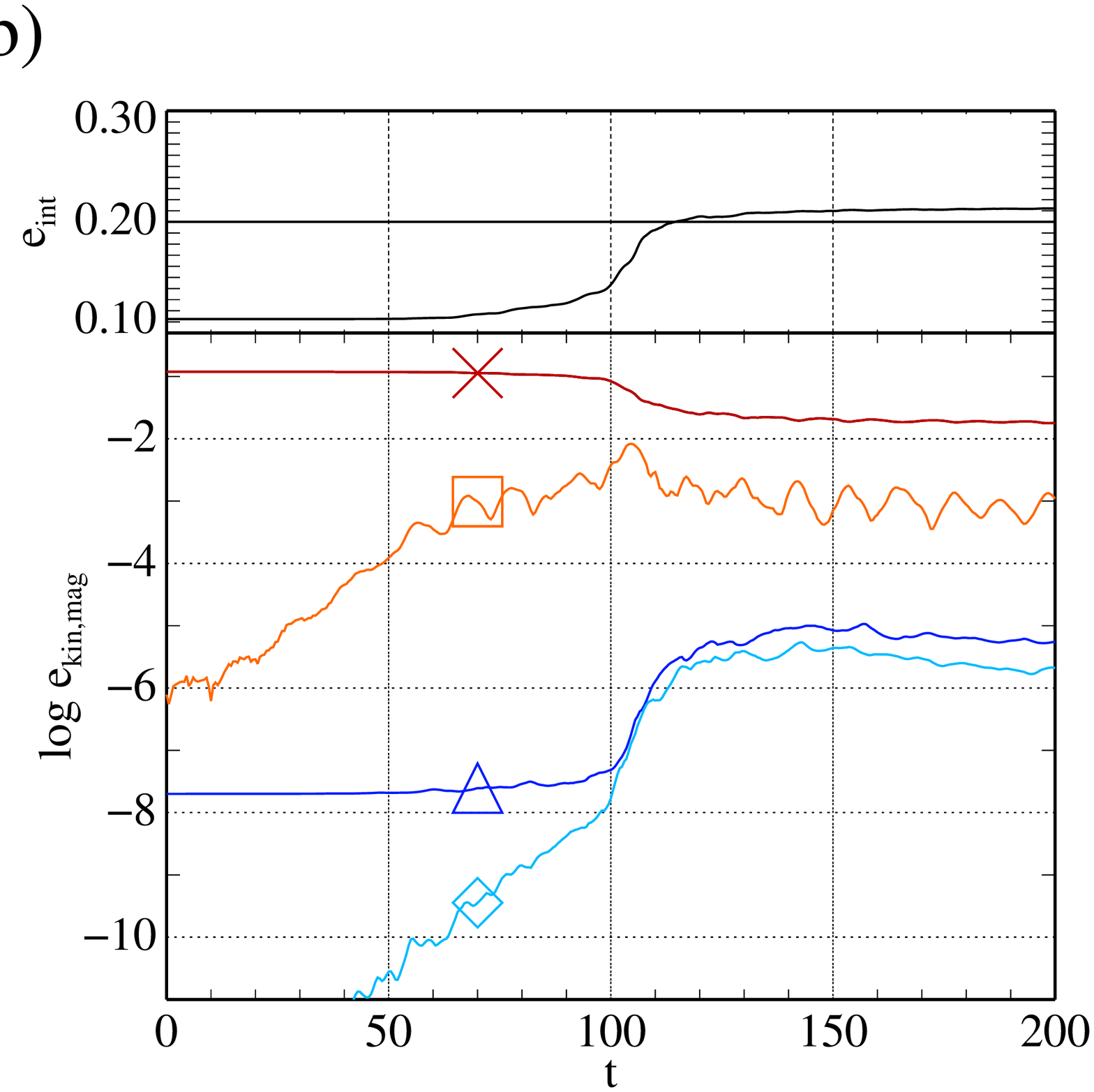}
  \caption{ \textit{Panel (a)}: Evolution of the model with $\Mach =
    1$ and $\Alfv = 125$, computed on a grid of $2048^2$ zones.  The
    top panel shows the internal energy density as a function of time.
    The bottom panel shows the logarithms of the volume-averaged
    kinetic energy densities $e^x_\mathrm{kin}$ (dark red line,
    $\times$ signs) and $e^y_\mathrm{kin}$ (orange line, squares), and
    of the volume-averaged magnetic energy densities
    $e^x_\mathrm{mag}$ (dark blue line, triangles) and
    $e^y_\mathrm{mag}$ (light blue line, diamonds), respectively.  The
    two vertical lines indicate the end of the KH growth (left) and
    kinematic amplification (right) phase, respectively.  \newline
    \textit{Panel (b)}: Same as panel (a), but for a supersonic model
    with $\Mach = 4.4$ and $\Alfv = 5000$.  Because of the model's
    completely different dynamics, the lines indicating the end of the
    growth phases are omitted.  }
  \label{Fig:M1-MA125--sat-tevo}
\end{figure}

As a typical example, we illustrate the evolution of the partial
energies of the model with $\Mach = 1$ and $\Alfv = 125$ in
\figref{Fig:M1-MA125--sat-tevo}.  After the end of the kinematic
amplification phases both the kinetic energy $\propto v_x^2$ (shear
component) and $\propto v_y^2$ (transverse component) decrease, while
the internal energy increases.  The magnetic energy remains roughly at
the level it has reached at the end of the kinematic amplification
phase.  In the final state, the transverse kinetic energy is less than
the total magnetic energy, and equal to the transverse magnetic
energy.

\begin{figure}
  \centering
  \includegraphics[width=7cm]{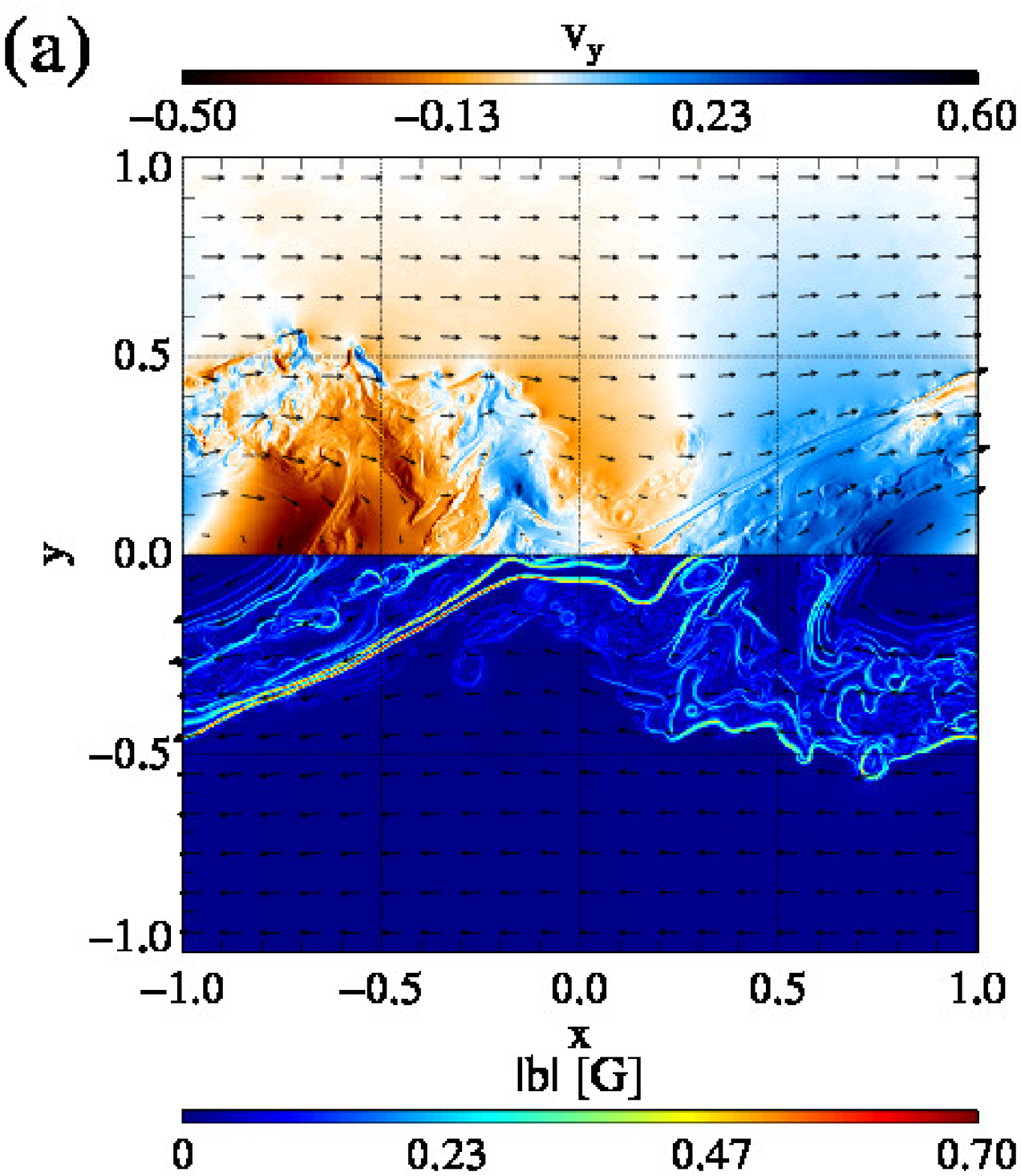}
  \includegraphics[width=7cm]{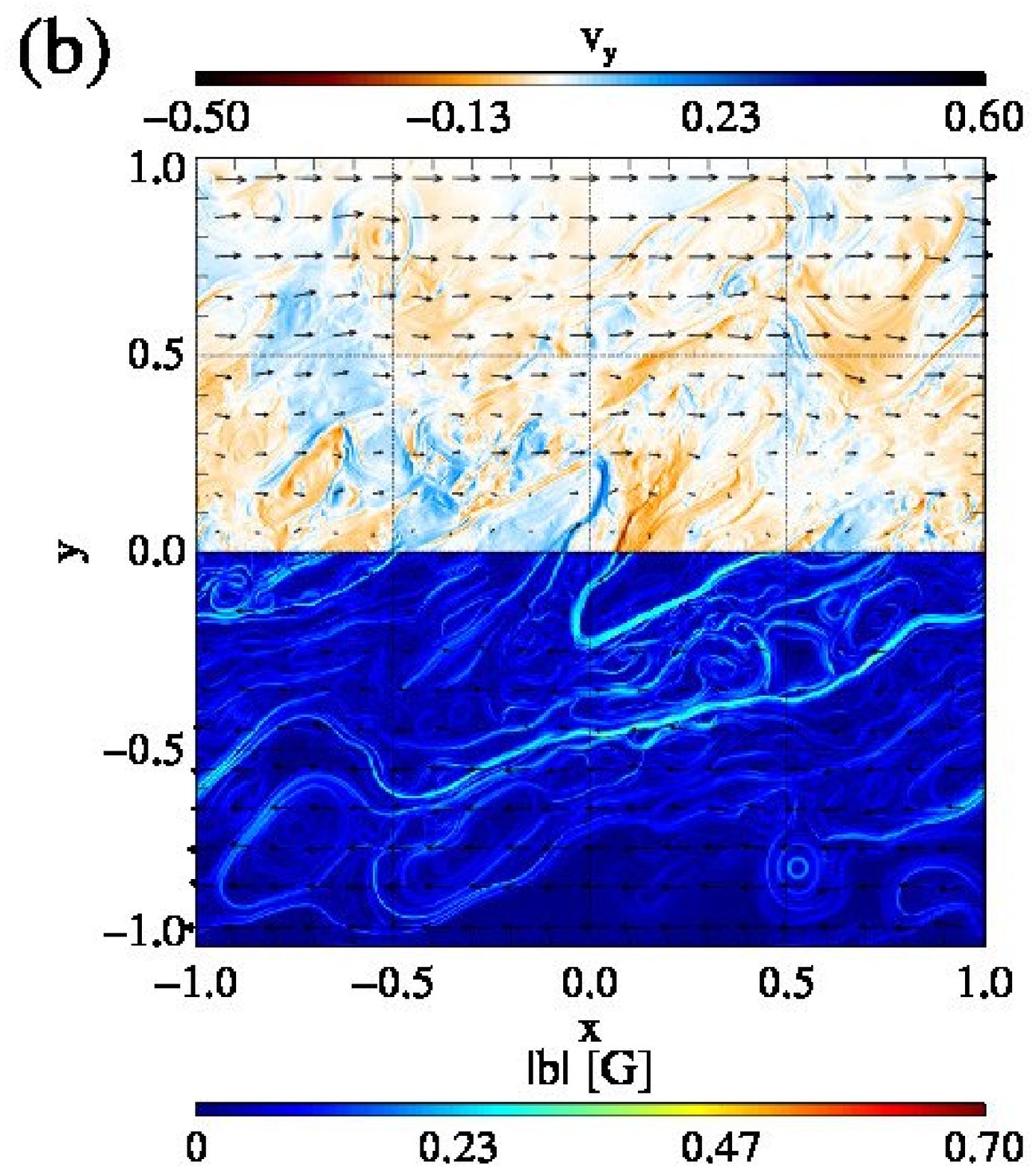}
  \caption{Structure of the model with $\Mach = 1$ and $\Alfv = 125$
    computed on a grid of $2048^2$ zones near the beginning of the
    saturation phase at $t = 34.4$ (upper panel), and at $t = 81.5$
    (lower panel), respectively.  The top and bottom half of each
    panel shows the $y$-component of the velocity and the modulus of
    the magnetic field, respectively.  The flow field is illustrated
    by the black arrows.}
  \label{Fig:kh2--M1-MA125--sat2d}
\end{figure}

To understand these results, we compare the model structure at the
beginning of the saturation phase with that near the end of the
simulation. According to \figref{Fig:kh2--M1-MA125--sat2d} the model
exhibits clear signs of disruptive dynamics
\citep[see][]{Jones_etal__1997__ApJ__MHD-KH-2d-2}.  The KH vortex is
still visible as a coherent pattern at $t = 34.4$, i.e., shortly after
the end of the kinematic amplification phase (panel (a)).  At $t =
81.5$ the vortex is disrupted, the flow field is dominated by a broad
transition region separating oppositely directed shear flows, and the
$y$-component of the velocity shows small-scale structures (see patchy
colors in upper part of \figref{Fig:kh2--M1-MA125--sat2d}, panel b).
The magnetic field is concentrated into a multitude of thin flux
sheets with a typical length scale $l_b \approx \Delta_g$.  Due to
magnetic reconnection the sheets possess a complex topology.  Several
closed field loops have formed that are stabilized by a combination
magnetic loop tension and total pressure ($P + \vec b^2 /2$).  The
flux sheet pattern is imprinted onto the flow field and the gas
pressure distribution. Although the gas pressure is reduced inside the
sheets there is sufficient magnetic pressure to keep the flux sheets
in pressure equilibrium with their surroundings. That explains why the
distribution of the total pressure is rather featureless.

As visible in \figref{Fig:kh2-M1-MA125-kinemat-term} (panel b) and
\figref{Fig:kh2--M1-MA125--sat2d} (panel a), the resistive
instabilities responsible for the termination of the kinematic
amplification phase spread along the flux sheets leading to a complex
field topology and inhibiting further growth of the field not only
locally but in the entire volume.

Locally, i.e. inside the flux sheets, the magnetic field is in
equipartition with the velocity field (globally it is still an order
of magnitude weaker). In resistive instabilities magnetic energy is
converted into internal one.  Since the magnetic field has been built
up previously at the expense of the kinetic energy, the instabilities
actually mediate the transformation of kinetic energy into internal
energy, hence acting akin to a hydrodynamic viscosity.  Eventually, a
steady state (in a statistical sense) develops where the magnetic
energy, and thus the effective viscosity, becomes time-independent,
while kinetic energy is converted into internal one at a constant
rate.  After the disruption of the KH vortex, the transverse velocity
reflects the turbulence resulting from the resistive instabilities,
i.e., $e^y_\mathrm{kin}$ is a measure (like the magnetic field
strength) of the intensity of turbulence. Consequently,
$e^y_\mathrm{kin}$ remains constant at saturation, and the disruption
of the KH vortex can be identified by the instance when
$e^y_\mathrm{kin}$ $\approx e^y_\mathrm{mag}$.

\begin{figure}
  \centering
  \includegraphics[width=7cm]{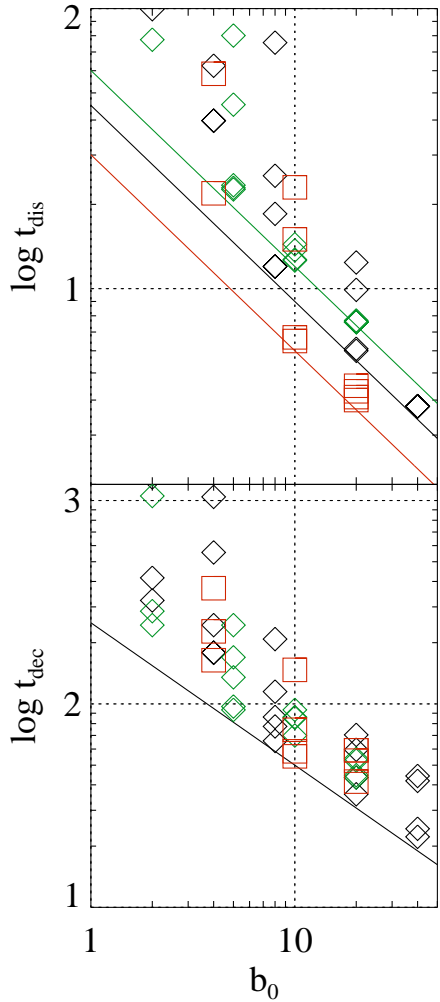}
  \caption{Time scales for the disruption of the KH vortex
    $t_{\mathrm{dis}}$ (upper panel) and deceleration of the flow,
    $t_{\mathrm{dec}}$ (lower panel) as a function of the initial
    field strength $b_0$.  The various models are represented by
    different symbols: black diamonds, green diamonds, and red squares
    correspond to models with $\Mach = 1$ and $a = 0.05$, $\Mach =
    0.5$ and $a=0.05$, and $\Mach = 1$ and $a = 0.15$, respectively.
    The vertical spread of identical symbols reflects different grid
    resolutions finer resolution yielding smaller values of
    $t_{\mathrm{dis}}$ and $t_{\mathrm{dec}}$, respectively . The
    lines $\propto b_0 ^ {-0.7}$ indicate the approximate scaling of
    the time scales with $b_0$.}
  \label{Fig:nKH2--M1-disrtime}
\end{figure}

The saturation level of the magnetic field, and thus the effective
viscosity, is set by its level at the termination of the kinematic
amplifictaion phase.  This level decreases with decreasing initial
field strength, i.e. the weaker the initial field the slower is the
resistive disruption of the KH vortex.  To quantify this effect, we
define a disruption time, $t_{\mathrm{dis}}$, as the time when
$e^y_\mathrm{kin}$ falls below $e^y_\mathrm{mag}$, and a deceleration
rate $\sigma_{\mathrm{dec}} \equiv \partial_{t} \log
E^{x}_{\mathrm{kin}} = 1/ t_{\mathrm{dec}}$
\footnote{We also considered alternative definitions of
  $\sigma_{\mathrm{dec}}$ that, however, do not change the arguments
  in the discussion below.}.
Both quantities are listed in \tabref{Tab:kh2--Abbremsung}, and
$t_{\mathrm{dis}}$ and $t_{\mathrm{dec}}$ are also shown as a function
of the initial Alfv\'en number in \figref{Fig:nKH2--M1-disrtime}.  For
models with very weak deceleration, the evolution of the kinetic
energy is dominated by large oscillations.  Thus, the determination of
the value of $\sigma_{\mathrm{dec}}$ is uncertain to some degree in
these cases, and the numerical values quoted in
\tabref{Tab:kh2--Abbremsung} should be taken with care.

Depending on the initial field strength, the models require a certain
minimum resolution to obtain converged values for $t_{\mathrm{dis}}$
and $t_{\mathrm{dec}}$, respectively.  If the resolution is too low,
the disruption of the vortex and the deceleration of the shear flow
proceed too slow due to an insufficient amount of field amplification.
In the following, we will focus on converged or nearly converged
models.

The disruption and deceleration time scale with the initial field
strength roughly as $b_0^{-0.7}$.  Comparing these times for models with
different initial shear profiles, we find that the disruption time
depends sensitively on both $\Mach$ and the initial shear layer width
$a$.  The larger the amplification factor of the magnetic energy
$f^{e}$ is for a given shear profile (see
\figref{Fig:kh2--fe-fb-scaling}), the faster is the disruption of the
vortex.  The deceleration time, on the other hand, shows a weaker
dependence on $\Mach$ and $a$.  Even for $a = 0.2$, which implies a
much slower KH growth and a very low saturation level of
$e_{\mathrm{kin}}^y \sim 10^{-3}$, the deceleration time is very
similar to that of the models discussed above, although the magnetic
field strength is much smaller.

For weaker fields, whose growth ends due to passive instead of
resisto-dynamic termination (i.e., non-converged models), the kinetic
energy decreases much more slowly.  Resistive instabilities grow much
slower in such models, because of the growth of their field strength
is restricted by numerical resolution.  Hence, the effective viscosity
is much lower in these models than in well resolved ones.

\subsection{Supersonic shear flows}
\label{Sek:supson}
%
We simulated supersonic shear flows with a Mach number $\Mach = 4.4$
using the same velocity profile as for the model with $\Mach = 1$, but
a reduced gas pressure of $P = 0.0375$.  In the following, we compare
models with very large ($\Alfv = 5000$) and small ($\Alfv = 25, 50$)
Alfv\'en numbers.  For the simulations we used grids with a resolution
between $128^2$ and $2048^2$ zones.  As the main result we find that
the growth rate of the magnetic field is lower in supersonic shear
flows than in sonic and subsonic shear flows.

For $\Alfv = 5000$, none of the simulations shows an effect of the
magnetic field on the flow.  For all grid resolutions the early
evolution of the magnetic model (shock formation and interaction) is
similar to that of the non-magnetic one.  Until $t \sim 70$ the
transverse kinetic energy increases roughly exponentially before
leveling off (\figref{Fig:M1-MA125--sat-tevo}, panel b).  The magnetic
field in $y$-direction is amplified at a similar rate as the kinetic
energy until $t \approx 100$, when the amplification rate increases
strongly.  This phase of efficient field growth, lasting until $t
\approx 130$, corresponds to the formation of large regions of
subsonic flow where most of the field amplification occurs.  The
magnetic field is concentrated in thin sheets.  While dominated by a
multitude of shock waves during early phases, the model shows a
subsonic vortical flow in the final state, similarly to the models
discussed in the previous subsection.  The kinetic energy has
decreased by a factor of four during the entire evolution.  Most of
this deceleration has occurred during the early saturation phase of
the KH instability when the magnetic field is amplified most strongly.

Comparing the evolution of the magnetic energy for simulations with
different grid resolution, we find trends similar to those of subsonic
models with dynamically negligible fields.  Stronger magnetic fields
are obtained for finer grids the explanation for this behavior being
the same as that for the resisto-dynamic termination for subsonic
shear flows: amplification ceases when the width of a flux sheet
becomes comparable to the grid spacing.

In models with $\Alfv = 25$ and $\Alfv = 50$ the magnetic field
modifes the dynamics.  In early stages, a number of weak shock waves
form.  Interacting with magnetic flux sheets close to the shearing
interface, theses shocks are disrupted.  Spreading away from the
interface in positive and negative $y$-direction, a wide region of
subsonic flow forms.  Both its geometry and formation differ from
those of subsonic shear flows. In barely magnetized models, a subsonic
flow possessing a considerable transversal extent results from the
interaction of oblique shocks (see \secref{sSek:2dhydro}), whereas in
more strongly magnetized models the magnetic field enforces a
subsonic region elongated along the $x$-direction.  We find
convergence with respect to the saturation level of the magnetic
energy, whose value is in general below the value of subsonic models.
At late times, we observe equipartition between the transverse kinetic
energy and the magnetic energy.  The deceleration times of the flow
are fairly similar to those of the non-magnetic models.

\subsection{Anti-parallel initial fields}
\label{Sek:banti}
%
We have recomputed a number of models with anti-parallel initial
fields, i.e., an initial field $b^x = b_0^x\, \mathrm{sign}(y)$.
Similar simulations were performed previously by
\cite{Keppens_etal__1999__PP__MHD-KH}, whose results we confirm.

For strong initial fields, corresponding to an initial Alfv\'en number
$\Alfv = 5$, we observe in accordance with
\cite{Keppens_etal__1999__PP__MHD-KH} a destabilization of the shear
layer with respect to the non-magnetic case.  

The qualitative dynamics of initially weakly magnetized shear flows
with is anti-parallel fields is similar to the case of parallel
initial fields, evolving through the three phases described in
\secref{Sek:weakfields}.  There are, however, quantitative differences
concerning, e.g., the saturation value of the magnetic energy or the
deceleration rate.  The KH growth phase is similar for both field
configurations, as is the growth rate of the magnetic field during the
kinematic amplification phase. The termination of the latter phase
depends, however, on the initial field orientation: for the same
initial Alfv\'en number, a model with anti-parallel initial field
experiences less amplification than one with parallel magnetic fields.
The modes of termination of the kinematic amplification phase are the
same as in the case of parallel fields (passive or resisto-dynamic
termination), but due to the presence of oppositely directed flux
sheets right from the beginning of the evolution reconnection of field
lines is enhanced. This leads to earlier termination, i.e.,lower
termination field strengths.  As a consequence, the magnetic
deceleration of the KH vortex is less efficient in case of initially
anti-parallel field.  The disruption times and deceleration timescales
are a factor of $\sim 2...3$ larger than those measured for
parallel-field models.

\section{Three-dimensional models}
\label{Sek:3d-mod}
%
In the following section, we study the evolution of KH instabilities
in three-dimensional shear flows.  Obviously, the numerical resolution
we can afford in 3D is much worse than in our best resolved 2D models.
This prevented us from performing a study as detailed as in the
two-dimensional case.  The 3D models we have simulated are listed in
\tabref{Tab:nkh3-models}.

\subsection{Subsonic shear flows, parallel magnetic field}
\label{sSek:3d-subson-p}

\subsubsection{Non-magnetic models}
%
In 3D the KH vortex is unstable against (purely) hydrodynamic
instabilities \cite{Ryu_etal__2000__ApJ__MHD-KHI-3d}: coherent vortex
tubes near the main KH vortex exert non-axial stresses on the vortex,
and fluid elements are prone to the so-called \emph{elliptic
  instability}, an instability caused by time-dependent shear forces,
which act on fluid elements while they orbit the vortex on elliptic
trajectories.  The result is isotropic decaying turbulence. 

As in 2D, we seeded the KH instability with small perturbations of the
$y$-component of the velocity varying sinusoidally in $x-direction$
(see Eq.\,\ref{Gl:KH-2d--perturb}).  To break the translational
symmetry in $z$-direction, we added a small random perturbation
$v_\mathrm{rndm}$ to all velocity components, where 
\begin{equation}
  v_\mathrm{rndm} = \xi_\mathrm{rnd} v^0_y\qquad \mathrm{with}\quad
                   \xi_\mathrm{rnd} \in [10^{-4},1] \,.
\label{eq.rnd}
\end{equation}

\begin{figure}
  \centering
  \includegraphics[width=7cm]{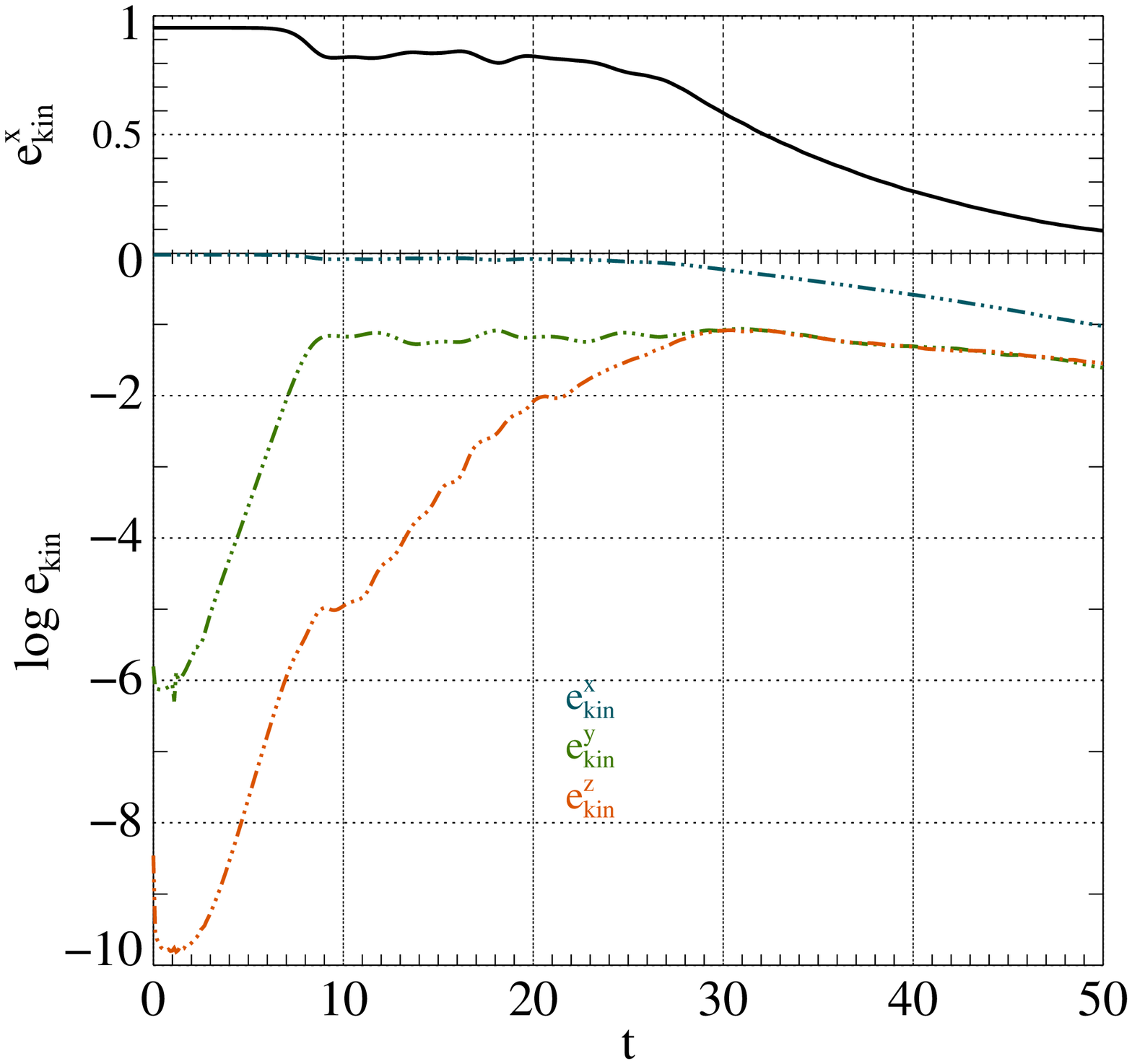}
  \includegraphics[width=7cm]{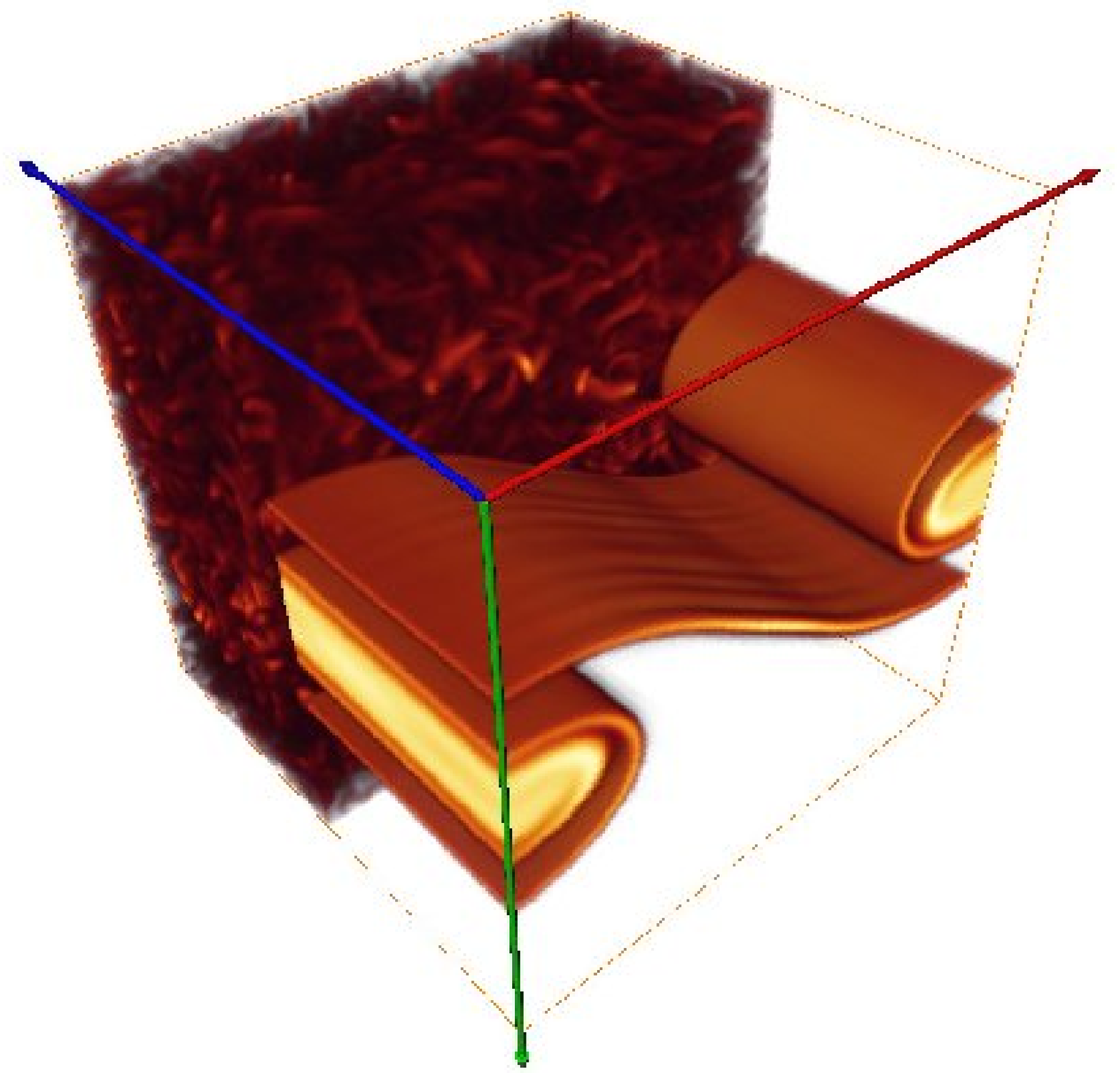}
  \caption{ \textit{Upper panel:} Temporal evolution of the
    volume-averaged kinetic energy densities defined in \eqref{kiny}
    for a 3D non-magnetized model.  The upper part shows
    $e^x_\mathrm{kin}$, and the lower one the logarithm of
    $e^x_\mathrm{kin}$ (blue), $e^y_\mathrm{kin}$ (green), and
    $e^z_\mathrm{kin}$ (red lines), respectively.
    \newline \textit{Lower panel:} Volume rendering of the modulus of
    the vorticity, $|\vec \nabla \times \vec v|$, of the same
    simulation at two different times.  The computational box (red,
    green, and blue arrows point into $x$, $y$, and $z$-direction,
    respectively) is divided into two halves: the front half shows
    $|\vec \nabla \times \vec v|$ at $t = 10$ when the KH vortex tube
    is still fully intact, and the back half at $t = 50$ after the
    complete disruption of the vortex tube by secondary instabilities.
  }
  \label{Fig:KH3--M1-MAunendlich--tevo}
\end{figure}

In the non-magnetized reference model a KH vortex tube elongated in
$z$-direction forms during the exponential growth of the instability.
The vortex tube is clearly visible in the (front part of the) lower
panel of \figref{Fig:KH3--M1-MAunendlich--tevo}, which shows the
vorticity distribution at $t= 10$ (shortly after the termination of
the growth of the instability), and at $t = 50$ (in the non-linear
phase), respectively.  

The temporal behavior of the volume-averaged kinetic energy densities
defined in \eqref{kiny} reflects the evolution of the flow
(\figref{Fig:KH3--M1-MAunendlich--tevo}).  Up to $t \approx 7$,
$e^x_\mathrm{kin}$ is practically constant. Then it starts to drop by
about 20\% within two time units when the forming vortex tube extracts
kinetic energy from the shear flow. Afterwards $e^x_\mathrm{kin}$
stays again approximately constant until the elliptic instability
begins to destroy the vortex tube at $t\approx 20$.  The transverse
kinetic energy densities, $e^y_\mathrm{kin}$ and $e^z_\mathrm{kin}$,
show exponential growth before saturating at the same level. Note that
$e^z_\mathrm{kin}$ saturates about 20 time units later than
$e^y_\mathrm{kin}$ (at $t \approx 30$), because it starts growing from
an initial value that is a factor of $10^4$ smaller. In addition, its
growth rate, which is similar to that of the magnetic energy during
the kinematic amplification phase of 2D models, decreases after the
end of the KH phase ($t \approx 9$) when the elliptic instability
developing along the vortex tube takes over ($t \ga 12$). The latter
saturates when the vortex tube is disrupted, and $e^z_\mathrm{kin}
\approx e^y_\mathrm{kin}$.  Subsequently, turbulence develops (see
vorticity pattern at $t=50$ in \figref{Fig:KH3--M1-MAunendlich--tevo},
lower panel), and the shear flow is strongly decelerated as indicated
by the decrease of $e^x_\mathrm{kin}$
(\figref{Fig:KH3--M1-MAunendlich--tevo}, upper part of upper panel).
The deceleration is considerably faster than in the case of weakly
magnetized 2D models.

\subsubsection{Weak-field models}
%
For weak-field models, the 3D KH vortex is subject to two different
instabilities competing for its disruption: the purely hydrodynamic
one discussed in the previous subsection, and the resistive ones
analyzed in \secref{Sek:weakfields}.  Which of these instabilities is
most efficient depends the importance of 3D effects, which in turn is
determined by the initial amplitude of the random perturbations.
Independently of the purely hydrodynamic instabilities, if there
exists a (weak) magnetic field, it may also disrupt the vortex. In the
latter case, the post-disruption flow shows a larger degree of
organization than a non-magnetized one due to the prevalence of flux
tubes and flux sheets where the magnetic and flow field are aligned.

\begin{figure}
  \centering
  \includegraphics[width=7cm]{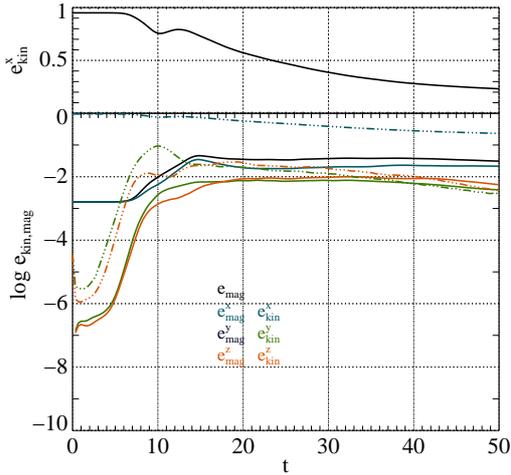}
  \caption{Temporal evolution of various energy densities of a 3D KH
    model having an initial Mach and Alfv\'en number of $\Mach = 1$
    and $\Alfv = 50$, respectively. The amplitude of the imposed
    random perturbation was comparable to that of the sinusoidal one,
    i.e., $\xi_\mathrm{rnd} = 1$ (see \eqref{eq.rnd}).  The top
    panel shows the evolution of $e_{kin}^x$. The bottom panel
    illustrates the evolution of the volume-averaged total magnetic
    energy density (black solid line), and of the magnetic energy
    densities corresponding to the three field components: $e_{mag}^x$
    (blue solid line), $e_{mag}^y$ (green solid line), and $e_{mag}^z$
    (red solid line), respectively. The dash-triple-dotted lines show
    the corresponding kinetic energy densities $e_{kin}^x$,
    $e_{kin}^y$, and $e_{kin}^z$ using the same color coding. }
  \label{Fig:KH3--M1-MA50-yz--tevo}
\end{figure}

\begin{figure}
  \centering
  \includegraphics[width=7cm]{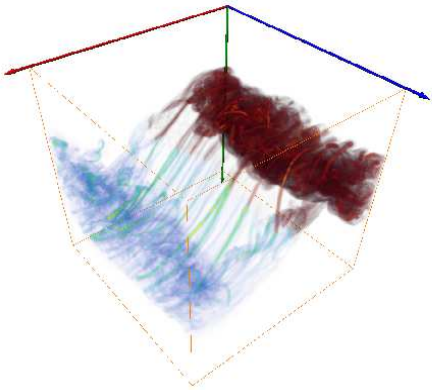}
  \caption{Volume rendered magnetic field strength, $|\vec b|$
    (blue-green) and modulus of the vorticity, $|\vec \nabla \times
    \vec v|$ (red-yellow) of a 3D KH model with initial Mach and
    Alfv\'en numbers $\Mach = 1$ and $\Alfv = 50$, respectively. The
    snapshot is taken during the kinematic amplification phase ($t =
    9.21$).  The amplitude of the imposed random perturbation was
    comparable to that of the sinusoidal one, i.e., $\xi_\mathrm{rnd}
    = 1$ (see \eqref{eq.rnd}).  The computational domain is given by
    the thin red box. The red, green, and blue arrows indicate the
    $x$, $y$, and $z$ coordinate axes, respectively. }
  \label{Fig:KH3--M1-MA50-yz--d0030}
\end{figure}

For a model with $\Alfv = 50$ and a strong random perturbation, i.e.,
comparable to the sinusoidal one ($\xi_\mathrm{rnd} \approx 1$; see
\eqref{eq.rnd}), the flow field shows considerable variations in
$z$-direction already during the formation of the KH vortex tube
(\figref{Fig:KH3--M1-MA50-yz--tevo}).  During the kinematic
amplification phase, we observe a pattern of thin vorticity tubes
arising from magnetic flux tubes wound up around the dominant 3D
vortex tube (located near the edge of the computational domain
$x$-direction; see \figref{Fig:KH3--M1-MA50-yz--d0030}).  The KH
vortex tube is disrupted until the end of the kinematic amplification
phase. At $t \approx 15$ the volume-averaged transverse kinetic energy
densities $e^y_\mathrm{kin}$ and $e^z_\mathrm{kin}$ reach
equipartition (see \figref{Fig:KH3--M1-MA50-yz--tevo}).  Magnetic
field amplification ceases at that point.  The subsequent deceleration
of the shear flow is mediated mainly by the hydrodynamic instabilities
active also in non-magnetized models (see previous subsection). Hence,
deceleration occurs with similar efficiency, but ceases when the
transverse kinetic energy densities drop below the magnetic ones at $t
\sim 50$ and the magnetic field begins to suppress the hydrodynamic
instabilities.  The final state of the model consists of decaying
volume filling turbulence.  Since deceleration is incomplete, the
model retains a slower, smooth shear flow.  The velocity and the
magnetic field are dominated by their respective $x$-components,
leading to considerably anisotropic turbulent fields.

Decreasing the amplitude of the random perturbation to
$\xi_\mathrm{rnd} = 10^{-2}$ or even $\xi_\mathrm{rnd} = 10^{-4}$
(see \eqref{eq.rnd}) while keeping the initial magnetic field fixed,
the shear flows evolves very differently.  For small random
perturbations field amplification and overall dynamics proceed
similarly as in 2D models during the KH growth and kinematic
amplifictaion phases regarding the formation of a flux sheet.  Indeed,
the $z$-variation of all physical quantities is very small, The
dynamics of weak-field models is very similar to that of
non-magnetized ones, too. During the KH growth phase, a vortex tube
forms, which is oriented in $z$-direction.

As in 2D models the initial KH growth phase is followed by a kinematic
amplification phase. This phase terminates, as in 2D, depending on
$\Alfv$ and the grid resolution either passively or dynamically by the
back-reaction onto the flow via Maxwell stresses and resistive
instabilities.  The kinematic amplification factor of the magnetic
energy, $f^b$, is the same as in 2D.

\begin{figure}
  \centering
  \includegraphics[width=7cm]{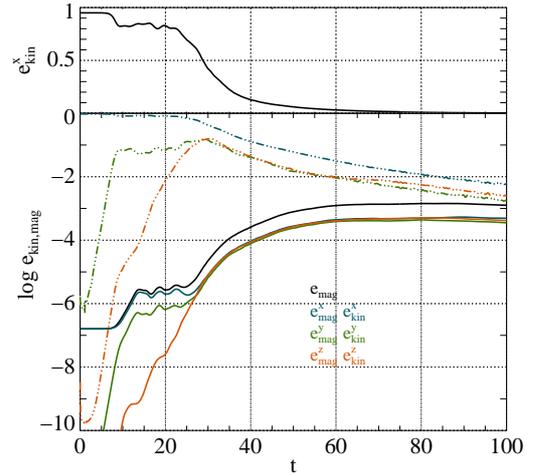}
  \caption{Same as \figref{Fig:KH3--M1-MA50-yz--tevo}, but for a model
    with an initial Alfv\'en number $\Alfv = 5000$.  }
  \label{Fig:KH3--M1-MA5000-y--tevo}
\end{figure}

For an initial Alfv\'en number $\Alfv = 5000$ we find passive
termination of the kinematic field amplification phase
(\figref{Fig:KH3--M1-MA5000-y--tevo}).  Since the magnetic field
remains far too weak to affect the evolution, the dynamics resembles
that of a non-magnetized model.  Until $t \approx 30$, 3D hydrodynamic
instabilities disrupt the KH vortex tube.  Indicative for the
development of these instabilities is the rise of $e_\mathrm{kin}^{z}$
until it reaches equipartition with $e_\mathrm{kin}^{y}$ at $t \approx
28$, growing at a rate comparable to the kinematic growth rate of the
magnetic field.  The volume-averaged total magnetic energy density,
and both $e^x_\mathrm{mag}$ and $e^y_\mathrm{mag}$ remain constant
during this phase, only $e^z_\mathrm{mag}$ increases exponentially.
After termination of the 3D instabilities all volume-averaged magnetic
energy densities are equal, growing slowly during the remaining
evolution. Turbulence spreads across the entire computational volume
and decelerates the shear flow with the same efficiency as in the
non-magnetized model.

For stronger initial fields (or finer grid resolution) the resistive
instabilities terminating the kinematic amplification phase are
accompanied by a rapid growth of the $z$-component of the velocity and
the magnetic field.  For models with $\Alfv = 50$ and $\Alfv = 25$,
this happens at $t \approx 15$.  Despite this rapid growth, the
influence of 3D effects remains moderate. At $t = 15$ close to the end
of the strong rise of $e_\mathrm{mag}^z$ and $e_\mathrm{kin}^z$, the
topology of the velocity field and magnetic field is still dominated
by a large planar structure resembling the flux sheet of 2D
simulations.  

\begin{figure*}
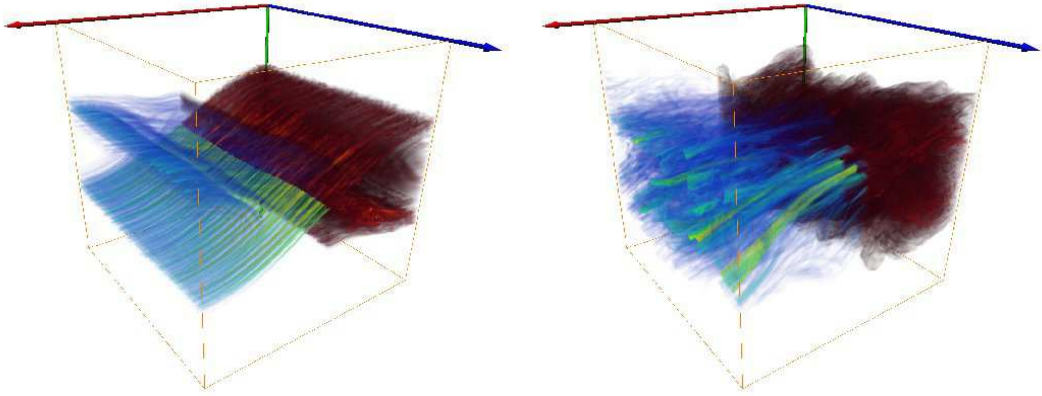

  \centering
  \includegraphics[width=7cm]{./fig/./KH3--M1-MA50-y2-G512.d0150.b_enstrophy.eps}
  \includegraphics[width=7cm]{./fig/./KH3--M1-MA50-y2-G512.d0250.b_enstrophy.eps}
  \caption{Same as \figref{Fig:KH3--M1-MA50-yz--d0030}, but for a
    model where the amplitude of the imposed random perturbation was
    much smaller than that of the sinusoidal one, i.e.,
    $\xi_\mathrm{rnd} << 1$ (see \eqref{eq.rnd}). The two snapshots
    are taken at $t = 15$ (left) and $t = 25$ (right), respectively.
  }
  \label{Fig:KH3--M1-MA50-y2-G512--3d}
\end{figure*}

This is a pronounced difference to the case of large random
perturbations (compare \figref{Fig:KH3--M1-MA50-y2-G512--3d} and
\figref{Fig:KH3--M1-MA50-yz--d0030}). Note, however, that there is
already some indication of the decay of the flux sheet into flux tubes
in the small random perturbation case, too.  After the
dynamic-resistive termination of the kinematic amplification phase,
the $z$-components of the magnetic field and the velocity start
growing again although at a smaller rate, while the $x$ and
$y$-components of the velocity are decelerated by the magnetic field.
The decay of the flux sheet into tubes is almost complete at $t = 25$
(right panel of \figref{Fig:KH3--M1-MA50-y2-G512--3d}) when
$e^y_\mathrm{kin} \approx e^z_\mathrm{kin}$ and $e^y_\mathrm{mag}
\approx e^z_\mathrm{mag}$ holds.  In the subsequent saturation phase,
turbulence develops, and the shear flow is decelerated at a rate
similar to that of the 2D models.

Comparing the properties of the turbulence and the deceleration rate
of 3D models with different initial field strength, different grid
resolution, and different initial perturbations, we find that the
intensity of the turbulent magnetic and velocity fields, and
consequently the deceleration of the shear flow, is determined by the
interplay of (3D) hydrodynamic instabilities, and (2D) magnetic
stresses and instabilities:
\begin{description}
\item[\emph{Hydrodynamic disruption:}] If field amplification is too
  weak to prevent the dominance of hydrodynamic over hydromagnetic
  instabilities during the early evolution, the KH vortex tube is
  disrupted and the shear flow is decelerated at a rate similar to
  that of the non-magnetic case.  The magnetic field is amplified or
  sustained in the turbulent velocity field provided by the
  hydrodynamic instabilities.  The evolution of this class of models
  tends towards isotropic decaying turbulence.
\item[\emph{Hydromagnetic disruption:}] If the magnetic field leads to
  the disruption of the KH vortex tube before hydrodynamic
  instabilities can set in, the deceleration of the shear flow is
  driven by magnetic fields.  In this case, the deceleration rate is
  similar to that of 2D flows, but it may also be smaller depending on
  the level of hydromagnetic turbulence, which is determined among
  other factors by the strength of the initial random perturbations,
  the grid resolution, etc.  The turbulent final state of such models
  is dominated by a larger $x$-component of the magnetic field, the
  transverse components of both fields being considerably smaller.
\end{description}

The two classes of hydrodynamic and hydromagnetic disruption roughly
correspond to the classes of models where $e^z_\mathrm{kin}$ does or
does not exceed $e_\mathrm{mag}$, respectively.  If
$e^z_\mathrm{kin}$ exceeds the volume-averaged total magnetic energy
density after reaching the saturation phase, deceleration enters the
more efficient hydrodynamic regime. Otherwise, deceleration is caused
by the magnetic field.  A given model can undergo a transition from
one class to the other one: for a weak initial field the early
evolution may be dominated by 3D hydrodynamic turbulence, leading to
an efficient deceleration of the shear flow and the magnetic field
remaining at the same level; but when the kinetic energy of the
turbulent flow decreases below that of the magnetic field, the
deceleration rate drops to the hydromagnetic value.

Hence, we can summarize the influence of physical and numerical
parameters on the turbulence and the deceleration as follows:
\begin{itemize}
\item Larger random perturbations favour 3D hydrodynamic
  instabilities.  Comparing for $\Alfv = 50$ a model with
  $\xi_\mathrm{rnd} = 10^{-2}$ and $\xi_\mathrm{rnd} = 10^{-4}$, we
  find significantly stronger magnetic fields and transverse
  velocities for the former model, indicating more vigorous turbulence
  and a faster deceleration of the shear flow.
\item In 2D, weaker initial fields lead to slower deceleration, while
  3D models exhibit a more complex dependence on the initial Alfv\'en
  number.  As discussed above, hydrodynamic instabilities of the KH
  vortex tube dominate in case of very weak fields, leading to very
  rapid deceleration (\figref{Fig:KH3--M1-MA5000-y--tevo}).  If the
  magnetic field is sufficiently strong, i.e., as long as
  $e_\mathrm{mag} > e^z_\mathrm{kin}$ holds, deceleration is initially
  similar to that in 2D for the same initial field strength, but drops
  strongly afterwards.  Due to the deceleration, the $y$ and
  $z$-components of the velocity field reach equipartition.
\item The dependence on the grid resolution is complementary to that
  on the initial Alfv\'en number.  Finer grids allow for a more
  efficient field amplification, and thus favor hydromagnetic over
  hydrodynamic deceleration.
\end{itemize}

According to the 2D simulations, a maximum kinematic amplification is
obtained for a sufficiently fine grid at a given initial field
strength, i.e., increasing the grid resolution does not enhance the
influence of the magnetic field.  Thus, we expect an upper limit for
the importance of magnetic vs. hydrodynamic deceleration corresponding
to the upper limit of the field amplification.  Even for infinite grid
resolution, kinematic amplification of the magnetic field may not lead
to a sufficiently fast field growth to compete with 3D hydrodynamic
instabilities, if the initial field is too weak.  Consequently, we
anticipate only a weak dependence on the magnetic field for large
initial Alfv\'en numbers.

Due to the lack of adequate numerical resolution in 3D, we do not give
any scaling laws for, e.g., $M_{x,y}^{\mathrm{max}}$ and
$t_\mathrm{dec}$ as a function of the initial Alfv\'en number or the
grid resolution.

\subsection{Supersonic shear flows}
\label{sSek:3d-supson}
\begin{figure}
  \centering
  \includegraphics[width=7cm]{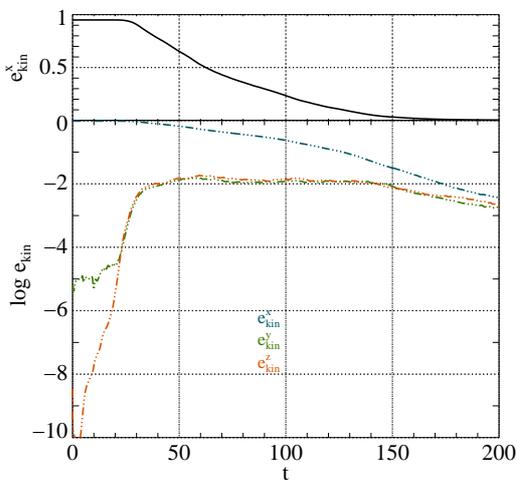}
  \caption{Same as the top panel of
    \figref{Fig:KH3--M1-MAunendlich--tevo}, but for a model with
    $\Mach = 4$ and $b_0 = 0$.  }
  \label{Fig:KH3--M4-MAunendlich--tevo}
\end{figure}

%
Three-dimensional supersonic shear flows show pronounced differences
with respect to 2D ones, the transverse kinetic energy densities
growing much faster in 3D (see \figref{Fig:KH3--M4-MAunendlich--tevo}
for the evolution of a non-magnetized model).  Furthermore, unlike for
subsonic models, 3D hydrodynamic instabilities disrupt supersonic
shear flows, i.e., they are not secondary instabilities feeding off a
KH vortex tube.  For the model shown in
\figref{Fig:KH3--M4-MAunendlich--tevo}, we find that
$e^z_\mathrm{kin}$ grows at a rate similar to the 2D one only until $t
\approx 20$ when it becomes comparable to $e^y_\mathrm{kin}$.
Subsequently, both energy densities grow at the same rate, which is
much faster than the corresponding 2D one.

The 3D instability prevents the shock-mediated formation of a KH
vortex (\secref{sSek:2dhydro}).  Instead of such a coherent
large-scale flow, a rather turbulent flow forms at the shearing
interface expanding in $y$-direction.  Similarly to the 2D case,
shocks develop at some distance from the interface, but these dissolve
when engulfed by the turbulent flow.  Unlike their 2D counterparts,
they play no role in the development of the instability.  During the
saturation phase, the kinetic energy decreases due to efficient
turbulent dissipation.

The interaction of shocks resulting from the usage of reflecting
boundaries is essential for the growth of the instability in 2D.  When
open boundaries allow shocks to leave the computational domain, our 2D
models are stable.  In three dimensions, on the other hand, the
instability does not depend on the presence of these shocks, i.e. the
instability also grows when open boundaries are imposed (in
$y$-direction). Hence, we find a good agreement between simulations of
supersonic models computed with either type of boundary condition.

\begin{figure}
  \centering
  \includegraphics[width=7cm]{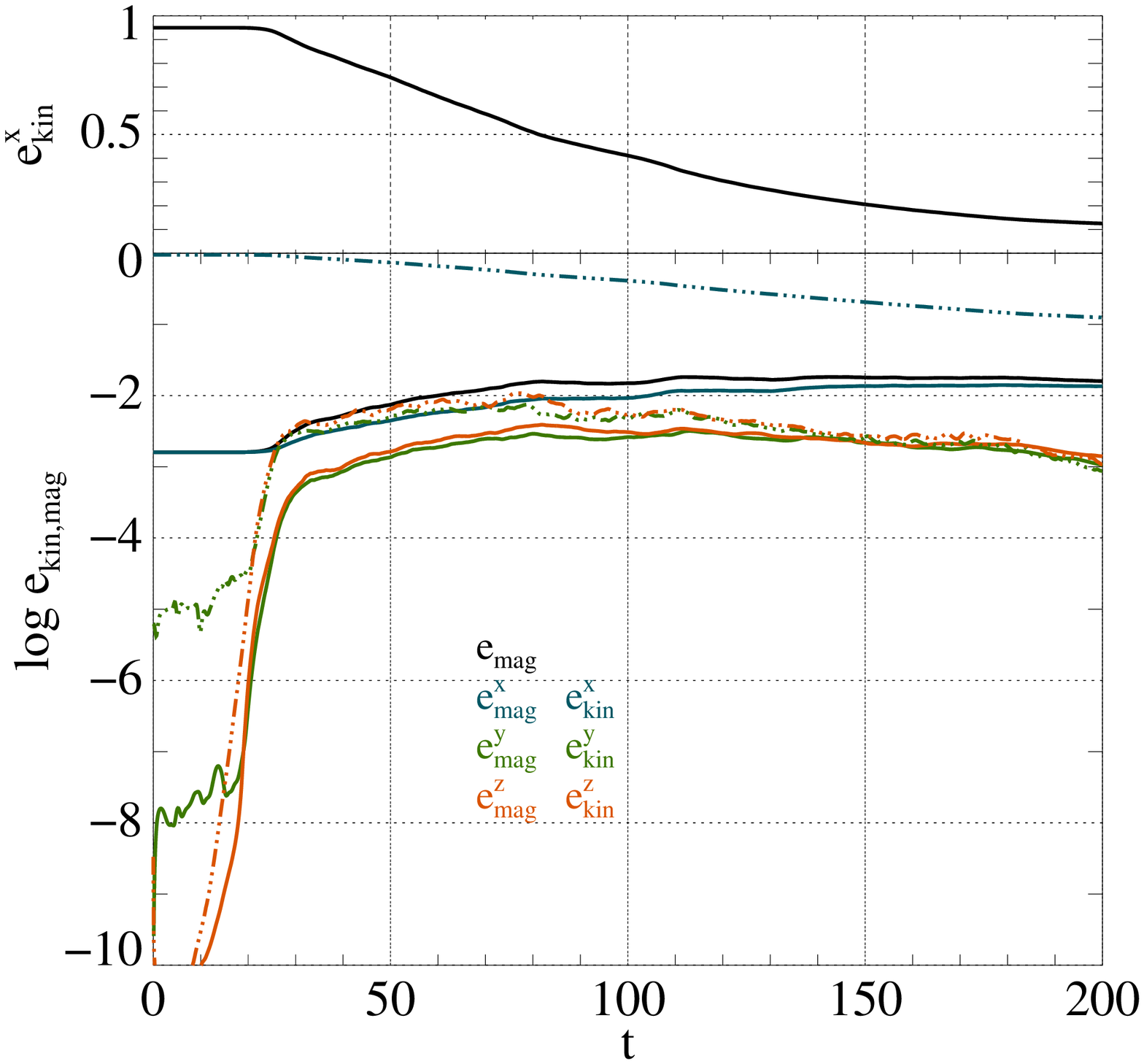}
  \caption{Same as \figref{Fig:KH3--M1-MA50-yz--tevo}, but for a model
    with $\Mach = 4$ and $\Alfv = 50$.  }
  \label{Fig:KH3--M4-MAa--tevo}
\end{figure}

A weak initial magnetic field is amplified at the same rate as the
kinetic energy when the instability develops.  The exponential
amplification ceases when the volume-averaged transverse kinetic
energy densities $e^y_\mathrm{kin}$ and $e^z_\mathrm{kin}$ saturate
(see \figref{Fig:KH3--M4-MAa--tevo}).  Afterwards ($30 \lesssim t
\lesssim 80$), we find only a very gradual growth of the magnetic
energy.  Typically, the volume-averaged transverse kinetic energy
density, $e^{yz}_\mathrm{kin} \equiv e^y_\mathrm{kin} +
e^z_\mathrm{kin}$, is reduced with respect to the non-magnetic case,
but when adding the volume-averaged transverse magnetic energy
density, $e^{yz}_\mathrm{mag} = e^y_\mathrm{mag} + e^z_\mathrm{magn}$,
the total transverse energy density $e^{yz}_\mathrm{kin} +
e^{yz}_\mathrm{mag}$ is at the same level as the transverse kinetic
energy of a non-magnetized model.

The deceleration rate of the shear flow depends, as in the subsonic
case, on the relative importance of hydrodynamic and hydromagnetic
turbulence.  There is, however, a physical difference to the subsonic
case: the supersonic instability is dominated by strong 3D
hydrodynamic turbulence already early on in the evolution, because it
does not result from coherent 2D flows such as a KH vortex.  Hence,
there is no efficient kinematic amplification, and the magnetic field
can become important only if it is maintained or slowly amplified by
the 3D turbulence responsible, at the same time, for a decrease of the
kinetic energy.

\begin{figure*}
  \centering
  \includegraphics[width=7cm]{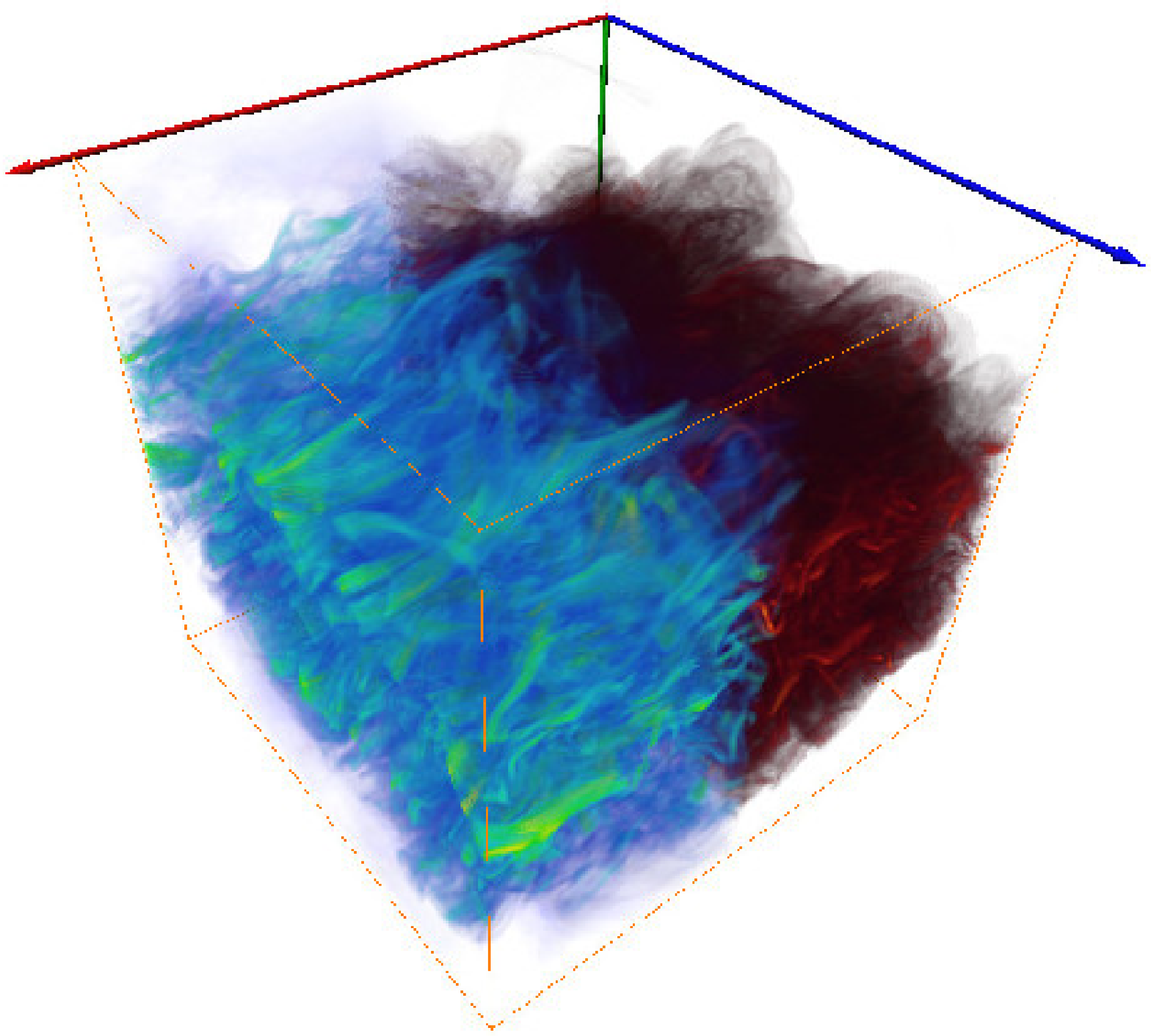}
  \includegraphics[width=7cm]{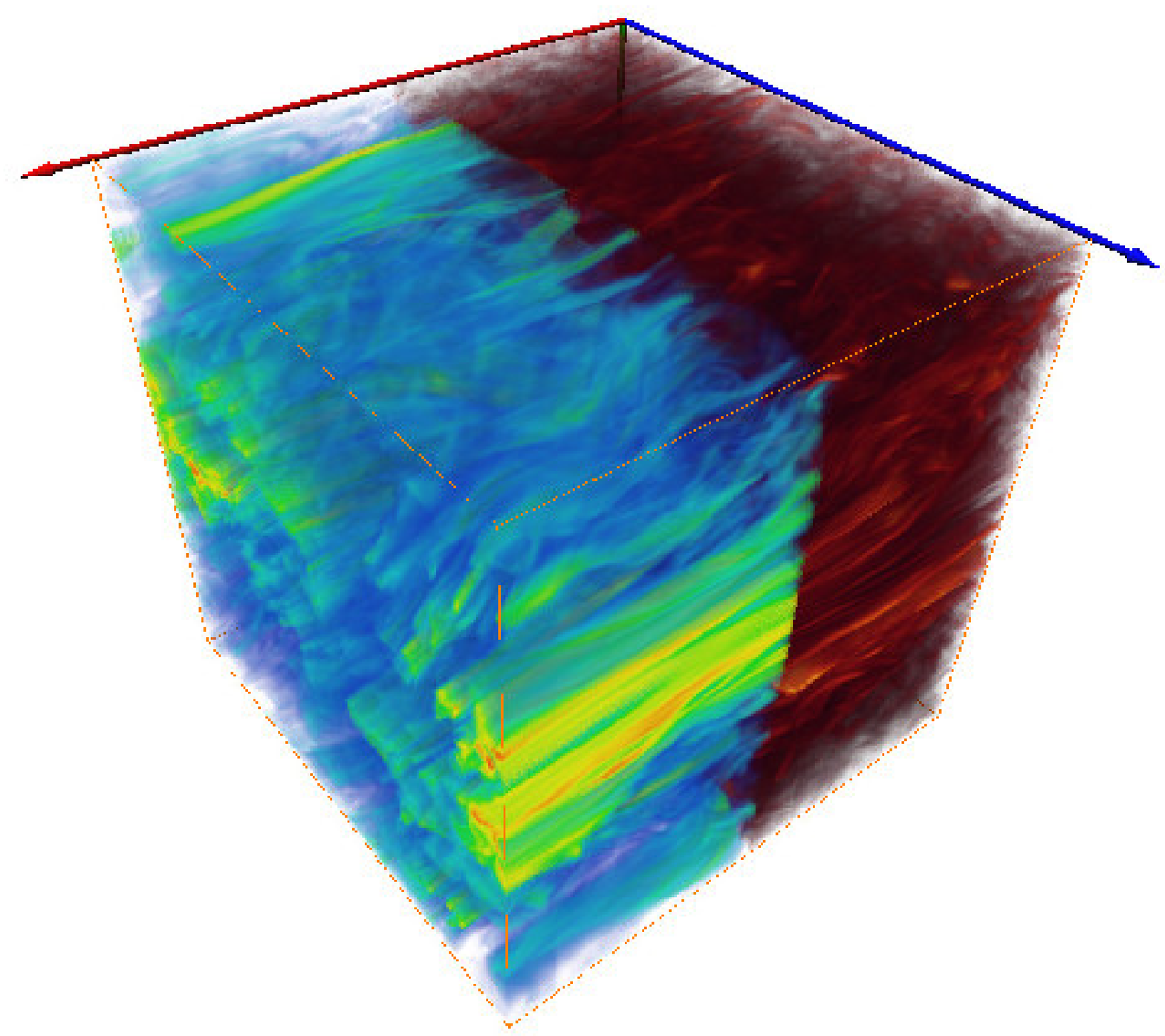}
  \caption{Structure of a model with $\Mach = 4$ and $\Alfv = 50$ (the
    same model as shown in \figref{Fig:KH3--M4-MAa--tevo}) at $t = 60$
    (left panel) and $t = 200$ (right panel).  The two panels show the
    same variables as the ones in
    \figref{Fig:KH3--M1-MA50-y2-G512--3d}, i.e., the volume-rendered
    magnetic field strength (foreground, blue-green) and the modulus
    of the vorticity (background, red).  }
  \label{Fig:KH3--M4-MAa--3dstruct}
\end{figure*}

At an intermediate stage, $t = 60$ (left panel of
\figref{Fig:KH3--M4-MAa--3dstruct}), the instability has not yet
affected the entire computational volume in $y$-direction.  Both the
velocity and the magnetic field of that model exhibit a pronounced
small-scale structure around the initial shearing layer.  No preferred
direction can be identified, and $e^{yz}_\mathrm{kin} >
e^{yz}_\mathrm{mag}$.  This has changed at $t=200$ (right panel), when
due to efficient turbulent deceleration the total kinetic energy
density has decreased by roughly an order of magnitude, similarly to
the transverse magnetic energy $e^{yz}_\mathrm{mag}$.  The
longitudinal magnetic energy density $e^x_\mathrm{mag}$, in contrast,
has remained at the same level with $e^x_\mathrm{mag} >
e^{yz}_\mathrm{mag}$.  The dominance of $e^x_\mathrm{mag}$ and hence
of $b^x$ exerts an ordering influence on the turbulent magnetic and
velocity fields, enforcing an alignment of the flow with the field,
similarly to the Alfv\'en effect of hydromagnetic turbulence.  As a
result, we find prominent coherent structures elongated in field
direction.

\subsection{Anti-parallel magnetic field}
\label{sSek:3d-antip}
%
We have simulated a few of the models discussed above also using
anti-parallel initial magnetic fields.  With the total flux through
surfaces $x = \mathrm{const.}$ vanishing, the $x$-component of the
magnetic field can decay to zero.  This will particularly happen for
weak fields. Stronger fields decay less efficiently because of
resistive instabilities.

For a large initial random perturbation, the evolution is very similar
to models with parallel initial fields.  The shear flow is decelerated
very efficiently, and kinetically dominated decaying turbulence with a
very weak degree of anisotropy develops.  Once the kinetic energy
density approaches the magnetic one, the deceleration rate decreases.
However, it does not tend to zero as in the parallel field case.
Instead of leveling off at a constant value, both the kinetic and
magnetic energy densities continue to decrease at a similar rate.

Models with a small initial random perturbation show, depending on the
initial field strength, hydrodynamic or hydromagnetic deceleration.
The field strength required for hydromagnetic to dominate over
hydrodynamic deceleration is higher than for parallel fields.  In
several models we find at late stages the same evolution as described
above: the kinetic and magnetic energy densities decay at a similar
rate.

\section{Merger-motivated models}
\label{Sek:mrgr}
%
After having discussed basic properties of magnetized shear layers, we
now address simulations mimicking the conditions of shear layers
arising in the merger of two magnetized neutron stars. We assume that
the merging neutron stars heat up so much that any solid crust they
may have developped during their pre-merger evolution has melted, and
the fluid approximation is valid in the shear layer.

\subsection{Physics, initial and boundary conditions}
\label{sSek:mrgr-init}
%
\subsubsection{Equation of state}
%
We employed a simple parametrised equation of state to describe the
thermodynamic properties of neutron star matter
\citep{Keil_Janka_Mueller__1996__ApJL__NS-Convection}.  This hybird
equation of state assumes that the total gas pressure, $P$, is given
by the sum of a barotropic part, $P_{\mathrm{b}}$, and a thermal part,
$P_{\mathrm{th}}$:
\begin{equation}
  P =  P_\mathrm{th} + P_\mathrm{b} 
    \equiv ( \Gamma_\mathrm{th} - 1 ) \varepsilon_\mathrm{th} +
            \kappa \rho^{ \Gamma_\mathrm{b} } \,,
\label{Gl:hyb-EOS}
\end{equation}
where the thermal energy density, $\varepsilon_\mathrm{th}$, is given
by the (total) energy density, $\varepsilon$, and the energy density
of the polytropic component, $\varepsilon_\mathrm{b}$, according to
\begin{equation}
  \label{Gl:hyb-EOS-epsth}
    \varepsilon^{\mathrm{th}} = \varepsilon - \varepsilon_\mathrm{b} \,.
\end{equation}
The sound speed, required for the approximate Riemann solver and for
the determination of the time step, is given by
\begin{equation}
  \label{Gl:hyb-EOS-cs}
  c_\mathrm{s}^2 = \frac{\Gamma_\mathrm{b}  P_\mathrm{b}  +
                        \Gamma_\mathrm{th} P_\mathrm{th}  
                       }{\rho}.
\end{equation}
We used $\Gamma_\mathrm{b} = \Gamma_\mathrm{th} = 1.333$, appropriate
for dense matter whose pressure is dominated by relativistically
degenerate electrons.

\subsubsection{Initial conditions}

With presently available computational resources it is not possible to
perform global simulations of the close encounter or merging of two
magnetized neutron stars with a grid resolution sufficiently high to
resolve also the growth of KH instabilities in shearing magnetized
neutron star matter. Nevertheless one can study some aspects of this
phenomenon by means of local simulations covering only a small volume
around the shear layer.

To this end we consider a quadratic (2D) / cubic (3D) computational
domain in Cartesian coordinates assuming that the $x$-axis is parallel
to the direction of the shear flow, the $y$-axis parallel to the line
connecting the centers of the two neutron stars, and the $z$-axis (in
3D) perpendicular to that line.  As the edges of our computational
domain have a size of 200\,m only, i.e., they are much smaller than
the radius of a neutron star, we consider only homogeneous initial
states, i.e, initial models with constant density and pressure.
Besides the shear flow in $x$-direction the initial models are static,
too. This approximation is justified as the merging neutron stars move
much faster in $x$-direction than they approach each other in
$y$-direction due to the action of gravity.  Accordingly, we use
periodic boundary conditions in $x$ and $z$-direction, and reflecting
ones in $y$-direction.

The shear velocity $v_x$, corresponding to either a Mach number of
$\Mach = 1$ or $\Mach = 4$, has the same $\tanh$-profile as that used
in the simulations of the previous sections, and we also consider both
parallel and anti-parallel initial magnetic field configurations.  The
shear velocity is supposed to mimic the orbital velocity of the two
neutron stars.  We trigger the instability by applying similar
perturbations as in the previous sections, i.e., a combination of a
sinusoidal and a random velocity perturbation.

\subsection{Two-dimensional models}
\label{sSek:mrgr-2d}

A number of models (see \tabref{Tab:mKH2-models}) computed in two
dimensions confirm the basic results discussed in the previous
sections, i.e., the occurrence of three phases, namely KH growth,
kinematic amplification, and saturation. This also holds for the
dependence of the parameters characterizing these phases, e.g., the
termination values of the field strength and magnetic energ, on the
initial data and the grid resolution.

We performed simulations with up to $2048^2$ zones.  The width of the
shear layer was $a = 10\,$m, and the initial velocity $v^x_0 = 1.83$
or $7.2 \times 10^{9}\,$cms/s, for models with $\Mach = 1$ or $\Mach =
4$, respectively.  Due to the affordable grid resolution we employed
rather strong initial fields of the order of $b_0^x \sim 10^{14}\,$G,
corresponding to Alfv\'en numbers $\Alfv \approx 115 \left(
  10^{14}\,\mathrm{G} / b_0^x \right)$.  The initial field was either
parallel or anti-parallel to the shear flow.

We start the discussion with models with $\Mach = 1$.  The KH
instability developed within less than $0.05$\,msec, establishing one
large KH vortex.  Afterwards, the magnetic field is amplified
kinematically by the vortical flow.  The physics of KH growth
termination is the same as that described in \secref{Sek:weakfields}.
Hence, we also find a similar dependence for the field amplification
factor $f_{kin}$ on the initial field strength and the grid
resolution.
\begin{itemize}
\item On finer grids one can resolve the increasingly thin structures
  of the magnetic field better. Consequently, one finds more efficient
  amplification, until for a sufficiently fine grid convergence of
  the amplification factor is achieved.
\item Weaker initial fields are amplified by a larger amount, i.e.,
  the maximum value of the field strength at the end of the KH growth
  phase depends only weakly on the initial field (assuming numerical
  convergence). The total magnetic energy increases with increasing
  initial field strength due to the larger volume filling factor of
  magnetic flux tubes for stronger initial fields.
\end{itemize}

After termination of the kinematic amplification phase, the topology
of the subsequent turbulent saturation phase is dominated by a
multitude of thin flux sheets.  Due to deceleration by magnetic
stresses, the kinetic energy of the shear flow decreases at a rate
depending on the initial field strength.  Lacking a driving force, the
turbulence decays gradually.  At late stages it is dominated by the
parallel component of the magnetic field $b^x$, leading to a strong
alignment of the flux sheets in $x$-direction.

Models with $b_{0}^{x} = 5, \times 10^{13}\,$G and $10\times
10^{13}\,$G reach slightly fluctuating maximum field strengths around
$3 \times 10^{15}\,$G in the saturated state.  The volume filling
factor of the magnetic field, i.e., the relative volume occupied by
intense magnetic flux tubes, decreases with decreasing initial field
strength leading to a weaker mean magnetic field and consequently a
slower deceleration of the shear flow for weaker initial fields.  We
find mean fields of $\sim 5\times 10^{14}\,$G and $\sim 2.5\times
10^{14}\,$G for $b_0^x = 10^{14}\,$G and $b_0^x = 5\times 10^{13}\,$G,
respectively.  The time scale for deceleration of the shear flow is
less than 1 millisecond.  For a model with an initial field of $2
\times 10^{14}\,$G, the deceleration is sufficiently rapid to cause a
significant decay (by about an order of magnitude) of the turbulent
energy within $0.5\,$msec.

The evolution of the shear layer is affected by the choice of the
initial field configuration.  Parallel initial fields have, similarly
to our observations above, a somewhat larger impact on the dynamics of
the KH instability.  In this case, the non-vanishing magnetic flux
threading surfaces $x = const.$ is conserved due to the boundary
conditions, and gives rise to an effective driving force.  Apart from
lacking this additional driver, anti-parallel magnetic fields are
prone to stronger dissipation due to presence of stronger currents at
the boundaries between regions of opposite magnetic polarity.

The evolution of models with a supersonic shear flow ($\Mach = 4$) is
similar to that of their dimensionless conterparts discussed
previously.  With initial fields between $10$ and $40 \times 10^{13} ~
\mathrm{G}$, the initial Alfv\'en numbers of the shear flow are
between $\sim 110$ and $\sim 440$, i.e., in the range covered in
\secref{Sek:supson}.  The dynamics is the same: pressure waves steepen
into oblique shocks, and the dissipation of kinetic into thermal
energy in these shocks creates a broad transition layer between the
two regions of positive and negative $v_x$.  The shear flow is
decelerated very efficiently even for very weak fields.  We find
3field amplification up to $5 ... 10 \times 10^{15} ~ \mathrm{G}$ for
the maximum field strength and $1 ... 2 \times 10^{15} ~ \mathrm{G}$
for the volume-averaged r.m.s.~value of the field, $\sqrt {1
  /\mathcal{V} \int \mathrm{d} \mathcal{V}\, \vec b^2}$, i.e., of the
same order as in the case $\Mach = 1$ but systematically higher by a
factor $\sim 2$, with considerably higher values for parallel than for
anti-parallel initial fields.

Hence, the results and in particular their dependence on the physical
and numerical parameters of the models explored in \secref{Sek:KH2},
are robust with respect to the described variations of the initial
conditions.  Consequently, we can expect them to apply to merger
systems without too strong modifications.

\subsection{Three-dimensional models}
\label{sSek:mrgr-3d}

One of the main questions to be addressed by 3D simulations is whether
the dynamics of these models is dominated by magnetic flux tubes or by
3D hydrodynamic instabilities.  As we have seen in the previous
sections, this has a distinct influence on, e.g., the magnetic field
strength achieved at saturation.

For the 3D simulations we used grids of up to $m_x \times m_y \times
m_z = 256^3$ zones.  The initial field strength was between $5$ and
$40 \times 10^{13}\,$G.  We again applied different combinations of
sinusoidal and random velocity perturbations to the shear layer.

The models (see \tabref{Tab:mKH3-models}) show the same overall
dynamics and the same evolutionary phases as the corresponding models
discussed in \secref{Sek:KH2}.  We find the initial KH growth phase,
the kinematic amplification phases, followed by the development of
parasitic instabilities leading to a non-linear saturated state.  The
flow during the first two phases is very similar to that in 2D, and
field amplification follows the same trends with initial field and
grid resolution as outlined above.  The further evolution depends, as
discussed above, strongly on the relative amplitude of random and
sinusoidal perturbations.

When a small random perturbation is imposed, field amplification
proceeds through the first two growth phases the field strength being
limited by its back reaction onto the flow.  These models suffer (if
resolved well on a sufficiently fine grid) hydromagnetic instabilities
of the flux sheet, leading to the break-up of the KH vortex tube and
the deceleration of the shear flow.  For a well-resolved model with
$b_0^x = 2\times 10^{13}\,$G the maximum magnetic field strength is
$\approx 9 \times 10^{15}\,$G, while the r.m.s.~maximum is only
$\approx 9 \times 10^{14}\,$G.

For models with large random perturbations and for models with very
weak initial fields the disruption of the KH vortex tube is
predominantly due to hydrodynamic instabilities leading to a very
efficient deceleration of the shear flow.  These instabilities grow on
a very short time scale, causing a strong growth of the
volume-averaged transverse kinetic energy densities $e^y_\mathrm{kin}$
and $e^z_\mathrm{kin}$, as well as of all volume-averaged magnetic
energy densities ($e^x_\mathrm{mag}, e^y_\mathrm{mag}$, and
$e^z_\mathrm{mag}$).

The amplification factors are similar for all components of the field,
leading to equipartition among them at peak magnetic energy.  The
amplification rate is at first very large but decreases strongly as
the parasitic instabilities saturate.  Eventually, the magnetic energy
reaches a maximum, and then starts to decrease again.  This maximum
depends either on the grid resolution (for the most weakly magnetized
models), or on the dynamic back-reaction of the field onto the flow.
In the latter case, field amplification ceases once the
volume-averaged transverse kinetic energy densities (decaying from
their maximum values at saturation of the parasitic instabilities)
decrease to roughly the level of the volume-averaged magnetic energy
density.

The magnetic field is amplified during all three growth phases: at the
KH growth rate during the KH growth phase, at a (smaller) rate
determined by the overturning velocity of the KH vortex tube during
the kinematic amplification phase, and during the growth of the
parasitic instabilities.  Since the magnetic field starts to decrease
shortly after saturation of the parasitic instabilities feeding off
the shear flow, the maximum field strength is reached at that moment.
The magnetic field energy can reach at most equipartition with the
(decaying) transverse kinetic energy, which typically has a value of
$\sim 10^{43}\,$erg. For a model with $b_0^x = 4 \times 10^{14}\,$G
the corresponding root mean square saturation field is $\sim 1.6\times
10^{15}\,$G.  The weaker the initial field, the smaller is the maximum
magnetic energy, since the achievable amplification factor is limited
by the duration of the three field amplification phases.  The maximum
field strength reached anywhere in the computational domain depends
only weakly on the initial field, and has a value between $6$ and $10
\times 10^{15}\,$G.

The decay of the turbulence (measured by the transverse kinetic and
magnetic energy densities) as well as that of the shear flow starts at
a similar rate for all models, and the magnetic field decreases much
faster than it does in the corresponding 2D models.  Shortly after
($\sim 0.05\,$msec) the r.m.s.\ field strength as well as the total
field strength reach their peak values early during the saturation
phase, the kinetic energy densities decay very rapidly, and much
faster than the magnetic energy density.  The decay slows down shortly
after $e^y_\mathrm{kin} + e^z_\mathrm{kin}$ has decreased below the
value of $e_\mathrm{mag}$.  Afterwards, all transverse energy
densities decay at a similar rate.  In the more strongly magnetized
models this happens when $e^x_\mathrm{kin}$ (which can undergo a phase
of particularly fast decay) is still larger than $e^y_\mathrm{kin} +
e^z_\mathrm{kin}$, whereas it is usually the other way round for
weaker initial fields (a similar effect can be observed for
under-resolved models where insufficient grid resolution limits the
field amplification).  During the further evolution the relative sizes
of $e^x_\mathrm{kin}$, $e^y_\mathrm{kin}$, and $e^z_\mathrm{kin}$
remain unchanged.
 
\begin{figure*}
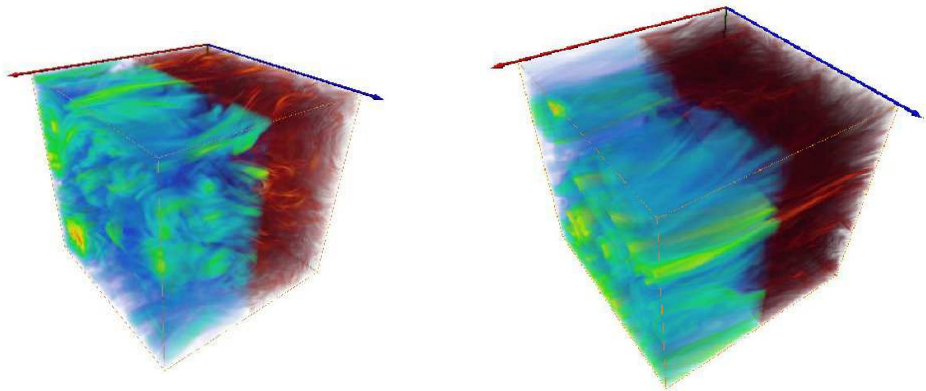

  \centering
  \includegraphics[width=7.5cm,bb=0 0 575 368]{./fig/./mKH3--nox-bx05--g256--d1000.b_vor.eps}
  \includegraphics[width=6cm]{./fig/./mKH3--nox-bx20--g256--d1000.b_vor.eps}
  \caption{3D structure of the final turbulent state of models with
    $b_0^x = 5\times 10^{13}\,$G (left panel) and $b_0^x = 20 \times
    10^{13}\,$G (right panel) at $t = 1\,$msec, respectively.  The
    panels show the volume-rendered magnetic field strength (front
    half; blue-green-yellow-red corresponding to increasing values of
    $|\vec b|$) and enstrophy (rear half; red-yellow corresponding to
    increasing values of $(\vec{\nabla} \times \vec v)^2$).  The red
    and blue long arrows mark the $x$ and $z$-direction, respectively.
  }
  \label{Fig:mKH3--3dstruct}
\end{figure*}

For weak initial fields, ($b_0^x = 1$, $5$, and $10 \times
10^{13}\,$G, the kinetic energy density dominates the magnetic one by
a factor $\approx 2.7$ in the final state.  Concerning the
volume-averaged kinetic energy densities we find $e^x_\mathrm{kin}
\approx e^z_\mathrm{kin} \approx 2.4 e^y_\mathrm{kin}$. This relation
also holds for the volume-averaged magnetic energy densities,
indicating a relatively high degree of isotropy of the turbulence.
The final state for $b_0^x = 5\times 10^{13}\,$G is shown in the left
panel of \figref{Fig:mKH3--3dstruct}. Obviously, neither the flow nor
the magnetic field show any preferred direction. Instead, one
recognizes a complex pattern of tangled small-scale flux tubes.

For sufficiently strong initial fields ($b_0^x = 20$, and $40 \times
10^{13}\,$G), the final state is more strongly magnetized.  As the
turbulent energy decays more rapidly than the $x$-component of the
magnetic field, $b_x$ dominates the dynamics after $t \approx
15\,$msec leading to a slower deceleration of the shear flow and a
more pronounced alignment of flow features (flux and vorticity tubes)
in $x$-direction (see \figref{Fig:mKH3--3dstruct}, right panel).

Similarly to the 2D models discussed above, a parallel initial
magnetic field has a stronger influence on the dynamics than an
anti-parallel one: the field strength reaches a higher maximum value,
and the influence of hydrodynamic instabilities is slightly less.  At
late stages, such models may exhibit a phase of hydromagnetic
deceleration, in contrast to the roughly constant value of
$e^x_\mathrm{kin}$ in models with strong anti-parallel initial fields.

For of supersonic shear flows with $\Mach = 4$, the evolution is
similar to that of the dimensionless models discussed in
\secref{sSek:3d-supson}.  We find a fast growth of 3D hydrodynamic
instabilities disrupting the shear flow before the shock-mediated
mechanism working in 2D can operate.  Turbulence sets in quickly
without the intermediate development of a KH vortex, and the shear
flow is decelerated very efficiently.  The maximum magnetic fields we
find are of the order of $1 \times 10^{16} ~ \mathrm{G}$ for the
absolute maximum, and $3 \times 10^{15} ~ \mathrm{G}$ for the
r.m.s.~field, respectively.  These values are rather insensitive to
the initial field strength and geometry.

\section{Summary and conclusions}
\label{Sek:Summary}

Global simulations indicate that the contact layer between two merging
neutron stars is a site of very efficient field amplification.  The
layer is prone to the Kelvin-Helmholtz instability, and thus,
exponential growth of any weak seed field is possible, as observed by
\cite{Price_Rosswog__2006__Sci__NS-NS-merger-B-amplif} (see also
\cite{Giacomazzo_Rezzolla_Baiotti__2009__PRL__NS_mergers_MHD,
  Anderson_etal__2008__PRL__NS_mergers_MHD_GW,
  Liu_etal__2008__prd__GRMHD_NS_mergers}).  The limitations of their
simulations, mainly concerning grid resolution, did not allow these
authors to determine the saturation level of the instability firmly
and accurately. Thus, the implications of magnetic fields for the
merger dynamics remains unclear.  On the basis of energetic arguments,
the instability might lead to a field in equipartition with the
kinetic or the internal energy of the shear flow, corresponding to
field strengths of the order of $10^{16}\,$G or $10^{18}\,$G,
respectively.

We reassessed these arguments by means of local high-resolution
simulations of magnetized shear layers in two and three spatial
dimensions.  To this end we performed more than 220 simulations
focusing on properties of the hydromagnetic KH instability in general
as well as on the contact surfaces of merging neutron stars.  We refer
to these two classes of simulations as \emph{dimensionless} and
\emph{merger-motivated} models, respectively.

We employed a recently developed multi-dimensional Eulerian
finite-volume ideal MHD code based on high-order spatial
reconstruction techniques and Riemann solvers of the MUSTA-type
\citep{Obergaulinger__2008__PhD__RMHD,
  Obergaulinger_etal__2009__AA__Semi-global_MRI_CCSN}.

We set up a KH-unstable shear flow in Cartesian coordinates in a
quadratic (2D) and cubic (3D) computational domain imposing periodic
boundary conditions in the direction of the shear flow and reflecting
or open ones in the transverse directions.  Focussing on the effects
of a magnetic field on the instability, we used a simplified equation
of state (ideal gas EOS and a hybrid barotropic/ideal gas EOS for the
dimensionless and the merger-motivated models, respectively) and
neglected additional physics, e.g., such as neutrino transport.

Under these simplifications, the shear flows are characterized by two
parameters, the initial Mach number $\Mach$ and the initial Alfv\'en
number $\Alfv$, measuring the magnitude of the jump in shear velocity
in units of the sound speed and the Alfv\'en velocity, respectively.

Analytic considerations and previous simulations of non-magnetized
shear flows show that the growth rate of the KH instability as well as
its saturation level (i.e., the kinetic energy of the circular KH
vortex formed by the instability) increase with increasing $\Mach$ for
subsonic shear flows .  A magnetic field is known to reduce the growth
rate and potentially, i.e., for $\Alfv < 2$, even to suppress the
instability \citep{Chandrasekhar__1961__Buch__HD-HM-stab,
  Miura_Pritchett__1982__JGR__MHD-KH-stability,
  Keppens_etal__1999__PP__MHD-KH}.  However, less is known about the
saturation level and the dynamic back-reaction, in particular for weak
initial fields.

\cite{Frank_etal__1996__ApJ__MHD-KH-2d-1,Jones_etal__1997__ApJ__MHD-KH-2d-2,
  Jeong_etal__2000__Apj__MHD-KH-2d-3, Ryu_etal__2000__ApJ__MHD-KHI-3d}
studied the evolution of the hydromagnetic KH instability in two and
three dimensions.  In 2D, models undergo a transition from
\emph{non-linear stabilization} of the KH vortex to its violent
\emph{disruption} or more gradual \emph{dissipation} when the initial
field strength is reduced, while in 3D purely hydrodynamic
\emph{elliptic} instabilities of the vortex tube may dominate over MHD
effects.

The study of these non-linear effects is hampered by high requirements
on the grid resolution that is necessary to follow the development of
increasingly thin magnetic flux sheets and tubes.  This limits the
range of Alfv\'en numbers for which numerical convergence can be
achieved to rather modest values. It also reduces the predictive power
for merger systems, where rather weak initial fields are expected.
This limitation can be overcome only when using large grids in
combination with a highly accurate code.  We evolved subsonic,
transsonic, and supersonic shear flows with $\Mach \in [0.5; 1; 4]$,
while using the maximum Alfv\'en numbers for which convergence is
achievable.  The resulting broad range of Alfv\'en numbers covered by
our simulations allows us to establish scaling laws governing the
field amplification as a function of the initial field strength.

The main results of our simulations are:
\begin{enumerate}
\item In 2D, we confirm the results of analytic work (in the linear
  regime) and previous simulations concerning the growth rate and the
  saturation of the transverse kinetic energy densities for strong
  initial fields due to
  \cite{Chandrasekhar__1961__Buch__HD-HM-stab,Miura_Pritchett__1982__JGR__MHD-KH-stability,Keppens_etal__1999__PP__MHD-KH}.
  This agreement supports the viability of our numerical approach for
  the problem at hand.
\item For subsonic shear flows ($\Mach = 0.5, 1$) we explored a wide
  range of initial field strengths covering Alfv\'en numbers up to
  $\Alfv = 5000$ in 2D.
  \begin{enumerate}
  \item For intermediate and weak fields, we distinguish two phases:
    the KH growth phase during which the field grows at the KH growth
    rate, and after formation of a KH vortex, a phase of kinematic
    field amplification by the overturning vortex.  The growth rate
    during the latter phase depends on the velocity of the vortex.
    The field is highly intermittent and concentrated in flux sheets,
    which are stretched by the flow leading to an exponential growth
    of the field strength while the sheet width decreases.
  \item The termination of the kinematic amplification phase occurs
    either numerically, when the flux sheets get too thin to be
    resolved on a given computational grid, or dynamically by
    back-reaction of the field onto the flow.  The most important mode
    of back-reaction is the growth of secondary resistive
    instabilities feeding off the magnetic energy of the flux sheets.
    These instabilities terminate the kinematic field growth and
    initiate the non-linear saturation phase during which the KH
    vortex is destroyed by the ensuing MHD turbulence and the shear
    flow is gradually decelerated.  This scenario is equivalent to
    that of the \emph{disruption} models of
    \cite{Frank_etal__1996__ApJ__MHD-KH-2d-1}.
  \item We quantified the amount of field amplification during the
    kinematic amplification phase by computing the ratio of the
    volume-averaged Maxwell stress component $M_{xy}$ at the beginning
    and at the end of that phase.  The amplification factor scales
    with the initial \Alfven\,number as $\Alfv^{3/4}$, corresponding
    to a scaling of the maximum Maxwell stress with the initial field
    strength as $b_0^{5/4}$.  If the simulation is under-resolved, the
    amplification factor is reduced by a factor $\propto m^{7/8}$ ($m$
    being the number of zones per dimension).  The maximum local field
    strength corresponds to a \emph{local} equipartition between the
    magnetic energy density of a flux sheet and the kinetic energy
    density of the shear flow; it depends only weakly on the initial
    field.
  \item The secondary resistive instabilities observed in our
    simulations are triggered by numerical resistivity instead of a
    physical one.  The numerical resistivity, which is a function of
    the grid resolution $\Delta$, is important only for small thin
    structures having a spatial size of the order of $\Delta$ or less.
    In our simulations, it causes current sheets to become unstable
    when their width approaches the grid spacing $\Delta$. Although
    only simulations with arbitrarily high resolution can sustain
    arbitrarily thin and intense current sheets, we observe
    nevertheless convergence: the field amplification becomes
    independent of the grid resolution, if $\Delta$ is smaller than
    some threshold which depends on the initial field strength.  The
    reason for this independence is the fact that the most unstable
    current sheets do not consist of individual flux sheets but of
    pairs or triples of coalescing flux sheets.  Thus, decreasing the
    distance between flux sheets does not lead to a stronger field
    (which would be the case, if a single flux sheet is compressed in
    transverse direction).
  \item The disruption of the vortex and the efficient dissipation set
    these \emph{resisto-dynamic} models apart from the class of
    \emph{dissipation} models with even weaker initial fields where
    the KH vortex remains intact, and only very slow dissipation is
    provided by turbulence.  In the simulations of
    \cite{Frank_etal__1996__ApJ__MHD-KH-2d-1}, secondary instabilities
    do not modify the flow field qualitatively.  Our simulations
    indicate that this is, partially at least, a resolution effect.
    If a simulation is under-resolved and the field growth is not
    limited by dynamic back-reaction but by the resolvable width of
    flux sheets, no disruption will occur, and the deceleration time
    of the shear flow is very long.  Converged simulations show, on
    the other hand, the disruption of the KH vortex by secondary
    magnetic instabilities when the magnetic field strength approaches
    a local maximum close to equipartition with the kinetic energy
    density of the shear flow. This happens in all converged models,
    but for weak initial fields, the deceleration time can be very
    long.
  \item Models with initially anti-parallel and parallel magnetic
    fields, but otherwise identical, give qualitatively similar
    results, the above discussed effects being somewhat less
    pronounced in case of the former field configuration.
  \end{enumerate}
\item The contact layer of merging neutron stars resembles supersonic
  shear flows.  In principle, these are stable.  We find, however,
  that an exponentially growing instability may occur when closed
  boundary conditions are imposed in the direction transverse to the
  shear flow.  The instability is mediated by shock waves traveling
  through the computationl domain. The corresponding growth rates are
  much smaller than for subsonic shear flows.  The effects of a
  magnetic field on a supersonic shear flow are qualitatively similar
  to those on subsonic shear flows.
\item In 3D the disruption of the KH vortex tube can be induced by a
  purely hydrodynamic secondary so-called \emph{elliptic} instability
  as discussed, e.g., by \cite{Ryu_etal__2000__ApJ__MHD-KHI-3d}.  It
  leads to a very rapid growth of the kinetic energy densities
  corresponding to all components of the flow velocity once the KH
  vortex tube forms, and decelerates the shear flow more efficiently
  than the MHD mechanisms outlined above.  Which of the two possible
  disruption mechanisms, elliptic or hydromagnetic, operates depends
  on the initial field strength $b_0$ and the value of volume-averaged
  kinetic energy density $e^z_\mathrm{kin}$.  The magnetic mechanism
  will dominate only if $e_\mathrm{mag} > e^z_\mathrm{kin}$, i.e., as
  long as the magnetic energy density exceeds the transverse kinetic
  energy in $z$-direction.  Due to the very fast growth of the
  elliptic instability, this may be the case only for a short time, if
  at all.  A rather strong initial field and small perturbations in
  $z$-direction are required for a hydromagnetic disruption.
\item 2D and 3D simulations of shear flows with merger-motivated
  initial conditions performed in a cubic computational domain of
  constant density and pressure having an edge size of $200\,$m show
  the same overall dynamics as corresponding dimensionless models.
  The initial Mach number of the shear flow was chosen to be $\Mach =
  1$ and $\Mach = 4$ corresponding to a density of $10^{13}\,$g
  cm$^{-3}$, and shear velocities of $1.83 \times 10^{9}\,$cm/s, and
  $7.2 \times 10^{9}\,$cm/s, respectively. The initial magnetic field
  strength was varied between $5\times 10^{13}\,$G and $4 \times
  10^{14}\,$G.
  \begin{enumerate}
  \item The instability grows rapidly: saturation occurs within
    $\lesssim 0.1\,$msec, and the disruption and deceleration times
    are much less than $1\,$msec.
  \item The dynamics is the same as that of the dimensionless models.
    Field amplification leads to a maximum field strength $\lesssim
    10^{16}\,$G, and a r.m.s.\ value of $\lesssim 1.6 \times
    10^{15}\,$G.  These values are the same for 3D models suffering
    hydrodynamic and hydromagnetic disruption.
  \end{enumerate}
\end{enumerate}

From our results, we may draw a few conclusions concerning the growth
and the influence of magnetic fields in neutron-star mergers.  The
foremost implication is that the maximum field strength, independent
whether it refers to a single point or a spatial average, is not
amplified to equipartition with the thermal energy density.  We can,
hence, exclude saturation fields of the order of $10^{18}\,$G in the
contact layers of neutron star mergers.

Instead, local equipartition with the kinetic energy density is
reached with corresponding maximum fields $\sim 10^{16}\,$G, as
speculated by \cite{Price_Rosswog__2006__Sci__NS-NS-merger-B-amplif}.
Due to the high degree of intermittency in the case of weak initial
fields, the (r.m.s.) average of the field strength is smaller, i.e,
its direct dynamic impact (e.g., disruption of the KH vortex tube or
deceleration of the shear flow) on the flow is probably rather
limited.  This is even more the case if the geometry of the system and
the perturbations resulting from the merger dynamics enhance the
importance of purely hydrodynamic instabilities.  More indirect
effects can, however, not be excluded, e.g., whether magnetic flux
tubes created at the shear layer are transported rapidly far away by
large-scale flows.  The short period of time during which the magnetic
field stays close to its maximum value and its fast decay impose
severe constraints on the impact that the amplified fields may have on
any hydromagnetic or electromagnetic jet-launching mechanism in a
merger of two neutron stars.  We note that magnetically driven
relativistic outflows may need much longer time scales ($\sim$ a few
msec) to tap the rotational energy of either the black hole or the
accretion disk resulting after the merger.

Though these results limit the prospect for magnetic effects to play a
dynamic role in neutron star mergers, their proper inclusion in
current and forthcoming simulations may be necessary, because magnetic
fields influence the dissipation rates in the shear layer, i.e., their
neglect may lead to an underestimation of the temperature in the shear
layer, and hence in the accretion disk.  Given the resolution
requirements imposed by weak initial fields, a more sophisticated
treatment of the problem probably also has to abandon the assumption
of ideal MHD and to consider the formulation of a turbulence model for
unresolved magnetic field structures.

\begin{acknowledgements}
%
  This research has been supported by the Spanish {\it Ministerio de
    Educaci\'on y Ciencia} (grants AYA2007-67626-C03-01,
  CSD2007-00050), and by the Collaborative Research Center on {\it
    Gravitational Wave Astronomy} of the Deutsche
  Forschungsgemeinschaft (DFG SFB/Transregio 7).  MAA is a Ram\'on y
  Cajal fellow of the {\em Ministerio de Educaci\'on y Ciencia}.  Most
  of the simulations were performed at the Rechenzentrum Garching
  (RZG) of the Max-Planck-Society. We are also thankful for the
  computer resources, the technical expertise, and the assistance
  provided by the Barcelona Supercomputing Center - Centro Nacional de
  Supercomputaci\'on.  Parts of this article have been written during
  M.O.'s visit to the Departamento de Astronom{\'i}a y Astrof{\'i}sica
  of the Universidad de Valencia. He wants to express his gratitude
  for the kind hospitality experienced there.
\end{acknowledgements}

\appendix

\section{Tables of models}
\label{Sek:Tables}

We provide tables listing the parameters and important properties of
the models computed:
\begin{description}
\item[\tabref{Tab:n2d-grow-models}] lists the parameters of models
  which we computed to compare the growth rates obtained numerically
  with theoretical predictions, serving as code validation.
\item[\tabref{Tab:n2d-HD-models}] lists 2D hydrodynamic models of
  transonic and supersonic shear flows.
\item[\tabref{Tab:n2d-weak-models}] and \tabref{Tab:kh2--Abbremsung}
  list the amplification factors of the magnetic field and the
  disruption and deceleration rates of models with weak initial
  fields, respectively.
\item[\tabref{Tab:nkh3-models}] lists the initial data of 3D
  dimensionless models.
\item[\tabref{Tab:mKH2-models}] and \tabref{Tab:mKH3-models} list the
  initial conditions of 2D and 3D merger-motivated models,
  respectively.
\end{description}

\begin{table*}[htbp]
  \centering
  \caption{Summary of models computed to compare numerical growth
    rates with theoretical predictions.  The colums give the model
    name, the size of the domain ($l_x$, $l_y$), the initial pressure,
    $P_0$, the velocity shear, $U_0$, the corresponding Mach number
    $\Mach = U_0 / c_\mathrm{s}$, the initial magnetic field $\vec
    b_0$, the initial width of the shear flow, $a$, the corresponding
    wave number, $k_x$, the growth rate, $\Gamma_\mathrm{MP}$,
    obtained from \cite{Miura_Pritchett__1982__JGR__MHD-KH-stability},
    and an estimate of the numerical growth rate,
    $\Gamma_\mathrm{num}$.
  }
  \begin{tabular}{l|cc|c|cccc|c|c|cc}
    \hline\hline
    name & $l_x$ & $l_y$ & 
    $m_x \times m_y$ 
    & $P_0$ & $U_0$ & $\Mach$
    & $a$ 
    & $\vec b_0$
    & $k_x$
    & $\Gamma_\mathrm{MP}$ & $\Gamma_\mathrm{num}$
    \\
    \hline\hline
    grw-1 
    & 1 & 2 & $ 50 \times 100$ 
    & 1 & 1.29 & 1 
    & 0.05 
    & $(0,0,0)$
    & $2\pi$ 
    & 1.73 & 1.64
    \\
    grw-2 
    & 1 & 2 & $100 \times 200$ 
    & 1 & 1.29 & 1 
    & 0.05 
    & $(0,0,0)$
    & $2\pi$ 
    & 1.73 & 1.74
    \\
    grw-3 
    & 1 & 2 & $200 \times 400$ 
    & 1 & 1.29 & 1 
    & 0.05 
    & $(0,0,0)$
    & $2\pi$ 
    & 1.73 & 1.75
    \\
    grw-4 
    & 1 & 2 & $400 \times 800$ 
    & 1 & 1.29 & 1 
    & 0.05 
    & $(0,0,0)$
    & $2\pi$ 
    & 1.73 & 1.75
    \\
    \hline
    grw-5 
    & 1 & 2 & $200 \times 400$ 
    & 1 & 1.29 & 1 
    & 0.025 
    & $(0,0,0)$
    & $2\pi$ 
    & 2.4 & 2.44
    \\
    grw-6 
    & 1 & 2 & $200 \times 400$ 
    & 1 & 1.29 & 1 
    & 0.1
    & $(0,0,0)$
    & $2\pi$ 
    & 0.66 & 0.68
    \\
    \hline
    grw-7 
    & 1 & 2 & $200 \times 400$ 
    & 1 & 0.645 & 0.5 
    & 0.05 
    & $(0,0,0)$
    & $2\pi$ 
    & 1.09 & 1.07
    \\
    grw-8 
    & 1 & 2 & $200 \times 400$ 
    & 1 & 1.843  & $10/7$ 
    & 0.05 
    & $(0,0,0)$
    & $2\pi$ 
    & 1.77 & 1.79
    \\
    \hline
    grw-9 
    & 1 & 2 & $200 \times 400$ 
    & 1 & 0.645 & 0.5 
    & 0.05 
    & $(0,0,0)$
    & $4\pi$ 
    & 1.36 & 1.35
    \\
    \hline
    grw-10
    & 1 & 2 & $200 \times 400$ 
    & 1 & 1.29 & 1
    & 0.05 
    & $(0.129,0,0)$
    & $2\pi$ 
    & 1.69 & 1.70
    \\
    grw-11
    & 1 & 2 & $200 \times 400$ 
    & 1 & 1.29 & 1
    & 0.05 
    & $(0.258,0,0)$
    & $2\pi$ 
    & 1.56 & 1.54
    \\
    \hline\hline
  \end{tabular}
  \label{Tab:n2d-grow-models}
\end{table*}

\begin{table*}[htbp]
  \centering
  \caption{Summary of 2D hydrodynamic supersonic models.  The table
    entries are the same data as Tab.\,\ref{Tab:n2d-grow-models} with
    the following exceptions: the column $\vec b_0$ is skipped, and we
    do not list a theoretical value of the growth rate. Instead, we
    give our choice of boundary conditions in the transverse direction
    in column ``BC''.  In the last column, we indicate models for
    which the instability grows oscillatory by a confirmation mark,
    $\surd$.  Note that model grw-3 of \tabref{Tab:n2d-grow-models}
    corresponds to model HD2r-0 with open boundaries.
  }

  \begin{tabular}{l|cc|c|cccc|c|c|cc}
    \hline\hline
    name & $l_x$ & $l_y$ & 
    $m_x \times m_y$ 
    & $P_0$ & $U_0$ & $\Mach$ 
    & $a$ 
    & $k_x$
    & BC
    & $\Gamma_\mathrm{num}$ & oscillations
    \\

    \hline\hline
    HD2o-1-l
    & 1 & 4 & $200 \times 800$ 
    & 1 & 2.322 & 1.8
    & 0.05
    & $2\pi$
    & open
    & 0.97 & 
    \\
    HD2o-1
    & 1 & 2 & $200 \times 400$ 
    & 1 & 2.322 & 1.8
    & 0.05
    & $2\pi$
    & open
    & 0.96 & 
    \\
    HD2o-1-i
    & 1 & 1 & $200 \times 200$ 
    & 1 & 2.322 & 1.8
    & 0.05
    & $2\pi$
    & open
    & 0.73 &
    \\
    HD2o-1-s
    & 1 & 0.5 & $200 \times 100$ 
    & 1 & 2.322 & 1.8
    & 0.05
    & $2\pi$
    & open
    & 0.16 & $\surd$
    \\
    HD2o-2
    & 1 & 2 & $200 \times 400$ 
    & 1 & 2.451 & 1.9
    & 0.05
    & $2\pi$
    & open
    & 0.30 & $\surd$
    \\
    HD2o-3
    & 1 & 2 & $200 \times 400$ 
    & 1 & 2.5155 & 1.95
    & 0.05
    & $2\pi$
    & open
    & 0.26 & $\surd$
    \\
    HD2o-4
    & 1 & 2 & $200 \times 400$ 
    & 1 & 2.58 & 2
    & 0.05
    & $2\pi$
    & open
    & 0
    \\
    HD2o-5
    & 1 & 2 & $200 \times 400$ 
    & 1 & 5.16 & 4
    & 0.05
    & $2\pi$
    & open
    & 0
    \\

    \hline
    HD2r-0
    & 1 & 2 & $200 \times 400$ 
    & 1 & 1.29 & 1
    & 0.05
    & $2\pi$
    & reflecting
    & 1.73 & 
    \\
    HD2r-1
    & 1 & 2 & $200 \times 400$ 
    & 1 & 2.322 & 1.8
    & 0.05
    & $2\pi$
    & reflecting
    & 0.96 & 
    \\
    HD2r-1-i
    & 1 & 1 & $200 \times 200$ 
    & 1 & 2.322 & 1.8
    & 0.05
    & $2\pi$
    & reflecting
    & 0.56 & 
    \\
    HD2r-1-s
    & 1 & 0.5 & $200 \times 100$ 
    & 1 & 2.322 & 1.8
    & 0.05
    & $2\pi$
    & reflecting
    & 0.56 & $\surd$
    \\
    HD2r-1-S
    & 1 & 0.25 & $200 \times 50$ 
    & 1 & 2.322 & 1.8
    & 0.05
    & $2\pi$
    & reflecting
    & 0.35 & $\surd$
    \\
    HD2r-4
    & 1 & 2 & $200 \times 400$ 
    & 1 & 2.58 & 2
    & 0.05
    & $2\pi$
    & reflecting
    & 0.46 & $\surd$
    \\
    HD2r-4-HR
    & 1 & 2 & $400 \times 800$ 
    & 1 & 2.58 & 2
    & 0.05
    & $2\pi$
    & reflecting
    & 0.44 & $\surd$
    \\
    HD2r-5
    & 1 & 2 & $200 \times 400$ 
    & 1 & 5.16 & 4
    & 0.05
    & $2\pi$
    & reflecting
    & 0.52 & $\surd$
    \\
    \hline\hline

  \end{tabular}

  \label{Tab:n2d-HD-models}
\end{table*}

\begin{table*}[htbp]
  \caption{Parameters of the weak-field models: the columns give the
    initial Mach number, $\Mach$, the shear-layer width, $a$, the
    initial magnetic field strength, $b_0^x$, the corresponding
    \Alfven~number, $\Alfv$, and the amplification factors $f^e$ (for
    the magnetic energy) and $f^b$ (for the field strength),
    respectively.  The models were simulated on grids of $m = 256,
    \ldots, 4096$ zones per dimension.  }
\centering
  \begin{tabular}{lc|rr||rr|rr|rr|rr|rr}
    \hline\hline
    $\Mach$
    & $a$
    & $b_0^x$ & $\Alfv$
    & \multicolumn{2}{c|}{256}
    & \multicolumn{2}{c|}{512}
    & \multicolumn{2}{c|}{1024}
    & \multicolumn{2}{c|}{2048}
    & \multicolumn{2}{c}{4096}

    \\

    &
    & $\left[ 10^{-4} \right]$ &
    & $f^{e}$ & $f^{b}$
    & $f^{e}$ & $f^{b}$
    & $f^{e}$ & $f^{b}$
    & $f^{e}$ & $f^{b}$
    & $f^{e}$ & $f^{b}$
    
    \\
    \hline\hline

    0.5 & 0.05 & $200$ & 25
    & 20.2 & 29.4
    & 22.9 & 30.6
    & 25.9 & 29.3
    & 27.7 & 28.3

    \\

    0.5 & 0.05 & $100$ & 50
    & 24.4 & 40.4
    & 33.5 & 57.2
    & 39.8 & 66.2
    & 43.5 & 64.3
    & 46.3 & 63.3

    \\

    0.5 & 0.05 & $50$ & 100
    & 27.0 & 50.0
    & 41.0 & 75.6
    & 55.3 & 102.2
    & 66.8 & 123.7
    & 73.3 & 125.3

    \\

    0.5 & 0.05 & $20$ & 250
    & 35.0 & 51.0
    & 44.4 & 95.0
    & 70.4 & 146.4
    & 105.3 & 213.0
    & & 

    \\
    \hline

    1 & 0.10 & $200$ & 50
    & 25.2 & 36.5
    & 33.6 & 50.2
    & 46.0 & 46.5
    & 45.0 & 49.2
    & & 

    \\

    1 & 0.10 & $40$ & 250
    & 18.2 & 37.6
    & 49.3 & 83.8
    & 74.5 & 132.2
    & 113.9 & 201.3
    &  & 

    \\
    \hline

    1 & 0.15 & $200$ & 50
    & 17.2 & 29.3
    & 27.8 & 39.7
    & 30.7 & 40.0
    & 35.9 & 46.4
    &  & 

    \\

    1 & 0.15 & $100$ & 100
    & 19.6 & 34.8
    & 35.0 & 56.3
    & 54.9 & 76.0
    & 61.2 & 81.5
    &  & 

    \\

    1 & 0.15 & $40$ & 250
    & 21.3 & 46.6
    & 40.2 & 69.5
    & 65.3 & 106.3
    & 103.9 & 152.4
    &  & 

    \\
    \hline

    1 & 0.20 & $200$ & 50
    & 5.8 & 14.2
    & 8.0 & 28.0
    & 12.8 & 26.0
    & 22.5 & 36.3
    &  & 
 
    \\

    1 & 0.20 & $40$ & 250
    & 6.4 & 35.7
    & 11.8 & 41.3
    & 18.1 & 62.2
    & 33.0 & 106.6
    &  & 

    \\
    \hline

    1 & 0.05 & $400$ & 25
    & 16.8 & 23.8
    & 19.6 & 25.9
    & 22.0 & 26.6
    & 23.3 & 25.4
    &  & 

    \\

    1 & 0.05 & $200$ & 50
    & 19.4 & 45.7
    & 27.5 & 46.2
    & 32.0 & 48.6
    & 36.1 & 51.5
    & 39.4 & 53.6

    \\

    1 & 0.05 & $80$ & 125
    & 20.2 & 35.6
    & 33.4 & 70.2
    & 50.1 & 96.6
    & 61.6 & 117.3
    & 67.3 & 118.9

    \\

    1 & 0.05 & $40$ & 250
    & 20.9 & 50.8
    & 37.0 & 88.0
    & 59.9 & 127.7
    & 83.6 & 178.5
    & 104.1 & 210.9

    \\

    1 & 0.05 & $20$ & 500
    & 21.2 & 55.1
    & 39.4 & 103.1
    & 63.0 & 153.1
    & 101.1 & 236.4
    & 145.6 & 330.8

    \\

    1 & 0.05 & $8$ & 1250
    & 21.2 & 56.4
    & 40.1 & 127.2
    & 67.7 & 187.8
    & 109.8 & 288.7
    & 169.0 & 444.4

    \\

    1 & 0.05 & $2$ & 5000
    & 21.2 & 55.2
    & 40.3 & 136.4
    & 68.6 & 218.5
    & 112.4 & 314.7
    & 182.9 & 515.2

    \\
    \hline\hline

  \end{tabular}

  \label{Tab:n2d-weak-models}
\end{table*}

\begin{table*}
  \centering
  \caption{Same as \tabref{Tab:n2d-weak-models}, but instead of the
    amplification factors we give the disruption time of the KH
    vortex, $t_\mathrm{dis}$, and the absolute value of the
    deceleration rate, $\sigma_{\mathrm{dec};3} = |\sigma_\mathrm{dec}
    / 10^{-3}|$, for simulations with $m = 256, \ldots, 4096$ zones
    per dimension.  We indicate simulations where no disruption is
    observed by a hyphen in the column for $t_\mathrm{dis}$,
    simulations where the determination of $\sigma_\mathrm{dec}$ is
    very inaccurate by a $\sim$ sign preceding the value of
    $\sigma_{\mathrm{dec}; 3}$, and simulations where we found no
    measurable deceleration by a hyphen in the column for
    $\sigma_{\mathrm{dec}; 3}$.  }

  \begin{tabular}{lc|rr||rr|rr|rr|rr|rr}
    \hline\hline
    $\Mach$
    & $a$
    & $b_0^x$ & $\Alfv$
    & \multicolumn{2}{c|}{256}
    & \multicolumn{2}{c|}{512}
    & \multicolumn{2}{c|}{1024}
    & \multicolumn{2}{c|}{2048}
    & \multicolumn{2}{c}{4096}

    \\
    \hline

    &
    & $\left[ 10^{-4} \right]$ &
    & $t_{\mathrm{dis}}$ & $\sigma_{\mathrm{dec};3}$
    & $t_{\mathrm{dis}}$ & $\sigma_{\mathrm{dec};3}$
    & $t_{\mathrm{dis}}$ & $\sigma_{\mathrm{dec};3}$
    & $t_{\mathrm{dis}}$ & $\sigma_{\mathrm{dec};3}$
    & $t_{\mathrm{dis}}$ & $\sigma_{\mathrm{dec};3}$
    
    \\
    \hline\hline

    0.5 & 0.05 & $200$ & 25
    & 7.6 & 19.0
    & 7.6 & 23.0
    & 7.6 & 18.6
    & 7.6 & 22.4

    \\

    0.5 & 0.05 & $100$ & 50
    & 14.4 & 10.7
    & 13.7 & 11.7
    & 12.7 & 14.3
    & 12.6 & 11.6
    &  & 

    \\

    0.5 & 0.05 & $50$ & 100
    & 80.1 & 4.1
    & 45.4 & 5.9
    & 23.4 & 7.4
    & 22.9 & 10.7
    & 22.6 & 10.3

    \\

    0.5 & 0.05 & $20$ & 250
    & -- & $\sim 0.4$
    & -- & $\sim 0.9$
    & -- & 3.5
    & 77.4 & 4.1

    \\
    \hline

    1 & 0.15 & $200$ & 50
    & 4.5 & 19.0
    & 4.3 & 16.9
    & 4.0 & 16.9
    & 4.1 & 24.1
    &  & 

    \\

    1 & 0.15 & $100$ & 100
    & 23.0 & 6.8
    & 15.0 & 13.4
    & 6.5 & 17.1
    & 6.7 & 18.1
    &  & 

    \\

    1 & 0.15 & $40$ & 250
    & -- & $\sim 0.17$
    & -- & 2.7
    & 58.5 & 4.4
    & 21.9 & 6.1
    &  & 

    \\
    \hline

    1 & 0.05 & $400$ & 25
    & 3.8 & 23.9
    & 3.8 & 22.6
    & 3.8 & 45.0
    & 3.8 & 41.1
    &  & 

    \\

    1 & 0.05 & $200$ & 50
    & 12.4 & 16.8
    & 9.9 & 14.1
    & 6.1 & 27.8
    & 6.0 & 23.0
    &  & 

    \\

    1 & 0.05 & $80$ & 125
    & 75.6 & 4.8
    & 25.3 & 8.7
    & 18.5 & 11.6
    & 12.0 & 15.2
    & 12.0 & 12.9

    \\

    1 & 0.05 & $40$ & 250
    & -- & $\sim 0.9$
    & -- & 1.8
    & 62.5 & 4.1
    & 39.8 & 5.6
    & 39.8 & 5.6

    \\

    1 & 0.05 & $20$ & 500
    & -- & --
    & -- & --
    & -- & $\sim 0.8$
    & -- & 2.4
    & 99.5 & 3.1

    \\

    1 & 0.05 & $8$ & 1250
    & -- & --
    & -- & --
    & -- & --
    & -- & $\sim 0.5$
    & -- & $\sim 0.8$

    \\

    1 & 0.05 & $2$ & 5000
    & -- & --
    & -- & --
    & -- & --
    & -- & --
    & -- & --

    \\
    \hline\hline

  \end{tabular}
  \label{Tab:kh2--Abbremsung}
\end{table*}

\begin{table*}
  \centering
  \caption{List of 3D models: the columns give the initial shear
    velocity, $U_0$, Mach number, $\Mach$, magnetic field strength,
    $b_0$, and Alfv\'en number $\Alfv$. The models were simulated on
    grids of $128^3$ to $512^3$ zones using parallel ( $+$ sign) and
    anti-parallel ($\pm$ sign) initial field configurations,
    respectively.  Most of the models were simulated several times
    using different initial perturbations.  }
  \label{Tab:nkh3-models}

  \begin{tabular}{cc|rr||cccc}

    \hline\hline

    $U_0$ & $\Mach$
    & \multicolumn{1}{c}{$b_0^x$} & \multicolumn{1}{c||}{$\Alfv$}
    & 128 & 256 & 512
    
    \\

    &
    & $10^{-4}$ &
    & & & 
    
    \\
    \hline

    1 & 1 & 0 & $\infty$
    & $+$ & & 
    \\

    1 & 1 & 400 & 25
    & $+$, $\pm$ & $+$, $\pm$ & 
    \\

    1 & 1 & 200 & 50
    & $+$, $\pm$ & $+$, $\pm$ & $+$
    \\

    1 & 1 & 2 & 5000
    & $+$ & &
    \\

    \hline

    1 & 4 & 0 & $\infty$
    & $+$ & $+$ &
    \\

    1 & 4 & 400 & 25
    & $+$ & $+$ &
    \\

    1 & 4 & 200 & 50
    & $+$ & $+$ &
    \\

    1 & 4 & 100 & 100
    & $+$ & $+$ &
    \\

    1 & 4 & 20 & 500
    & $+$ & $+$ &
    \\

    \hline\hline
    
  \end{tabular}
\end{table*}

\begin{table*}
  \centering
  \caption{List of 2D merger-motivated models simulated on grids of
    $1024^2$ and $2048^2$ zones, respectively.  Each simulated model
    is indicated by a $\surd$ sign.  The initial shear profile had a
    maximum velocity of $v_0^x = 1.83 \times 10^{9}\,$cm/s (i.e., the
    Mach number of the shear flow is $\Mach = 1$), and a width of $a =
    20\,$m.  The first column lists the initial field strength,
    marking models with anti-parallel initial fields by a $\pm$
    preceding the numerical value.  Most models were simulated using
    different initial perturbations, and most models were additionally
    simulated on coarser grids.  }
 \begin{tabular}{r||cc|cc}
    \hline\hline
    \multicolumn{1}{c||}{$b_0^x$ [$10^{13}\,$G]} 
    & \multicolumn{2}{c|}{$\Mach = 1$} 
    & \multicolumn{2}{c}{$\Mach = 4$} 
    \\
    \multicolumn{1}{c||}{} & $1024^2$ & $2048^2$ & $1024^2$
    \\
    \hline

    $\pm 5$ & $\surd$ & & 
    \\
    $10$ & $\surd$ & $\surd$
    \\
    $\pm 10$ & $\surd$ & $\surd$ & $\surd$ 
    \\
    $20$ & $\surd$ &  & $\surd$
    \\
    $\pm 20$ & $\surd$ & $\surd$ & $\surd$
    \\
    $40$ & & & $\surd$
    \\
    $\pm 40$ & & & $\surd$
    \\
    \hline\hline

  \end{tabular}
  \label{Tab:mKH2-models}
\end{table*}

\begin{table*}
  \centering
 \caption{Same as \tabref{Tab:mKH2-models} but for the 3D
   merger-motivated models simulated on grids of $128^3$ and $256^3$
   zones, respectively. }
   \begin{tabular}{r||cc|cc}
    \hline\hline
    \multicolumn{1}{c||}{$b_0^x$ [$10^{13}\,$G]} 
    & \multicolumn{2}{c|}{$\Mach = 1$} 
    & \multicolumn{2}{c}{$\Mach = 4$} 
    \\
    \multicolumn{1}{c||}{} & $128^3$ & $256^3$ & $128^3$ & $256^3$
    \\
    \hline
    $5$ & $\surd$ & $\surd$
    \\
    $\pm 5$ & $\surd$ & $\surd$
    \\
    $10$ & $\surd$ & $\surd$ & $\surd$ & $\surd$
    \\
    $\pm 10$ & $\surd$ & $\surd$ & $\surd$ & $\surd$
    \\
    $20$ & $\surd$ & $\surd$ & $\surd$ & $\surd$
    \\
    $\pm 20$ & $\surd$ & $\surd$ & $\surd$ & $\surd$
    \\
    $40$ &         & $\surd$ & $\surd$ & $\surd$
    \\
    $\pm 40$ &         & $\surd$ & $\surd$ & $\surd$
    \\
    \hline\hline

  \end{tabular}
  \label{Tab:mKH3-models}
\end{table*}

\bibliographystyle{aa}
\bibliography{./biblio.bib}

\begin{thebibliography}{35}
\expandafter\ifx\csname natexlab\endcsname\relax\def\natexlab#1{#1}\fi

\bibitem[{{Agertz} {et~al.}(2007){Agertz}, {Moore}, {Stadel}, {Potter},
  {Miniati}, {Read}, {Mayer}, {Gawryszczak}, {Kravtsov}, {Nordlund}, {Pearce},
  {Quilis}, {Rudd}, {Springel}, {Stone}, {Tasker}, {Teyssier}, {Wadsley}, \&
  {Walder}}]{Agertz_etal__2007__MNRAS__SPH-grid-comparison}
{Agertz}, O., {Moore}, B., {Stadel}, J., {et~al.} 2007, \mnras, 380, 963

\bibitem[{{Anderson} {et~al.}(2008){Anderson}, {Hirschmann}, {Lehner},
  {Liebling}, {Motl}, {Neilsen}, {Palenzuela}, \&
  {Tohline}}]{Anderson_etal__2008__PRL__NS_mergers_MHD_GW}
{Anderson}, M., {Hirschmann}, E.~W., {Lehner}, L., {et~al.} 2008, Physical
  Review Letters, 100, 191101

\bibitem[{{Baty} {et~al.}(2003){Baty}, {Keppens}, \&
  {Comte}}]{Baty_etal__2003__PhysPlas__2dMHD-KH-compress}
{Baty}, H., {Keppens}, R., \& {Comte}, P. 2003, Physics of Plasmas, 10, 4661

\bibitem[{{Biskamp}(2000)}]{Biskamp__2000__Buch__Reconnection}
{Biskamp}, D. 2000, {Magnetic Reconnection in Plasmas} (Magnetic reconnection
  in plasmas, Cambridge, UK: Cambridge University Press, 2000 xiv, 387
  p.~Cambridge monographs on plasma physics, vol.~3, ISBN 0521582881)

\bibitem[{{Chandrasekhar}(1961)}]{Chandrasekhar__1961__Buch__HD-HM-stab}
{Chandrasekhar}, S. 1961, {Hydrodynamic and hydromagnetic stability}
  (International Series of Monographs on Physics, Oxford: Clarendon, 1961)

\bibitem[{{Evans} \& {Hawley}(1988)}]{Evans_Hawley__1998__ApJ__CTM}
{Evans}, C.~R. \& {Hawley}, J.~F. 1988, \apj, 332, 659

\bibitem[{{Frank} {et~al.}(1996){Frank}, {Jones}, {Ryu}, \&
  {Gaalaas}}]{Frank_etal__1996__ApJ__MHD-KH-2d-1}
{Frank}, A., {Jones}, T.~W., {Ryu}, D., \& {Gaalaas}, J.~B. 1996, \apj, 460,
  777

\bibitem[{{Gardiner} \& {Stone}(2005)}]{Gardiner_Stone__2005__JCP__Athena-code}
{Gardiner}, T.~A. \& {Stone}, J.~M. 2005, \jcop, 205, 509

\bibitem[{{Gardiner} \& {Stone}(2008)}]{Gardiner_Stone__2008__JCP__Athena-code}
{Gardiner}, T.~A. \& {Stone}, J.~M. 2008, \jcop, 227, 4123

\bibitem[{{Giacomazzo} {et~al.}(2009){Giacomazzo}, {Rezzolla}, \&
  {Baiotti}}]{Giacomazzo_Rezzolla_Baiotti__2009__PRL__NS_mergers_MHD}
{Giacomazzo}, B., {Rezzolla}, L., \& {Baiotti}, L. 2009, \mnras, 399, L164

\bibitem[{{Harten}(1983)}]{Harten__1983__JCP__HR_schemes}
{Harten}, A. 1983, Journal of Computational Physics, 49, 357

\bibitem[{{Iroshnikov}(1964)}]{Iroshnikov__1964__SovietAstronomy__Turbulence_o%
f_a_Conducting_Fluid_in_a_Strong_Magnetic_Field}
{Iroshnikov}, P.~S. 1964, Soviet Astronomy, 7, 566

\bibitem[{{Jeong} {et~al.}(2000){Jeong}, {Ryu}, {Jones}, \&
  {Frank}}]{Jeong_etal__2000__Apj__MHD-KH-2d-3}
{Jeong}, H., {Ryu}, D., {Jones}, T.~W., \& {Frank}, A. 2000, \apj, 529, 536

\bibitem[{{Jones} {et~al.}(1997){Jones}, {Gaalaas}, {Ryu}, \&
  {Frank}}]{Jones_etal__1997__ApJ__MHD-KH-2d-2}
{Jones}, T.~W., {Gaalaas}, J.~B., {Ryu}, D., \& {Frank}, A. 1997, \apj, 482,
  230

\bibitem[{{Keil} {et~al.}(1996){Keil}, {Janka}, \&
  {M{\"u}ller}}]{Keil_Janka_Mueller__1996__ApJL__NS-Convection}
{Keil}, W., {Janka}, H.-T., \& {M{\"u}ller}, E. 1996, \apjl, 473, L111+

\bibitem[{{Keppens} {et~al.}(1999){Keppens}, {T{\'o}th}, {Westermann}, \&
  {Goedbloed}}]{Keppens_etal__1999__PP__MHD-KH}
{Keppens}, R., {T{\'o}th}, G., {Westermann}, R.~H.~J., \& {Goedbloed}, J.~P.
  1999, Journal of Plasma Physics, 61, 1

\bibitem[{{Kraichnan}(1965)}]{Kraichnan__1965__PhysicsofFluids__Inertial-Range%
_Spectrum_of_Hydromagnetic_Turbulence}
{Kraichnan}, R.~H. 1965, Physics of Fluids, 8, 1385

\bibitem[{{LeVeque}(1992)}]{LeVeque_Book_1992__Conservation_Laws}
{LeVeque}, R.~J. 1992, Numerical Methods for Conservation Laws, 2nd edn.,
  Lectures in mathematics - ETH Z\"urich (Birkh\"auser)

\bibitem[{{Levy} {et~al.}(2002){Levy}, {Puppo}, \&
  {Russo}}]{Levy_etal__2002__SIAM_JSciC__WENO4}
{Levy}, D., {Puppo}, G., \& {Russo}, G. 2002, SIAM J.~Sci.~Comput., 24, 480

\bibitem[{{Liu} {et~al.}(1994){Liu}, {Osher}, \& {Chan}}]{Liu_etal__1994__WENO}
{Liu}, X.-D., {Osher}, S., \& {Chan}, T. 1994, \jcop, 115, 200

\bibitem[{{Liu} {et~al.}(2008){Liu}, {Shapiro}, {Etienne}, \&
  {Taniguchi}}]{Liu_etal__2008__prd__GRMHD_NS_mergers}
{Liu}, Y.~T., {Shapiro}, S.~L., {Etienne}, Z.~B., \& {Taniguchi}, K. 2008,
  \prd, 78, 024012

\bibitem[{{Londrillo} \& {del
  Zanna}(2004)}]{Londrillo_Del_Zanna__2004__JCP__UpwindCT}
{Londrillo}, P. \& {del Zanna}, L. 2004, \jcop, 195, 17

\bibitem[{{Miura} \&
  {Pritchett}(1982)}]{Miura_Pritchett__1982__JGR__MHD-KH-stability}
{Miura}, A. \& {Pritchett}, P.~L. 1982, \jgr, 87, 7431

\bibitem[{{Monaghan}(1992)}]{Monaghan__1992__ARAA__SPH}
{Monaghan}, J.~J. 1992, \araa, 30, 543

\bibitem[{{Obergaulinger}(2008)}]{Obergaulinger__2008__PhD__RMHD}
{Obergaulinger}, M. 2008, PhD thesis, Technische Universit{\"a}t M{\"u}nchen

\bibitem[{{Obergaulinger} {et~al.}(2009){Obergaulinger}, {Cerd{\'a}-Dur{\'a}n},
  {M{\"u}ller}, \& {Aloy}}]{Obergaulinger_etal__2009__AA__Semi-global_MRI_CCSN}
{Obergaulinger}, M., {Cerd{\'a}-Dur{\'a}n}, P., {M{\"u}ller}, E., \& {Aloy},
  M.~A. 2009, \aap, 498, 241

\bibitem[{{Oechslin} {et~al.}(2007){Oechslin}, {Janka}, \&
  {Marek}}]{Oechslin_etal__2007__AA__NS-NS-merger--EOS-1}
{Oechslin}, R., {Janka}, H.-T., \& {Marek}, A. 2007, \aap, 467, 395

\bibitem[{{Orszag} \& {Tang}(1979)}]{Orszag_Tang__1979__JFM__MHD_turb}
{Orszag}, S.~A. \& {Tang}, C.-M. 1979, Journal of Fluid Mechanics, 90, 129

\bibitem[{{Price} \&
  {Rosswog}(2006)}]{Price_Rosswog__2006__Sci__NS-NS-merger-B-amplif}
{Price}, D.~J. \& {Rosswog}, S. 2006, Science, 312, 719

\bibitem[{{Rosswog}(2007)}]{Rosswog__2007__RevMex__NS-merger}
{Rosswog}, S. 2007, in Revista Mexicana de Astronomia y Astrofisica Conference
  Series, Vol.~27, Revista Mexicana de Astronomia y Astrofisica, vol. 27,
  57--79

\bibitem[{{Ryu} \& {Jones}(1995)}]{Ryu_Jones__1995__ApJ__MHD_1d}
{Ryu}, D. \& {Jones}, T.~W. 1995, \apj, 442, 228

\bibitem[{{Ryu} {et~al.}(2000){Ryu}, {Jones}, \&
  {Frank}}]{Ryu_etal__2000__ApJ__MHD-KHI-3d}
{Ryu}, D., {Jones}, T.~W., \& {Frank}, A. 2000, \apj, 545, 475

\bibitem[{{Suresh} \& {Huynh}(1997)}]{Suresh_Huynh__1997__JCP__MP-schemes}
{Suresh}, A. \& {Huynh}, H. 1997, \jcop, 136, 83

\bibitem[{{Titarev} \& {Toro}(2005)}]{Titarev_Toro__2005__IJNMF__MUSTA}
{Titarev}, V.~A. \& {Toro}, E.~F. 2005, International Journal for Numerical
  Methods in Fluids, 49, 117

\bibitem[{{Toro} \& {Titarev}(2006)}]{Toro_Titarev__2006__JCP__MUSTA}
{Toro}, E.~F. \& {Titarev}, V.~A. 2006, \jcop, 216, 403

\end{thebibliography}

\end{document}